\documentclass[trackchanges, twocolumn]{aastex701}

\usepackage{subcaption}
\usepackage{CJKutf8}

\begin{document}

\title{Aarmed with Data: Bumps, Outflows, and Disk-like Emission in TDE 2025aarm}


\author[orcid=0000-0002-9085-8187,gname=Aysha, sname=Aamer]{Aysha Aamer}
\affiliation{Department of Astronomy and Steward Observatory, University of Arizona, 933 North Cherry Avenue, Tucson, AZ 85721-0065, USA}
\email[show]{aamer2@arizona.edu}


\author[gname=Kate D., sname=Alexander]{Kate D. Alexander}
\affiliation{Department of Astronomy and Steward Observatory, University of Arizona, 933 North Cherry Avenue, Tucson, AZ 85721-0065, USA}
\email{kdalexander@arizona.edu}

\author[gname=Charlotte R., sname=Angus]{Charlotte R. Angus}
\affiliation{Astrophysics Research Centre, School of Mathematics and Physics, Queen's University Belfast, Belfast BT7 1NN, UK}
\email{c.angus@qub.ac.uk}

\author[gname=Panos, sname=Charalampopoulos]{Panos Charalampopoulos}
\affiliation{Institute of Space Sciences (ICE-CSIC), Carrer de Can Magrans, 08193 Cerdanyola del Vallès, Barcelona, Spain}
\email{panos.charalampopoulos.astro@gmail.com}

\author[orcid=0000-0001-9570-0584, gname=Megan, sname=Newsome]{Megan Newsome}
\affiliation{Department of Astronomy, The University of Texas at Austin, 2515 Speedway, Stop C1400, Austin, TX 78712, USA}
\email{newsome.megane@gmail.com}

\author[orcid=0000-0001-7090-4898, gname=Iair, sname=Arcavi]{Iair Arcavi}
\affiliation{The School of Physics and Astronomy, Tel Aviv University, Tel Aviv 69978, Israel}
\email{arcavi@gmail.com}

\author[orcid=0000-0002-8597-0756, gname=Giorgos, sname=Leloudas]{Giorgos Leloudas}
\affiliation{DTU Space, Department of Space Research and Space Technology, Technical University of Denmark, Elektrovej 327, 2800 Kgs. Lyngby, Denmark}
\email{giorgos@space.dtu.dk}

\author[gname=Matt, sname=Nicholl]{Matt Nicholl}
\affiliation{Astrophysics Research Centre, School of Mathematics and Physics, Queen's University Belfast, Belfast BT7 1NN, UK}
\email{matt.nicholl@qub.ac.uk}

\author[orcid=0000-0003-0123-0062, gname=Jennifer, sname=Andrews]{Jennifer Andrews}
\affiliation{Gemini Observatory, 670 North A`ohoku Place, Hilo, HI 96720-2700, USA}
\email{Jennifer.Andrews@noirlab.edu}

\author[gname=Katie, sname=Auchettl]{Katie Auchettl}
\affiliation{School of Physics, The University of Melbourne, Parkville, VIC 3010, Australia}
\email{katie.auchettl@unimelb.edu.au}

\author[gname=Thomas, sname=de Boer]{Thomas de Boer}
\affiliation{Institute for Astronomy, University of Hawai'i, 2680 Woodlawn Drive, Honolulu, HI 96822, USA}
\email{tdeboer@hawaii.edu}

\author[orcid=0000-0002-4924-444X, gname=Azalee, sname=Bostroem]{K. Azalee Bostroem}
\affiliation{IPAC, Mail Code 100-22, Caltech, 1200 E. California Blvd., Pasadena, CA 91125}
\email{bostroem@ipac.caltech.edu}

\author[0000-0002-7714-493X]{Y.-Z. Cai \begin{CJK*}{UTF8}{gbsn}(蔡永志)\end{CJK*}}
\email{caiyongzhi789@gmail.com}
\affiliation{INAF - Osservatorio Astronomico di Padova, Vicolo dell'Osservatorio 5, 35122 Padova, Italy}
\affiliation{International Centre of Supernovae (ICESUN), Yunnan Key Laboratory of Supernova Research, Yunnan Observatories, Chinese Academy of Sciences (CAS), Kunming, 650216, China}

\author[gname=Kenneth C., sname=Chambers]{Kenneth C. Chambers}
\affiliation{Institute for Astronomy, University of Hawai'i, 2680 Woodlawn Drive, Honolulu, HI 96822, USA}
\email{chambers@ifa.hawaii.edu}

\author[orcid=0000-0003-0528-202X,gname=Collin,sname=Christy]{Collin~T.~Christy}
\email{collinchristy@arizona.edu}
\affiliation{Department of Astronomy and Steward Observatory, University of Arizona, 933 North Cherry Avenue, Tucson, AZ 85721-0065, USA}

\author[0009-0007-8485-1281, gname=Sara, sname=Faris]{Sara Faris}
\affiliation{The School of Physics and Astronomy, Tel Aviv University, Tel Aviv 69978, Israel}
\email{sarafaris452@gmail.com}

\author[orcid=0000-0003-4537-3575,gname=Noah,sname=Franz]{Noah Franz}
\email{nfranz@arizona.edu}
\affiliation{Department of Astronomy and Steward Observatory, University of Arizona, 933 North Cherry Avenue, Tucson, AZ 85721-0065, USA}

\author[orcid=0000-0001-6395-6702, gname=Sebastian, sname=Gomez]{Sebastian Gomez}
\affiliation{Department of Astronomy, The University of Texas at Austin, 2515 Speedway, Stop C1400, Austin, TX 78712, USA}
\email{sebastian.gomez@austin.utexas.edu}

\author[gname=Maider, sname=González-Bañuelos]{Maider González-Bañuelos}
\affiliation{Institute of Space Sciences (ICE-CSIC), Carrer de Can Magrans, 08193 Cerdanyola del Vallès, Barcelona, Spain}
\affiliation{Institut d’Estudis Espacials de Catalunya (IEEC), 08860 Castelldefels (Barcelona), Spain}
\email{maigb27@gmail.com}

\author[orcid=0000-0002-1650-1518 ]{Mariusz Gromadzki}
\affiliation{Astronomical Observatory, University of Warsaw, Al. Ujazdowskie 4, 00-478 Warszawa, Poland}
\email{marg@astrouw.edu.pl}

\author[gname=Paul J.,sname=Groot]{Paul J. Groot}
\email{p.groot@astro.ru.nl}
\affiliation{Department of Astronomy and Inter-university Institute for Data Intensive Astronomy, University of Cape Town, Private Bag X3, 7701 Rondebosch, South Africa}
\affiliation{South African Astronomical Observatory, P.O. Box 9, 7935 Observatory, South Africa}
\affiliation{Department of Astrophysics/IMAPP, Radboud University, P.O. Box 9010, 6500 GL, Nijmegen, The Netherlands.}

\author[orcid=0000-0003-2375-2064, gname=Claudia, sname=Guti\'errez]{Claudia Guti\'errez}
\affiliation{Institute of Space Sciences (ICE-CSIC), Carrer de Can Magrans, 08193 Cerdanyola del Vallès, Barcelona, Spain}
\affiliation{Institut d’Estudis Espacials de Catalunya (IEEC), 08860 Castelldefels (Barcelona), Spain}
\email{cgutierrez@ice.csic.es}

\author[orcid=0000-0002-9454-1742, gname=Brian, sname=Hsu]{Brian Hsu}
\affiliation{Department of Astronomy and Steward Observatory, University of Arizona, 933 North Cherry Avenue, Tucson, AZ 85721-0065, USA}
\email{bhsu@arizona.edu}

\author[gname=Mark,sname=Huber]{Mark Huber}
\email{mehuber7@hawaii.edu}
\affiliation{Institute for Astronomy, University of Hawai'i, 2680 Woodlawn Drive, Honolulu, HI 96822, USA}

\author[orcid=0000-0002-0440-9597,gname=Thomas L., sname=Killestein]{Thomas L. Killestein}
\affiliation{Department of Physics, University of Warwick, Gibbet Hill Road, Coventry, CV4 7AL, UK}
\email{tom.killestein@gmail.com}

\author[gname=Chien-Cheng,sname=Lin]{Chien-Cheng Lin}
\email{cclin33@hawaii.edu}
\affiliation{Institute for Astronomy, University of Hawai'i, 2680 Woodlawn Drive, Honolulu, HI 96822, USA}

\author[gname=Thomas B.,sname=Lowe]{Thomas B. Lowe}
\email{tlowe@hawaii.edu}
\affiliation{Institute for Astronomy, University of Hawai'i, 2680 Woodlawn Drive, Honolulu, HI 96822, USA}

\author[gname=Peter, sname=Lundqvist]{Peter Lundqvist}
\affiliation{The Oskar Klein Centre, Department of Astronomy, Stockholm University, AlbaNova, SE-10691, Stockholm, Sweden}
\email{peter@astro.su.se}

\author[gname=Dylan, sname=Magill]{Dylan Magill}
\affiliation{Astrophysics Research Centre, School of Mathematics and Physics, Queen's University Belfast, Belfast BT7 1NN, UK}
\email{dmagill14@qub.ac.uk}

\author[gname=Seppo, sname=Mattila]{Seppo Mattila}
\affiliation{Tuorla Observatory, Department of Physics and Astronomy, University of Turku, 20014 Turku, Finland}
\affiliation{School of Sciences, European University Cyprus, Diogenes Street, Engomi, 1516 Nicosia, Cyprus}
\email{sepmat@utu.fi}

\author[gname=Francesca,sname=Onori]{Francesca Onori}
\email{francesca.onori@inaf.it}
\affiliation{INAF - Osservatorio Astronomico di Roma Via Frascati, 33, 00078 Monte Porzio Catone (RM)}

\author[orcid=0000-0002-6639-6533,gname='Gregory', sname='Paek']{Gregory S. H. Paek}
\affiliation{Institute for Astronomy, University of Hawai'i, 2680 Woodlawn Drive, Honolulu, HI 96822, USA}
\email{gregorypaek94@gmail.com}

\author[gname=Jeniveve, sname=Pearson]{Jeniveve Pearson}
\affiliation{Department of Astronomy and Steward Observatory, University of Arizona, 933 North Cherry Avenue, Tucson, AZ 85721-0065, USA}
\email{jenivevepearson@arizona.edu}

\author[orcid=0000-0003-4663-4300,gname=Miika, sname=Pursiainen]{Miika Pursiainen}
\affiliation{Department of Physics, University of Warwick, Gibbet Hill Road, Coventry, CV4 7AL, UK}
\email{Miika.Pursiainen@warwick.ac.uk}

\author[orcid=0009-0006-0165-6986, gname=Niko, sname=Pyykkinen]{Niko Pyykkinen}
\affiliation{Department of Physics and Astronomy, University of Turku, FI-20014 Turku, Finland}
\email{njpyyk@utu.fi}

\author[gname=Conor, sname=Ransome]{Conor Ransome}
\affiliation{Department of Astronomy and Steward Observatory, University of Arizona, 933 North Cherry Avenue, Tucson, AZ 85721-0065, USA}
\email{cransome@arizona.edu}

\author[orcid=0009-0003-8153-9576, gname=Sara, sname=Romagnoli]{Sara Romagnoli}
\affiliation{School of Physics, The University of Melbourne, Parkville, VIC 3010, Australia}
\email{sara.romagnoli@student.unimelb.edu.au}

\author[0009-0008-3724-1824, gname=Jennifer, sname=Shi]{Jennifer Shi}
\affiliation{School of Physics, The University of Melbourne, Parkville, VIC 3010, Australia}
\email{jennifer.shi@student.unimelb.edu.au}
 
\author[gname=Manisha, sname=Shrestha]{Manisha Shrestha}
\affiliation{School of Physics and Astronomy, Monash University, Clayton, Victoria 3800, Australia}
\affiliation{The ARC Centre of Excellence for Gravitational Wave Discovery, Clayton, Victoria 3800, Australia}
\email{manisha.shrestha@monash.edu}

\author[orcid=0000-0001-5510-2424, gname=Nathan, sname=Smith]{Nathan Smith}
\affiliation{Department of Astronomy and Steward Observatory, University of Arizona, 933 North Cherry Avenue, Tucson, AZ 85721-0065, USA}
\email{nathans@as.arizona.edu}

\author[gname=Maximilian, sname=Stritzinger]{Maximilian Stritzinger}
\affiliation{Department of Physics and Astronomy, Aarhus University, Ny Munkegade 120, DK-8000 Aarhus C, Denmark}
\email{max@phys.au.dk}

\author[gname=Bhagya,sname=Subrayan]{Bhagya Subrayan}
\email{bsubrayan@arizona.edu}
\affiliation{Department of Astronomy and Steward Observatory, University of Arizona, 933 North Cherry Avenue, Tucson, AZ 85721-0065, USA}

\author[gname=Richard, sname=Wainscoat]{Richard Wainscoat}
\affiliation{Institute for Astronomy, University of Hawai'i, 2680 Woodlawn Drive, Honolulu, HI 96822, USA}
\email{rjw@hawaii.edu}

\author[orcid=0000-0002-0025-0179, gname=Ziyang, sname=Wang]{Ziyang Wang}
\email{wangziyang@ucas.ac.cn}
\affiliation{School of Astronomy and Space Science, University of Chinese Academy of Sciences, Beijing 100049, China}
\affiliation{National Astronomical Observatories, Chinese Academy of Sciences, Beijing 100101, China}

\author[gname=Simon, sname=de Wet]{Simon de Wet}
\affiliation{DTU Space, Department of Space Research and Space Technology, Technical University of Denmark, Elektrovej 327, 2800 Kgs. Lyngby, Denmark}
\email{sndwe@dtu.dk}

\author[gname=Hannah, sname=Wichern]{Hannah Wichern}
\affiliation{DTU Space, Department of Space Research and Space Technology, Technical University of Denmark, Elektrovej 327, 2800 Kgs. Lyngby, Denmark}
\email{hawic@space.dtu.dk}

\begin{abstract}

The origin of the optical emission in tidal disruption events (TDEs) remains one of the major outstanding questions in the field, in part due to the limited number of nearby events with high-cadence monitoring to track their evolving photometric and spectroscopic properties. We present multi-wavelength observations of the nearby ($z=0.01368$) TDE\,2025aarm, including near-daily spectroscopic coverage prior to the optical peak.
Its proximity makes it one of the brightest TDEs discovered, reaching a peak magnitude of $m_r\sim15.5$ ($M_r\sim-18$). The light curve deviates from a smooth evolution, exhibiting multiple rebrightening episodes visible in both the individual filter light curves and the bolometric luminosity. Blackbody modelling reveals that these rebrightenings are associated with an increase in temperature of $> 5,000-10,000$\,K, while the inferred photospheric radius remains approximately constant. Simultaneously, the H$\alpha$ line not only increases in blueshift but also broadens, suggesting a link between the continuum rebrightenings to changes in the kinematics of the line-forming gas. We identify a persistent absorption component at $\sim-3900$\,km\,s$^{-1}$ in multiple Balmer lines, providing further evidence for outflowing material. The H$\alpha$ profile also exhibits excess flux compared to a Gaussian on both sides of the line, inconsistent with simple scattering-dominated outflow models. Disk-profile modelling provides evidence for the emergence of a disk-like component least $\sim20$ days after peak, with substantial changes in the disk properties between $\sim50$ and 60 days.
The rebrightenings appear to trace episodes of enhanced energy injection into the emitting material. These observations highlight the complexity of TDE emission processes and demonstrate how dense multi-wavelength monitoring can disentangle the roles of accretion, reprocessing, and outflows in shaping TDE emission.

\end{abstract}



\section{Introduction} \label{sec:intro}

Tidal disruption events (TDEs) occur when a star approaches sufficiently close to a supermassive black hole (SMBH) and is torn apart by tidal forces \citep{Hills1975,Rees1988}. Roughly half of the stellar debris becomes unbound, while the remainder remains gravitationally bound to the SMBH and returns towards pericentre. The subsequent evolution of the bound debris gives rise to a luminous flare lasting months to years. TDEs provide a unique laboratory for probing quiescent black holes in otherwise inactive galaxies, as well as the onset of accretion.

Observationally, wide-field time-domain surveys over the past decade have significantly increased the number of known TDEs \citep[e.g.][]{Gezari2021}. These events have revealed a surprisingly diverse phenomenology, with peak luminosities spanning orders of magnitude, and a wide range of rise and decline timescales \citep{vanVelzen2021, Hammerstein2023, Yao2023}. Optically selected TDEs are often characterised by a strong blue continuum, with blackbody temperatures in the range $T \sim 2-5\times 10^{4}$\,K, and broad emission lines ($v \sim 10^{4}$km\,s$^{-1}$) (\cite{Arcavi2014, Charalampopoulos2022}, for a review see \cite{vanVelzen2020}).

TDEs have been observed across the electromagnetic spectrum, from radio to X-rays, providing key insights into the physical processes at play \citep{Franz2026}. X-ray emission is generally thought to originate from the innermost regions of the accretion flow near the SMBH \citep{Roth2016, Guolo2024}, whilst late-time plateau emission in the ultraviolet (UV) is believed to arise from the cooling and radial expansion of this newly formed accretion disk \citep{vanVelzen2019, Mummery2020}. In contrast, the origin of the optical and UV emission at peak remains an open question. If the TDE luminosity is powered by the accretion of returning stellar debris \citep{Rees1988}, the accretion rate would broadly tracing the fallback rate, thus the observed emission might naively be expected to originate from a compact accretion disk close to the SMBH. However, the blackbody radii inferred for optically selected TDEs are often substantially larger than expected for a compact accretion disk around a SMBH \citep{Hung2017, Wevers2017, Hinkle2020}. This has led to the proposal that the optical/UV emission is reprocessed by intervening material, in which high-energy radiation produced close to the black hole is absorbed and re-emitted at longer wavelengths by optically thick material surrounding the accretion flow \citep{Loeb1997,Guillochon2013,Metzger2016,Dai2018}, similar to the optical emission of accretion disks in low-mass X-ray binaries \citep{deJong1996}. Several alternative mechanisms have been proposed for powering the optical/UV emission. One possibility is that a significant fraction of the optical emission may instead arise from shocks generated as the stellar debris streams self-intersect and circularise prior to accretion disk formation \citep{Piran2015, Shiokawa2015, Bonnerot2021a}. Recent radiation-hydrodynamic simulations have also shown that shocks produced by the collision of the returning debris stream with an eccentric accretion disk could power the optical/UV emission near peak \citep{Steinberg2024}. Alternatively, the cooling-envelope model proposes that rapid dissipation of the returning debris forms an extended, pressure-supported envelope, with the optical/UV emission subsequently powered by its gradual gravitational contraction \citep{Metzger2022}. The efficiency and timescale of debris circularisation remain highly uncertain, with simulations suggesting that relativistic apsidal precession and black hole spin can influence this process \citep{Hayasaki2016}. Distinguishing between these scenarios has proven challenging, as both can reproduce many of the observed properties of TDEs. 

Because reprocessing models require high-energy radiation from an accretion flow, observational constraints on the timing of accretion disk formation provide an important link between theory and observations. Several events exhibit delayed X-ray brightening that has been interpreted as the emergence of an accretion disk once the surrounding reprocessing material becomes optically thin \citep{Gezari2017,vanVelzen2019,Holoien2019b,Metzger2022,Guolo2024,Franz2026}.One of the earliest examples was ASASSN-15oi, which exhibited a substantial increase in its soft X-ray luminosity at late times \citep{Gezari2017}. More recently, AT\,2019azh showed delayed X-ray brightening interpreted as evidence for inefficient early circularisation followed by later disk formation \citep{Liu2022}. Similarly, the double-peaked Balmer line profiles observed in AT,2018hyz were interpreted as emission from a nearly circular accretion disk \citep{Short2020}.

At the same time, growing evidence suggests that winds and outflows are also an important component of many optical TDEs. Broad emission lines, blueshifted absorption features, evolving line asymmetries, and radio detections have all been interpreted as signatures of outflowing material \citep[e.g.][]{Chornock2014, Hung2019, Roth2018,  vanVelzen2016, Alexander2016, Cendes2024, Alexander2026}. One of the clearest examples is AT\,2019qiz, where the optical peak was explained as being powered by an optically thick outflow launched during the early super-Eddington accretion phase \citep{Nicholl2020}. In that event, the evolution of the continuum temperature, line widths, and Bowen fluorescence features were consistent with a receding reprocessing layer surrounding the inner accretion flow. Similar behaviour has now been observed in several TDEs, suggesting that winds and outflows may play a fundamental role in shaping the observed optical properties of these events.

A further clue to the origin of the optical emission comes from the complex photometric evolution observed in many TDEs. While some events show smooth rises and declines, others display plateaus, rebrightening episodes, and short-timescale variability superimposed on the broader evolution \citep[e.g.][]{Holoien2019b, Charalampopoulos2022, Faris2024, Huang2024, Dgany2026}. The origin of this variability remains uncertain, but proposed explanations include changes in reprocessing efficiency, episodic accretion, evolving obscuration, or additional shocks within the debris streams \citep{Piran2015, Metzger2016, Roth2016, Dai2018, Bonnerot2022}. 
Establishing whether such behaviour is common, and how it relates to the formation of accretion disks and outflows, requires well-sampled photometric and spectroscopic observations spanning the full evolution of a TDE. However, the distances at which most TDEs are discovered and the detection limits of current wide-field surveys mean that observations are often concentrated around peak, with substantially poorer coverage as the transient fades. The resulting photometric uncertainties and sparse late-time sampling can make low-amplitude or short-term variability difficult to identify. Nearby TDEs therefore provide a particularly valuable opportunity to trace this variability over a much larger fraction of their evolution.

In this work, we present the detailed follow-up of the TDE\,2025aarm, hosted in the galaxy SDSS J043212.40$-$052239.6 at $z = 0.01368$. An initial study by \citet{Simongini2026} focused on the early optical, UV, and X-ray evolution of the event, finding evidence for a delayed accretion scenario in which the early emission is powered by circularisation shocks. Subsequently, \citet{Baldini2026} presented a six-month X-ray monitoring campaign, revealing the first low-hard to high-soft X-ray state transition observed in a thermal TDE, corresponding to an evolution from a power-law-dominated to a disk-dominated X-ray spectrum. Here, we build upon that work with extensive multi-wavelength photometric and spectroscopic monitoring spanning from the rise to late-time evolution ($-60$ to $+130$ rest-frame phase), providing one of the most comprehensive datasets obtained for a nearby TDE. The exceptional cadence enables us to investigate the relationship between short-timescale variability, spectroscopic evolution, and the emergence of accretion disk and outflow signatures.

The paper is structured as follows. In Section \ref{sec:observations}, we present the observations and data reduction. Section \ref{sec:host} describes the properties of the host galaxy. In Section \ref{sec:lc}, we present the photometric evolution of TDE\,2025aarm and modelling of its light curve. The spectroscopic evolution is presented in Section \ref{sec:spec_ev}. We discuss the implications of our findings in Section \ref{sec:discussion}, and conclude in Section \ref{sec:conclusions}.

\section{Observations} \label{sec:observations}

\subsection{Discovery and Classification}

TDE\,2025aarm was discovered on MJD 60949.17 by the Gravitational-wave Optical Transient Observer \citep[GOTO;][]{Steeghs2022, Dyer2024} under the internal name GOTO25iqe and  a magnitude of $m_L=18.96$\,mag \citep{Neill2025}. Other wide-field sky surveys then began to report this transient with BlackGEM \citep{Groot2024} reporting a magnitude of $m_q=17.38$ on MJD 60962.29, the Asteroid Terrestrial-impact Last Alert System (ATLAS) project \citep{Tonry2018} reporting $m_w=17.321$ on MJD 60960.11, and the Panoramic Survey Telescope and Rapid Response System \cite[Pan-STARRS;][]{Chambers2016} reporting $m_g=15.7$ on MJD 60986.46. It was classified as a TDE H+He on MJD 60978.68 \citep{Newsome2025, Faris2025}. This classification was based on the presence of a blue continuum, and broad H$\alpha$ and He II emission features. The transient is coincident with the galaxy SDSS J043212.40-052239.6 at $z = 0.01368$, supported by narrow Balmer and Na I D absorption lines at a redshift of $\sim$0.0135 in the classification spectrum.

\subsection{UV/Optical/IR Photometry}
High cadence photometric follow-up was triggered using the Las Cumbres Observatory \citep[Las Cumbres;][]{Brown2013} network in the $u,g,r,i,z,B,$ and $V$ filters. Within their network, several 1\,m telescopes were used across multiple observatories and images were reduced using the BANZAI pipeline \citep{curtis_mccully_2018_1257560}.
Aperture photometry was performed on the images using a 5" aperture and then host subtracted using synthetic photometry of the host spectral energy distribution (SED). We provide further details of this process in Section \ref{sec:host}. Flux subtraction was adopted in lieu of image subtraction as suitable reference images are not yet available, requiring the transient to fade sufficiently before they can be obtained. The resulting host-subtracted photometry is internally consistent and agrees well with independently host-subtracted photometry from other facilities, providing confidence in this approach.

Near UV and optical observations with the Neil Gehrels Swift Observatory \citep[\textit{Swift};][]{Gehrels2004} were triggered beginning MJD 60981.1 (PI's: Kuin, Stein, Miller, Charalampopoulos, Sun, Konno). UV observations were obtained with UVOT on Swift in the $UVW2$, $UVM2$, $UVW1$, $U$, $B$, and $V$ bands. For each filter and epoch, the individual exposures were stacked using \texttt{uvotimsum} within \textsc{HEASOFT}, before source photometry was extracted with \texttt{uvotsource} adopting a 5" aperture. The reduction and calibration followed standard UVOT procedures \citep{Poole2008,Brown2009}, and the resulting count rates were converted to AB magnitudes using the UVOT photometric zero points \citep{Poole2008, Breeveld2011}. 
While the host-galaxy contribution is expected to be negligible in the UV bands, host magnitudes were still subtracted from the data. The host magnitudes were derived from synthetic photometry of the host SED (see Section \ref{sec:host}). We note that \citet{Miller2025} report archival detections of the host galaxy in the $U$ and $UVW1$ bands. We remeasured these fluxes using the same 5" aperture adopted for the other UVOT observations to ensure a consistent photometric analysis. Nevertheless, we adopted the synthetic host magnitudes for all UVOT filters to ensure a consistent host-subtraction procedure, particularly for $UVW2$ and $UVM2$, for which no archival host measurements exist (see Section \ref{sec:host}).

Further optical follow-up of TDE\,2025aarm was triggered using Pan-STARRS in $g,r,i,z,$ and $y$ bands from MJD 60986.46 with a $\sim$weekly cadence. Serendiptious $w-$band photometry was obtained from the Pan-STARRS Near-Earth Object Survey. Magnitudes were extracted through forced photometry performed on difference imaging using the Pan-STARRS Image Processing Pipeline \citep{Magnier2020}.

Photometric measurements were also obtained from the ATLAS forced-photometry server \citep{Tonry2018,Smith2020,Shingles2021} and the Zwicky Transient Facility (ZTF) forced-photometry service \citep{Masci2019}. In both cases, the forced photometry was performed on difference images, yielding measurements in the ATLAS $c$- and $o$-bands, and the ZTF $g$- and $r$-bands. To improve the signal-to-noise ratio, the ATLAS observations were binned to a daily cadence.

Additional $u,g,r,i,z$ imaging was obtained with the Alhambra Faint Object Spectrograph and Camera (ALFOSC) mounted on the 2.56 m Nordic Optical Telescope (NOT) on La Palma, Spain. All imaging data were reduced with a custom pipeline, primarily using seeing-matched aperture photometry on difference images to mitigate issues with poor point spread function (PSF) reconstruction and sampling.

Follow-up photometry was also reported by the Black Hole Target and Observation Manager (BHTOM), but is not included in our analysis due to its limited temporal coverage \citep{Majumdar2026}.

All photometry was corrected for Milky Way extinction assuming $E(B-V)=0.0681$ \citep{Schlafly2011}. A correction for host-galaxy extinction was also applied (see Section~\ref{sec:host}). Throughout this work, we adopt a redshift of $z=0.01368$ \citep{Newsome2025}, and assume a flat $\Lambda$CDM cosmology with $H_{0}=70$\,km\,s$^{-1}$\,Mpc$^{-1}$ and $\Omega_{\Lambda}=0.7$, corresponding to a luminosity distance of 59.2\,Mpc.

\subsection{Spectroscopy}

Following discovery, an intense spectral follow-up campaign was triggered for this event through a number of collaborations. The earliest spectrum is from the Liverpool Telescope \citep[LT;][]{Steele2004} using the SPRAT instrument \citep{Piascik2014}. These spectra were reduced using the \texttt{PypeIt} reduction pipeline \citep{pypeit:joss_pub, pypeit:zenodo} with a custom recipe for LT/SPRAT.

Additional spectroscopic monitoring was carried out through the Arizona Transient Exploration and Characterization (AZTEC) programme. A total of seven spectra were obtained with the Binospec instrument on the MMT \citep{Fabricant2019}, together with four epochs using the Boller \& Chivens (B\&C) spectrograph on the Bok 2.3\,m Telescope, and a single spectrum from the MMT BlueChannel spectrograph \citep{Angel1979}. Observations taken with Binospec \citep{Fabricant2019} on MMT were reduced using \texttt{PypeIt}. \texttt{PypeIt} has built-in compatibility for Binospec data reduction. We used a standard configuration file in our semi-automated reductions along with a sensitivity function constructed using a standard star with the same central wavelength as each observation. This sensitivity function is then applied to the spectrum for flux calibration. The spectra taken with the B\&C and BlueChannel were reduced using standard IRAF routines \citep{Tody1986, Tody1993}.

Further follow-up observations were obtained with the robotic FLOYDS spectrographs on the Las Cumbres 2\,m telescopes located at Haleakala Observatory (OGG) and Siding Spring Observatory (COJ). The data were reduced and extracted using the FLOYDS pipeline\footnote{\url{https://github.com/svalenti/FLOYDS_pipeline}}.

Spectra were obtained with ALFOSC mounted on NOT as part of the NUTS2 (NOT Unbiased Transient Survey 2) collaboration and program 72-504 (PI: Charalampopoulos). Reductions were performed using Foscgui\footnote{Foscgui is a graphic user interface aimed at extracting SN spectroscopy and photometry obtained with FOSC-like instruments. It was developed by E. Cappellaro. A package description can be found at https://sngroup.oapd.inaf.it/foscgui.html}. 

Additional spectra of AT2025aarm were obtained using the Wide Field Spectrograph (WiFeS) which is mounted on the Australian National University (ANU)'s 2.3\,m telescope at Siding Springs Observatory \citep{Dopita2007, Dopita2010, Price2024}. These observations were taken using Nod and Shuffle mode and the B3000/R3000 gratings to cover the full wavelength range of the instrument (3400-9500\,\AA\,). Data were reduced using the updated \texttt{pyWiFeS} reduction pipeline \citep{Price2024, Childress2014}, while flux calibration was obtained using a spectrophotometric standard star that was observed on the same night.

A series of eight high-resolution spectra were obtained with the X-Shooter instrument on the Very Large Telescope (VLT), providing simultaneous wavelength coverage from the UV to the near-IR \citep{Vernet2011}. The data were reduced using the standard ESO Reflex workflows \citep{Freudling2013}. Flux calibration was carried out using spectrophotometric standard stars observed on the same night as the science exposures. Telluric absorption in the X-Shooter VIS arm was corrected using {\tt molecfit} \citep{Smette2015,Kausch2015}.

Additional spectra were obtained with Mookodi \citep{Erasmus2024} on the 1\,m Lesedi telescope at the South African Astronomical Observatory. These spectra were reduced using a Mookodi-specific pipeline which has been adapted from the python-based package Automated SpectroPhotometric REDuction \citep[ASPIRED;][]{Lam2023} toolkit.

In order to perform host subtraction, firstly the Las Cumbres photometry was re-extracted using a fixed 3" aperture, chosen to approximately match the slit widths used for the majority of the spectroscopic observations. Each spectrum was subsequently flux calibrated and linearly mangled to the contemporaneous photometry. Host subtraction was performed using an archival Dark Energy Spectroscopic Instrument  (DESI) spectrum of the host galaxy \citep{DESICollaboration2026}.  This spectrum was scaled to match archival Pan-STARRS imaging, extracted with a 3" aperture at the location of the TDE. An example of this process is shown in Figure \ref{fig:host_sub}. All spectra were additionally corrected for both Milky Way and host-galaxy extinction, as described in Section \ref{sec:host}. Continuum subtraction for plotting was done by fitting a cubic polynomial to relatively featureless regions of the spectrum at 3900-4000\,\AA, 4220-4280\,\AA, 5100-5550\,\AA, 6000-6350\,\AA, and 7100-9000\,\AA, and subsequently subtracting the fitted continuum. The full spectroscopic log is presented in Appendix \ref{sec:spec_obs}.

\begin{deluxetable}{cccc}
    \tablecaption{Archival host magnitudes and synthetic magnitudes calculated from the scaled host SED in the corresponding filter bandpasses. Magnitudes given in AB.\label{tab:host_mags}} 
    \tablehead{Filter & Telescope & Archival Magnitude & Synthetic Magnitude}
    \startdata
        \hline
        $UVW2$ & \textit{Swift} & & 18.69$\pm$0.70\\
        $UVM2$ & \textit{Swift} & & 18.73$\pm$0.70\\
        $UVW1$ & \textit{Swift} & 18.96$\pm$0.05 & 18.13$\pm$0.70\\
        $U$ & \textit{Swift} & 17.26$\pm$0.02& 17.02$\pm$0.70\\
        $B$ & \textit{Swift} & & 15.57$\pm$0.70\\
        $V$ & \textit{Swift} & & 14.88$\pm$0.70\\
        $g$ & Pan-STARRS & 15.3$\pm$0.1& 15.20$\pm$0.70\\
        $r$ & Pan-STARRS & 14.7$\pm$0.2& 14.61$\pm$0.70\\
        $i$ & Pan-STARRS & 14.3$\pm$0.1& 14.28$\pm$0.70\\
        $B$ & Las Cumbres & & 15.61$\pm$0.70\\
        $V$ & Las Cumbres & & 15.14$\pm$0.70\\
        $g$ & Las Cumbres & & 15.30$\pm$0.70\\
        $r$ & Las Cumbres & & 14.62$\pm$0.70\\
        $i$ & Las Cumbres & & 14.29$\pm$0.70\\
        \hline \hline
    \enddata
\end{deluxetable}

\section{Host} \label{sec:host}

The TDE 2025aarm is hosted by the galaxy SDSS J043212.40$-$052239.6, located at a redshift of $z = 0.01368$. Using the Pan-STARRS catalog search, the galaxy has an apparent Kron magnitude of $m_{g,\rm{host}} = 14.930 \pm 0.001$, and $m_{r,\rm{host}} = 14.275 \pm 0.001$. These magnitudes correspond to absolute magnitudes of $M_{g,\rm{host}} = -18.93$, and $M_{r,\rm{host}} = -19.59$, after correcting for Galactic extinction. K-correction at this distance is negligible ($\sim0.01\,$mag). The host $g-$band magnitude is consistent with the sample of 10 TDE host galaxies studied by \citet{Law-Smith2017}, who find a median value of $M_g = -19.81^{+1.01}_{-0.49}$. The host colour of $g-r = 0.579\pm0.001$ is also consistent with this sample with a median value of $g-r = 0.68^{+0.09}_{-0.11}$. 

Host galaxy properties were fitted using the \textsc{BLAST} SED fitting framework \citep{Wang_2023, Tejero-Cantero_2020sbi, Greenberg_2019, Astropy_2018, Harris_2020, Gagliano2025_Prost, Bradley_2024, Ginsburg_2019, Fernique_2015, Speagle_2020, Johnson_2021, Johnson_2021,Rodrigo_2020, Zonca_2019,Johnson_2024, NSFACCESS, Bokeh}. \textsc{BLAST} is a host galaxy SED fitting code that compiles archival imaging across a broad wavelength range. Specifically, it utilises data from \textit{GALEX}, SDSS, DES, Pan-STARRS, 2MASS, and \textit{WISE} to construct multi-wavelength photometry of the host galaxy \citep{Dey_2019,Skrutskie_2006, Martin_2005,Wright_2010, York_2000, Flewelling_2020, Chambers_2016, Magnier_2020, Blanton_2017}. BLAST uses aperture photometry extracted using both a fixed circular aperture of $7.16^{\prime\prime}$ and a larger elliptical aperture with semi-major and semi-minor axes of $17.1^{\prime\prime}$ and $9.1^{\prime\prime}$, respectively, corresponding to global measurements of the galaxy light.

The resulting global SED is automatically fit by \textsc{BLAST} using the \textsc{Prospector} SED fitting code \citep{Leja2017, Johnson2021}. We adopt the resulting stellar mass of $\log(M_{\star}/M_{\odot}) =10.02^{+0.08}_{-0.18}$ and a specific star formation rate of $\rm{sSFR}= 6.1^{+42.4}_{-0.1}\times 10^{-14}$\,yr$^{-1}$. The sSFR is at the lower end of values from the sample of hosts in \citet{Ramsden2026}. 

We can also estimate the mass of the central SMBH using the M$-\sigma$ relation from \citet{Kormendy2013}. To do this, we modelled the DESI host spectrum with the penalized pixel-fitting code \texttt{pPXF} \citep{Cappellari2023}. The DESI spectra have a wavelength dependent resolving power of $R\sim2000-5000$, corresponding to a velocity resolution of $\sim150-60$\,km\,s$^{-1}$ across the full spectral range \citep{DESICollaboration2022}. Prior to fitting, the spectrum was corrected to the rest frame and rebinned onto a logarithmic wavelength scale. The stellar continuum was fitted using the eMILES stellar population templates \citep{Vazdekis2016}, convolved to match the wavelength-dependent instrumental resolution of the DESI spectrum. The fit works by broadening and shifting the template spectra until they best reproduce the observed stellar absorption features, allowing the stellar velocity dispersion of the host galaxy to be measured. From this fit we measured an observed stellar velocity dispersion of $\sigma_{\rm obs}=101.9$\,km\,s$^{-1}$. Using the $M_{\rm BH}$-$\sigma_{\ast}$ relation, this corresponds to a black hole mass of $\log_{10}(M_{\rm BH}/M_{\odot})=7.45 \pm 0.29$ accounting for the intrinsic scatter in the relation. 

However, black hole mass estimates derived from the $M_{\rm BH}$-$\sigma_{\ast}$ relation remain uncertain for TDE host galaxies, particularly in the low-mass regime where the relation is less well constrained. Calibrations are dominated by more massive galaxies, and may therefore not accurately represent the properties of lower-mass, star-forming, or post-starburst systems commonly associated with TDEs \citep{Arcavi2014, Kormendy2013, French2020}. Recent work by \citet{Ramsden2026} suggests that the relation becomes shallower at lower masses when TDE host galaxies are included, which would lead to an overestimate for a given stellar velocity dispersion.

Our black hole mass estimate is consistent with previously reported values for TDE\,2025aarm. Using \texttt{pPXF}, \citet{Baldini2026} measured $\log_{10}(M_{\rm BH}/M_{\odot}) = 6.87 \pm 0.31$, while \citet{Simongini2026} inferred $\log_{10}(M_{\rm BH}/M_{\odot}) = 6.92 \pm 0.55$ from host-galaxy SED modelling.

To facilitate host subtraction of our transient photometry, we also derived synthetic photometry of the host SED to compute expected magnitudes in the \textit{Swift} UVOT filters ($UVW2$, $UVM2$, $UVW1$, and $U$) and the Las Cumbres filters ($B$, $V$, $g$, $r$ and $i$). To do this, image cutouts were obtained from the Pan-STARRS image cutout webpages, and magnitudes were extracted using a 5" aperture to remain consistent with the UVOT and Las Cumbres photometry. The inferred SED was then scaled to these fluxes. The synthetic magnitudes were subtracted from the transient photometry from \textit{Swift} and Las Cumbres to ensure that the measured light curve reflected only the TDE flux. The host SED is visually consistent with the archival DESI spectrum in the overlapping wavelength region, lending confidence to the resulting host subtractions. Table \ref{tab:host_mags} tabulates the resulting synthetic host magnitudes used. The quoted uncertainties incorporate the statistical uncertainty propagated from the fitted host SED, and the uncertainty associated with scaling the spectrum to match the archival photometry.

Although the synthetic magnitudes agree with the archival magnitudes for the Pan-STARRS optical bands, they disagree slightly with the archival \textit{Swift} UV bands. This is to be expected as the SED was scaled by a constant based on the optical ratios. Using the UV data would require tilting the SED which introduces more uncertainties. This difference could be caused by the SED fit using global photometry whereas the nucleus may have a slightly different shaped SED. The UV discrepancy may also arise from differences in the PSF of different instruments resulting in the same sized aperture capturing different fractions of the galaxy light. For this reason we use the synthetic magnitudes derived here for consistency.

The host galaxy extinction was estimated using the \textsc{Prospector} modelling. In this framework, dust attenuation is parametrised via the optical depth $\tau_{2}$, which represents the diffuse dust component affecting stellar populations. The dust treatment follows the two-component model of \citet{Charlot2000}, in which young stars experience additional attenuation from birth clouds. The host galaxy colour excess is computed as $E(B-V) = 0.268 \times \tau_{2}$, assuming a \citet{Calzetti2000} attenuation law. From this, we obtain a host extinction of $A_V = 0.04_{-0.02}^{+0.07}$.

As an independent check, we also measured the host extinction using the Na I D absorption features in the spectra. The equivalent width (EW) of the Na I D doublet was measured for each spectrum, and converted to extinction using the empirical relation of \citet{Poznanski2012}. The resulting values of $A_V$ were averaged to obtain a mean host extinction of $A_V = 0.084 \pm 0.021$. This is consistent with the extinction derived from the SED fitting within uncertainties.

We adopt a final host extinction of $A_V = 0.084 \pm 0.021$ for all subsequent analysis. All photometric and spectroscopic data were corrected for host extinction assuming a standard total-to-selective extinction ratio of $R_V = 3.1$.

\begin{figure*}
    \centering
    \includegraphics[width=2\columnwidth]{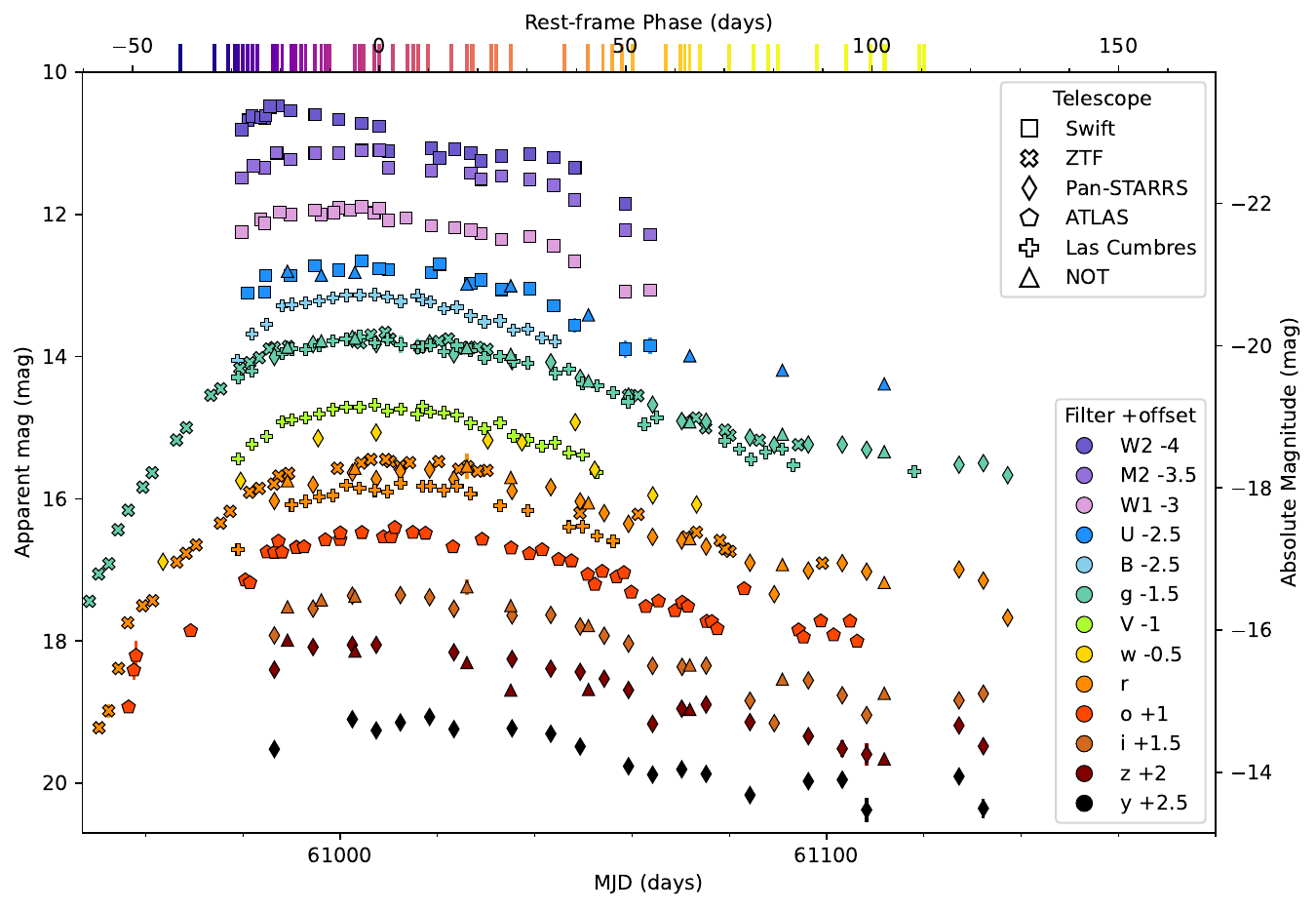}
    \caption{Light curve of TDE\,2025aarm including data from ATLAS, ZTF, Las Cumbres, NOT, Pan-STARRS and \textit{Swift}. Magnitudes are corrected for MW and host extinction, and presented in AB magnitudes. Coloured notches on the top correspond to epochs where optical spectra were obtained.}
    \label{fig:lc}
\end{figure*}

\section{Light Curves} \label{sec:lc}

\begin{figure}
    \centering
    \includegraphics[width=1\columnwidth]{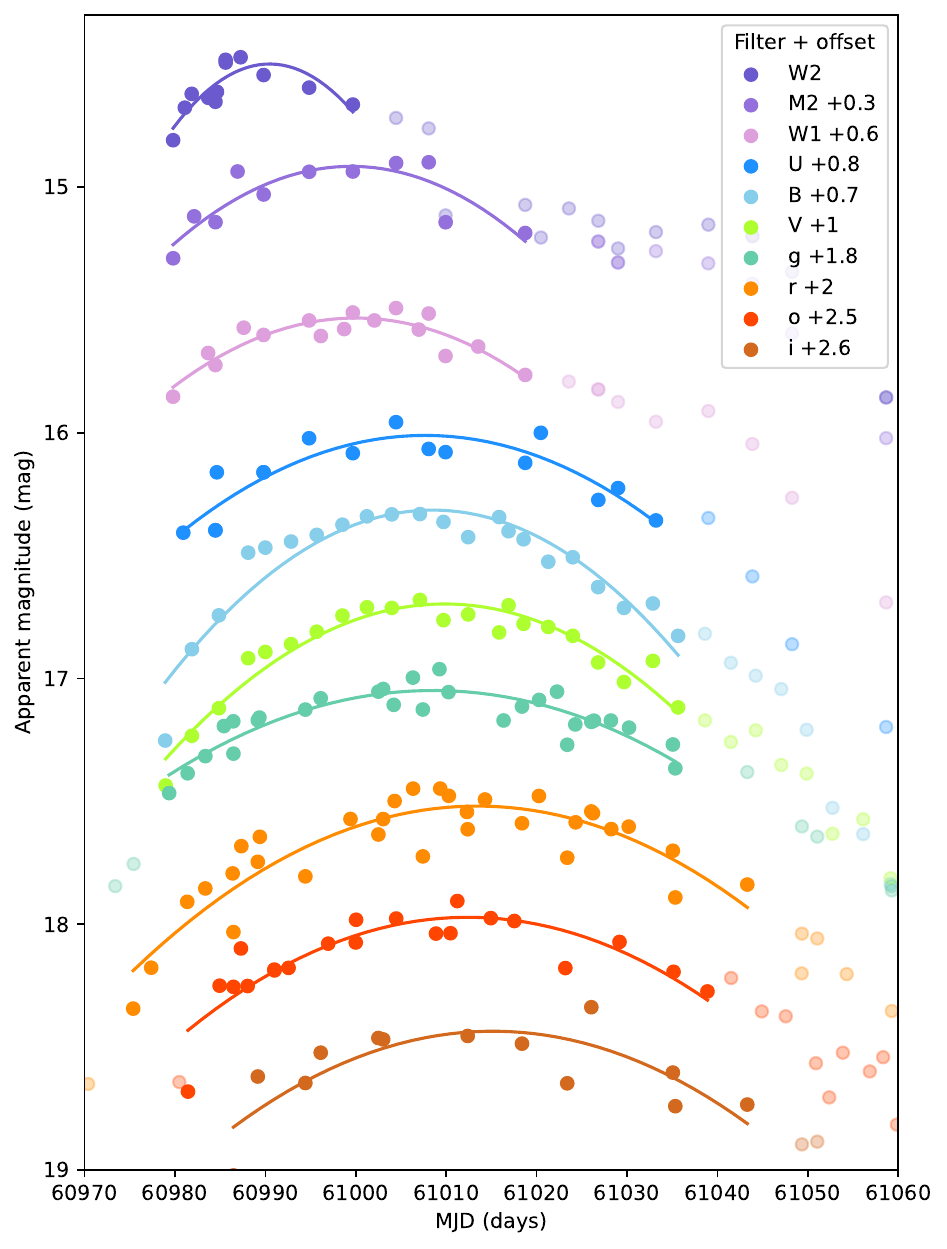}
    \caption{Quadratic fits to the light-curve maxima in each photometric band. The points show the observed photometry and the solid curves the best-fitting quadratic models used to determine the time of maximum light.}
    \label{fig:peak_fits}
\end{figure}

\begin{figure}
    \centering
    \includegraphics[width=1\columnwidth]{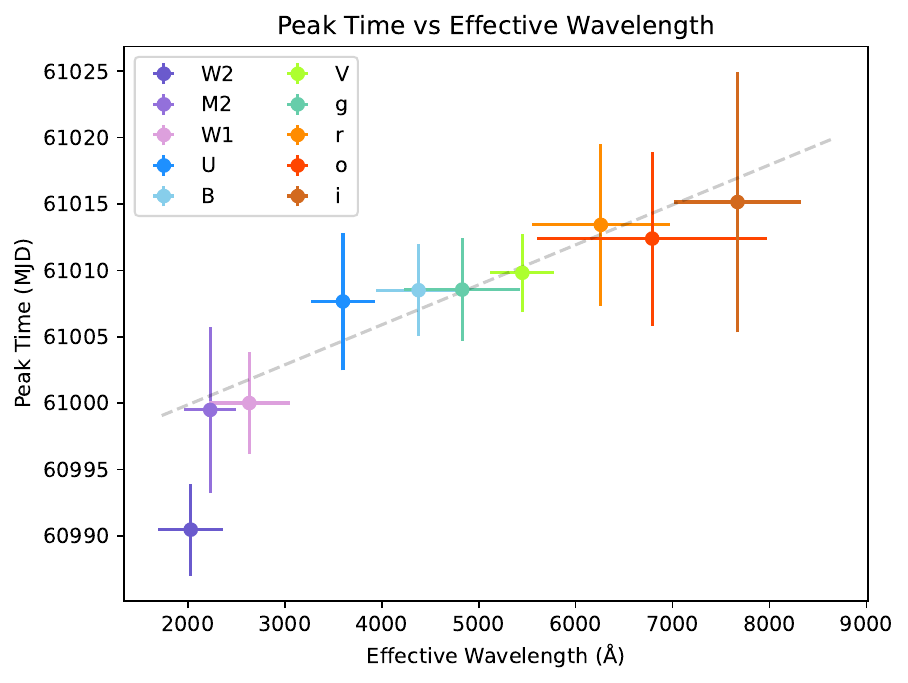}
    \caption{Time of maximum light as a function of the effective wavelength of each filter, derived from the quadratic fits shown in Figure \ref{fig:peak_fits}. The dashed line indicates the line of best fitthrough the points excluding $UVW2$. The transient peaks progressively later at longer wavelengths.}
    \label{fig:peak_lambda}
\end{figure}

The multi-band light curve of TDE\,2025aarm is shown in Figure \ref{fig:lc}. The dataset spans 13 photometric bands, covering wavelengths from the near-UV to the near-IR. We note that the UV observations from \textit{Swift} do not extend beyond MJD 61059 due to the halt of \textit{Swift} science operations. Data was collected until the target became sun-constrained by ground based telescopes. We note a spread in the $r$-band of $\sim$0.5\,mag between different telescopes. This may arise from a combination of differences in the filter transmission curves and host-subtraction methods, although the synthetic host magnitude used for Las Cumbres is consistent with the archival Pan-STARRS photometry measured within the same aperture.

The TDE reached a peak magnitude of $m_{g} = 15.162\pm0.004$\,mag and $m_{r} = 15.449 \pm 0.005$ in ZTF. This corresponds to absolute magnitudes of $M_{g} = -18.68$\,mag and $M_{r} = -18.40$\,mag. Although this falls within the broad range of peak $r$-band magnitudes ($\sim-18$ to $\sim-20$,mag) in the sample of \citet{Hammerstein2023}, TDE\,2025aarm lies towards the faint end of the distribution, with most events clustering closer to $M_r\sim-20$,mag. TDE\,2025aarm therefore represents a relatively low-luminosity event within the optical TDE population.

To determine the time of peak in each band, we fit low-order polynomials to the light curves around maximum light as shown in Figure \ref{fig:peak_fits}. We find a clear linear correlation between the time of peak and the effective wavelength of the filter passband, such that bluer bands peak earlier than redder bands (Figure \ref{fig:peak_lambda}). The only exception to this trend is the $UVW2$ band (see Section \ref{sec:bumps}). A similar wavelength-dependent delay has been observed in other TDEs such as AT\,2019azh \citep{Faris2024}, and is generally interpreted as a consequence of the cooling of the emitting region following disruption. For the remainder of this analysis, we adopt the $g$-band peak at MJD 61008 as the reference epoch.

The rise time in the $g-$band, defined as the interval between first light and peak, is measured to be $t_{\rm{rise}} = 57$\,days when converted to the rest-frame. We also determine the time scales taken for the light curve to rise or fade by half its maximum brightness ($t_{1/2,\rm{rise}}$ and $t_{1/2,\rm{decline}}$). To do this, we fit the $g-$band light curve with the a power-law rise and exponential decline as described in \citet{Yao2023}. From this we find the time taken from half the maximum luminosity to maximum luminosity to be $t_{1/2,rise} = 31.8\pm0.1$\,days and the decline timescale to reach half the luminosity is $t_{1/2,\rm{decline}}=39.0\pm0.1$\,days. This gives a total rest-frame duration above half-max light of $t_{1/2}=70.8\pm0.1$\,days. Studies in the literature have found a significant correlation between the light curve evolution and the black hole mass \citep{vanVelzen2020, Gezari2021, Hammerstein2023, Yao2023}. The empirical relation defined in \citet{Yao2023} for a sample of 28 TDEs with $z<0.24$ is:

\begin{equation}
\frac{t_{1/2}}{41.6^{+3.8}_{-3.5}\,\rm{days}} = \frac{M_{\rm{BH}}}{10^{6} M_{\odot}}^{0.14 \pm 0.04}
\end{equation}

where $M_{\rm{BH}}$ is the mass of the black hole. Using our time of $t_{1/2}$ gives $\log_{10}(M_{\rm BH}/M_{\odot}) = 7.44 \pm 0.53$, consistent with the mass estimated from the $M$-$\sigma$ relation.

\begin{figure*}
    \centering
    \includegraphics[width=2\columnwidth]{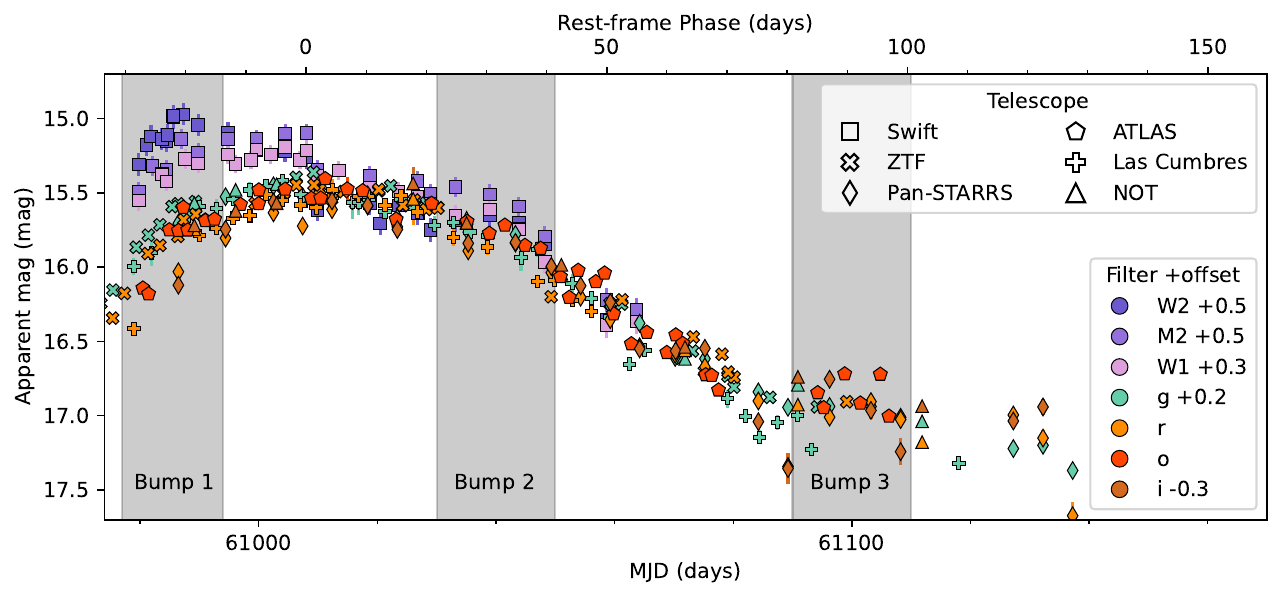}
    \caption{Light curve of TDE\,2025aarm including data from ATLAS, ZTF, LCO, NOT, Pan-STARRS and \textit{Swift}. Grey shaded regions denote where the bumps in the light curve are present. Filters have been vertically offset to overlap in regions without bumps. Magnitudes are corrected for MW and host extinction, and presented in AB magnitudes.}
    \label{fig:lc_stacked}
\end{figure*}

\subsection{Bumpy Light Curves} \label{sec:bumps}

Inspection of the multi-band light curves reveals the presence of significant deviations from a smooth monotonic evolution, particularly around peak and at later phases. This can be seen in Figure \ref{fig:lc_stacked} where the filters in which the bumps are most prominent have been overlaid for clarity. As discussed in Section \ref{sec:lc}, quadratic fits to the peak regions in each band are shown in Figure \ref{fig:peak_fits}. The premature peak of the $UVW2$ band with respect to other filters, coincides with a bump in the $B$ and $r$ bands around MJD 60987 ($\sim20$ days pre-peak), although it becomes progressively less prominent towards redder wavelengths. This behaviour suggests that the early peak in flux in $UVW2$ is not associated with the primary TDE peak, but instead represents an additional emission component superimposed on the underlying light curve. This bump will henceforth be referred to as Bump 1. Figure \ref{fig:lc_stacked} also shows tentative evidence for substructure within Bump 1, with the broader feature potentially comprising two distinct peaks.

In addition to the primary peak-associated structure, we identify a secondary blue bump beginning around MJD 61030, (Bump 2: +20 days post peak) most prominently observed in the bluer bands. A further significant rise in flux is observed from MJD 61090, particularly in the $r-$band and other redder filters (Bump 3: +80 days post peak). However, no contemporaneous UV observations are available at this latter epoch, preventing us from determining whether these features are also accompanied by enhanced UV emission. The phases of these bumps are shown in Figure \ref{fig:lc_stacked}.

To assess the significance of these bumps, we first defined three windows encompassing each bump. For each telescope and filter combination, these were masked off and a second-order spline was fit to the remaining photometry. Gaussian components were then fitted to the residuals within each window, provided that at least three data points were available. We calculated both the the Bayesian Information Criterion \cite[BIC;][]{Schwarz1978} and Akaike Information Criterion \cite[AIC;][]{Akaike1974} within each window for a spline fit alone, and for the spline+Gaussian fit, defining

\begin{equation}
    \Delta{\rm BIC} =
    {\rm BIC}_{\rm spline}-
    {\rm BIC}_{\rm spline+Gaussian}.
\end{equation}

and

\begin{equation}
\Delta{\rm AIC} =
{\rm AIC}_{\rm spline}-
{\rm AIC}_{\rm spline+Gaussian}.
\end{equation}

The second model contains three additional parameters: the amplitude, centre, and width of the Gaussian. Positive values of $\Delta{\rm BIC}$ and $\Delta{\rm AIC}$ therefore indicate that the improvement in the fit outweighs the penalty associated with adding these parameters. The results of these fits are presented in Table \ref{tab:BIC}.  We regard $\Delta{\rm BIC}>10$ as strong evidence for the presence of a bump, while negative values indicate the addition of the Gaussian does not strongly improve the fit. The AIC provides a complementary model comparison with a different penalty for additional parameters. Some of the best-sampled features yield very large $\Delta$ values because the improvement in the fit accumulates across many data points, whereas the BIC penalty increases only logarithmically with the number of fitted parameters. We therefore interpret values substantially above 10 simply as strong evidence for the feature, rather than attaching significance to their exact value.

\begin{deluxetable}{cc|cc|cc|cc}
    \tablecaption{BIC and AIC comparison for the candidate light-curve bumps. Within each bump window, we compare a fixed smooth-spline baseline with the same baseline plus a Gaussian component. We define $\Delta{\rm BIC}={\rm BIC}_{\rm spline}-{\rm BIC}_{\rm spline+Gaussian}$ and $\Delta{\rm AIC}={\rm AIC}_{\rm spline}-{\rm AIC}_{\rm spline+Gaussian}$, such that positive values favour the presence of a bump. Dashes indicate insufficient coverage. \label{tab:BIC}} 
    \tablehead{Filter & Telescope & \multicolumn{2}{c}{Bump 1} & \multicolumn{2}{c}{Bump 2} & \multicolumn{2}{c}{Bump 3} \\
     & & $\Delta$BIC & $\Delta$AIC & $\Delta$BIC & $\Delta$AIC & $\Delta$BIC & $\Delta$AIC }
    \startdata
        \hline
        $UVW2$ & \textit{Swift} & 21 & 22 & 50 & 49 & - & - \\
        $UVM2$ & \textit{Swift} & 0.7 & -1 & 7 & 5 & - & - \\
        $UVW1$ & \textit{Swift} & -3 & -5 & 1 & 0 & - & - \\
        $g$ & ZTF               & 222 & 221 & - & - & - & - \\
        $g$ & Pan-STARRS        & - & - & - & - & 16 & 15 \\
        $g$ & Las Cumbres       & -3 & -6 & 54 & 52 & - & - \\
        $r$ & ZTF               & 3171 & 3170 & - & - & - & - \\
        $r$ & Pan-STARRS        & - & - & - & - & -1 & 0 \\
        $o$ & ATLAS             & 545 & 545 & 408 & 407 & 5645 & 5645 \\
        $i$ & Pan-STARRS        & - & - & - & - & 9 & 9 \\ 
        \hline \hline
    \enddata
\end{deluxetable}

As seen in Table \ref{tab:BIC}, Bump 1 is strongly supported in all bands with sufficient coverage. The exceptions are the $UVW1$ band, where the inclusion of a Gaussian component does not significantly improve the fit, and the Las Cumbres $g-$band which was not well fit by the underlying spline. Bump 2 is likewise supported in all bands. Bump 3 is sampled only at redder wavelengths and is strongly supported in all bands with coverage, except for Pan-STARRS $r$, where the Gaussian component is marginally not preferred.

Such structure has been observed in a growing number of TDEs. AT\,2019azh displayed both pre-peak and a post-peak bumps \citep{Hinkle2021, Liu2022, Faris2024}. Other events have also shown departures from smooth light-curve evolution, including the UV undulations seen in AT\,2018fyk \citep{Wevers2019}, the multicolor bump in AT\,2018iih \citep{Hammerstein2023}, and the rapid cooling in the UV seen in AT\,2023clx \citep{Charalampopoulos2024}.

\subsection{Blackbody Parameters} \label{sec:bb_params}

\begin{figure}
    \centering
    \includegraphics[width=1\columnwidth]{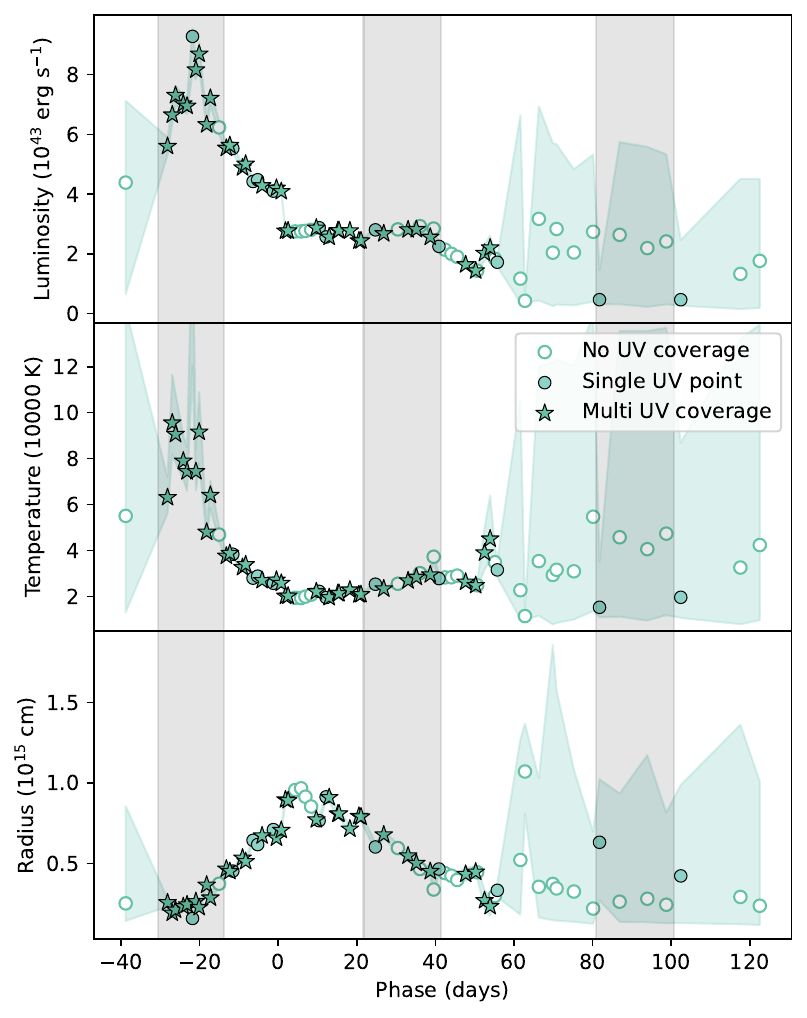}
    \caption{Evolution of the blackbody parameters from SED fitting. UV data were interpolated where photometry was unavailable before 60 days. Filled stars indicate where there were multiple UV points to anchor the fits, filled circles indicate where there was a single UV band, and unfilled circles show where there was no UV coverage at that epoch. Grey shaded regions indicate the locations of Bump 1, 2, and 3 respectively.  
    \textit{Top:} Bolometric luminosity. 
    \textit{Middle:} Blackbody temperature. 
    \textit{Bottom: }Blackbody radius.}
    \label{fig:BB_fits}
\end{figure}

To characterise the transient SED evolution, we fit blackbody models to the photometric data at individual epochs. Epochs were constructed by grouping all data points within a window of 1 day from the earliest unassigned observation, adopting the mean MJD of each group as the epoch time. Where more than one observation existed in a given telescope and filter combination, the weighted mean of the flux was taken. Each epoch was then fit using a Markov Chain Monte Carlo (MCMC) approach implemented with the python package \texttt{emcee} \citep{Foreman-Mackey2013} approach to fit the blackbody parameters. Fits were only performed where the SED covered at least 3 filters. For the remaining epochs lacking UV coverage, the three \textit{Swift} UV bands were linearly interpolated between the nearest bracketing measurements, up to a phase of 60 days, after which no \textit{Swift} coverage was available. We adopted uniform priors on the radius in the range $1\times10^{13} - 2\times10^{16}$\,cm, and on the temperature in the range $1\times10^{3} - 2\times10^{5}$\,K. \citet{Arcavi2022} showed that for optical-only SEDs, temperatures above $\sim3.5\times10^{4}$\,K become uncertain, while for temperatures above $\sim5\times10^{4}$\,K only lower limits can generally be obtained. The inclusion of UV coverage, however, allows higher temperatures to be constrained more reliably.

The results of the SED fitting are shown in Figure \ref{fig:BB_fits}. Individual blackbody fits are shown in Figure \ref{fig:BB_fits_grid}. To assess the sensitivity of our results to the adopted temperature prior and use of interpolation, we repeated the fitting procedure with an upper temperature limit of $5\times10^{4}$\,K and without any interpolation. These fits are presented in Figure \ref{fig:BB_fits_high}. Epochs where multiple UV filters were available are indicated by a filled star. Filled circles indicate where there was only 1 UV filter to anchor the SED fits and the other bands were interpolated, and open circles are epochs where no UV filters were available and all UV bands were interpolated. We can see that for epochs when interpolation was not possible and only optical bands were available, there is a much greater uncertainty associated with the SED fits, as expected. 

The bolometric luminosity reaches a maximum of $L_{\rm{bol}}= 8.69^{+0.39}_{-0.43} \times 10^{43}$\,erg\,s$^{-1}$ at $-20$ days using the non-interpolated points. Following this, the luminosity decreases until the peak of the $g-$band light curve, then plateaus at $\sim2-3\times 10^{43}$\,erg\,s$^{-1}$. The plateau corresponds to the duration of the broad optical peak, and Bump 2. Before the initial peak, there is also a preceeding bump in luminosity reaching $L_{\rm{bol}}= 7.31^{+0.32}_{-0.34} \times 10^{43}$\,erg\,s$^{-1}$, corresponding to the substructure within Bump 1 in the UV bands. Another rise in luminosity is seen around 50 days which corresponds to a small undulation seen in the optical light curve, suggesting there is a potential rebrightening in the UV that we did not observe. Although we see Bump 3 in the redder filters, it is difficult to constrain whether there is a corresponding rise in the blackbody luminosity due to the lack of UV coverage at these epochs.

The temperature evolution of TDE\,2025aarm is displayed in the middle panel of Figure \ref{fig:BB_fits}, with the temperature peaking at $T_{\rm{BB}} = 9.55^{+2.08}_{-1.35}\times10^{4}$\,K at $-27$\,days. Once again, this peak corresponds to the substructure in Bump 1. The temperature then begins to decrease before peaking again at $T_{\rm{BB}} = 9.16^{+1.73}_{-1.36}\times10^{4}$\,K in conjunction with the peak of the bolometric luminosity\footnote{There is a single interpolated temperature point reaching $\sim150,000$ K, however there is only a single UV data point to anchor the blackbody fit and therefore we exclude it from this analysis.}. Following this initial high-temperature phase, the temperature declines rapidly until the main optical peak. Similar rapid cooling has been seen in other TDEs such as AT\,2019qiz \citep{Nicholl2020}, which has been used to support the presence of outflows. Post optical peak, the temperature begins to gradually rise up to $\sim30,000$\,K. This secondary increase is broadly consistent with Bump 2 discussed in Section \ref{sec:bumps}. After a brief decline from 40 days onwards, the temperature sharply rises once more, reaching $\sim45,000$\,K by around 60 days post-peak.


Overall, TDE\,2025aarm exhibits temperatures at the hotter end of the distribution for optically selected TDEs. For example, the sample of 30 TDEs presented by \citet{Hammerstein2023} has peak temperatures below $40,000$\,K. However, the non-parametric light-curve model used by \citet{Hammerstein2023} fits the temperature at 30-day intervals, a cadence chosen to accommodate the available UV coverage across their sample. Meanwhile, the \citet{Yao2023} study of 33 TDEs assumes a  fixed temperature near peak when determining the peak blackbody properties. In contrast, the dense multi-band coverage of TDE\,2025aarm allows independent blackbody fits to individual SEDs at a much finer cadence, resolving substantial temperature evolution on timescales that would not be captured by these approaches. \citet{Hinkle2020} similarly performed epoch-by-epoch SED fitting, although their analysis adopted a flat temperature prior with an upper bound of $55,000$,K. TDE\,2025aarm was observed in the UV more than three weeks before maximum light, allowing us to trace the early cooling evolution that is typically poorly sampled in TDEs discovered near peak. This overall evolution is unusual compared to optically selected TDE samples, which typically show relatively constant temperatures with the evolution in bolometric luminosity driven primarily by changes in the blackbody radius \citep[e.g.][]{Hinkle2020,Hammerstein2023}. 

The bottom panel of Figure \ref{fig:BB_fits} shows the evolution of the blackbody radius which reaches a maximum of $R_{\rm{BB}}= 9.10^{+0.26}_{-0.27}\times10^{14}$\,cm at a phase of 12 days. The radius exhibits an overall rise followed by a gradual decline, with low-amplitude fluctuations superimposed on this long-term evolution. While these fluctuations may reflect genuine variability, they may also be the result of intrinsic scatter in the data. There is also an initial peak apparent around the time of optical peak. However, this feature is less well constrained, as some of the intervening blackbody fits rely on interpolated photometry.

\begin{figure*}
    \centering
    \includegraphics[width=2\columnwidth]{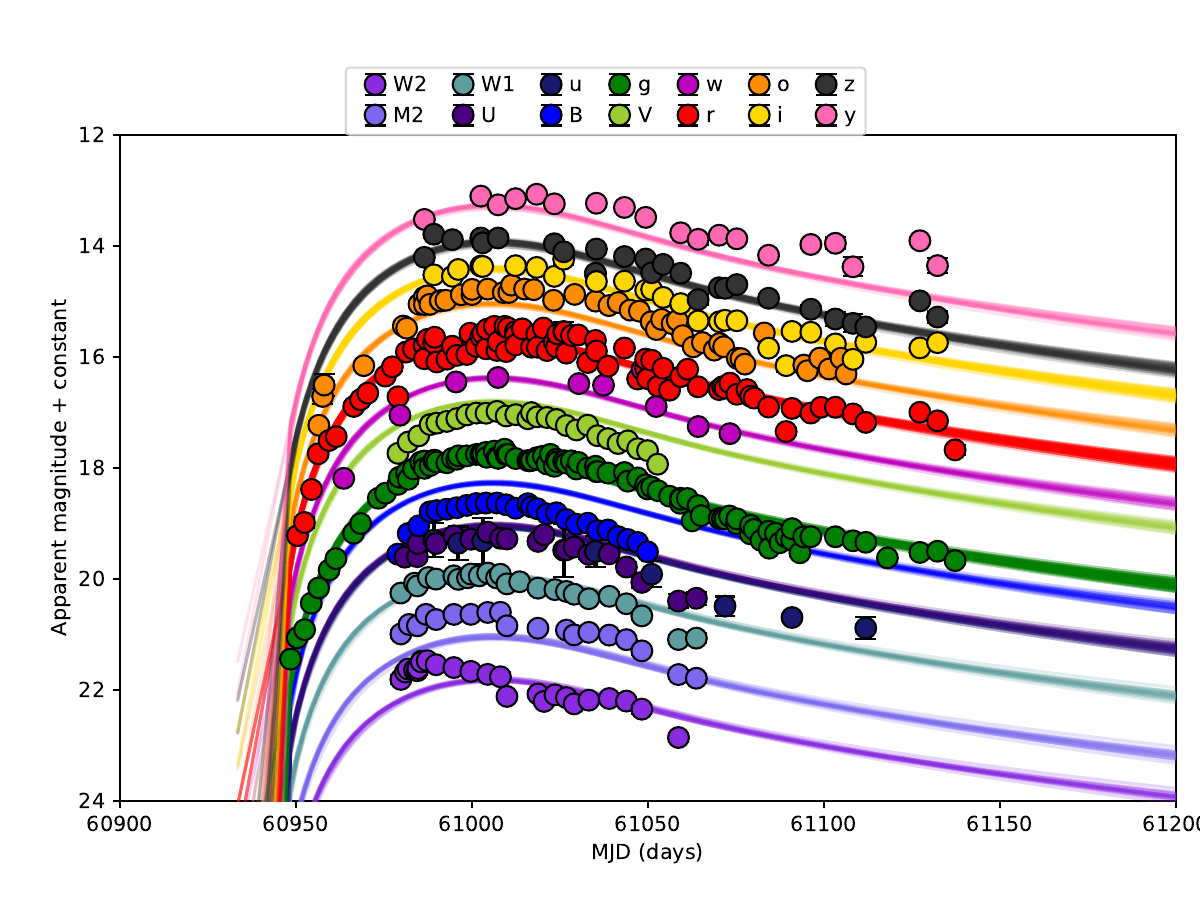}
    \caption{Light curve modelling of TDE\,2025aarm using the \texttt{tde} model in \textsc{MOSFiT}.}
    \label{fig:mosfit}
\end{figure*}

\begin{figure*}
    \centering
    \includegraphics[width=2\columnwidth]{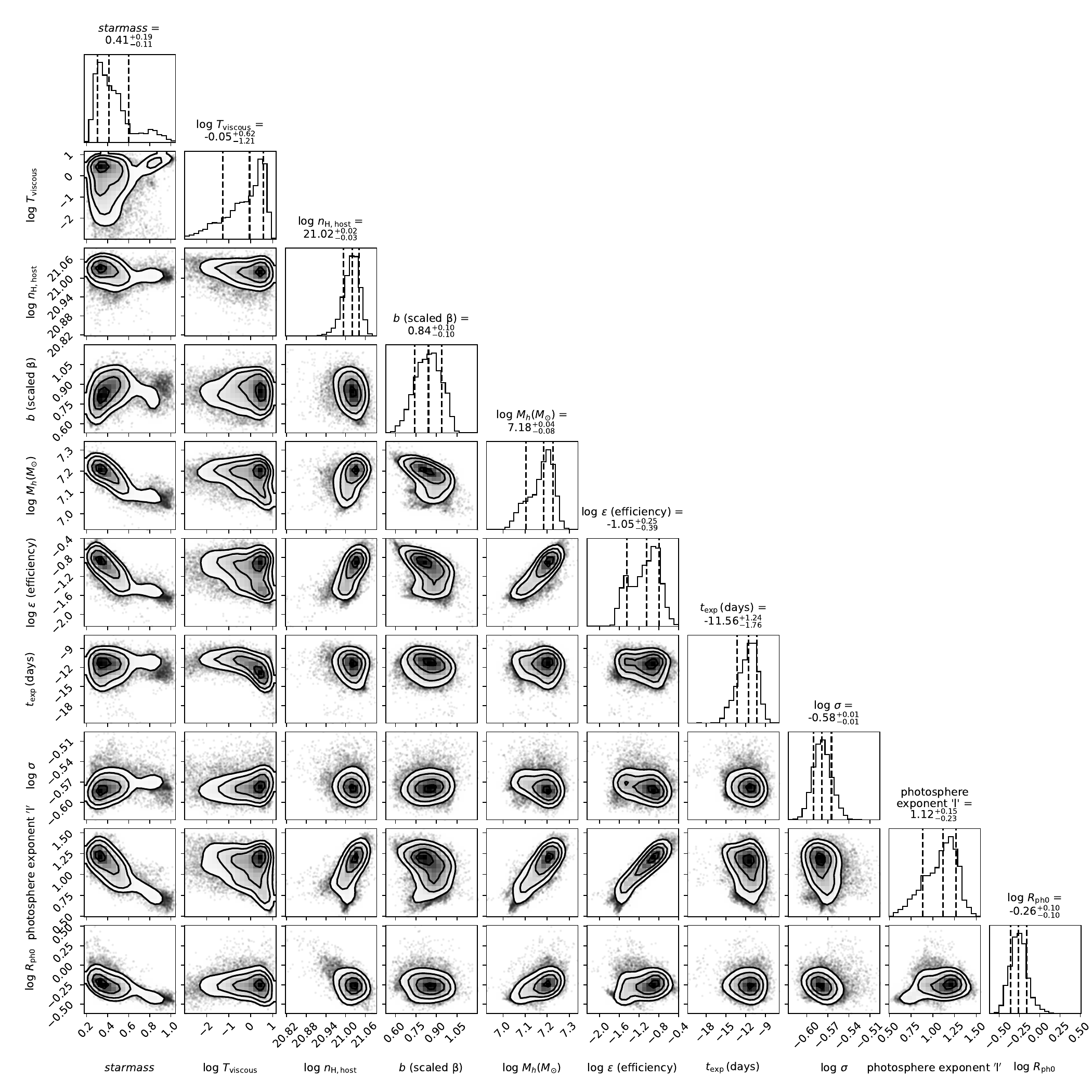}
    \caption{Corner plot of the posteriors for the \texttt{tde} \textsc{MOSFiT} model applied to the light curve of TDE\,2025aarm.}
    \label{fig:mosfit_corner}
\end{figure*}

\subsection{Light Curve Modelling} \label{sec:mosfit}

We model the multi-band light curve of TDE\,2025aarm using the Modular Open-Source Fitter for Transients \citep[\textsc{MOSFiT}][]{Guillochon2018}, adopting the standard \texttt{tde} model framework \citep{Mockler2019}. The fitting procedure simultaneously models the observed photometry across all bands, yielding posterior distributions for key physical parameters including the black hole mass, stellar mass, and impact parameter. We performed fits to both the full light curve and up to 80 days post peak using the default priors \citep{Mockler2019}. The latter restricts the fit to the peak and early decline before the TDE enters the plateau stage, where the emission is expected to more closely trace the fallback evolution. As the inferred parameters were consistent between the two fits, we adopt the results from the full light curve throughout the remainder of this work.

The best-fitting model is shown in Figure \ref{fig:mosfit} and provides a reasonable overall fit to the light curve evolution. The model successfully reproduces the rise and overall decline in the $g$, $r$, and $o$ bands, but struggles to capture some of the finer structure in the data. In particular, the model consistently overestimates the $B$-band luminosity by $\sim0.5$\,mag and under predicts the $UVM2$ evolution. It also struggles to reproduce the late-time luminosity in the $y$ band. This excess emission could potentially indicate an additional red emission component such as dust reprocessing, though this is typically observed in redder bands \citep[e.g.][]{Reynolds2026}. The model also does not reproduce the short timescale bumps seen in the light curve. This is to be expected, as \textsc{MOSFiT} assumes a relatively smooth evolution in the fallback and reprocessing of material, whereas the observed variability may arise from processes such as changes in the reprocessing layer, delayed circularisation, stream-stream collisions, or variability in outflow properties.

A further limitation of the model is its difficulty in reproducing the UV light curve near peak. Similar discrepancies have been seen in previous events such as AT\,2019qiz \citep{Nicholl2020} and AT\,2019azh \citep{Faris2024}, where the observed UV emission is often stronger than predicted by standard models. This behaviour is also seen in TDE\,2025aarm, where the model under predicts the $UVW2$ flux around peak. In this case, the discrepancy may be exacerbated by the UV bump (Bump 1) discussed in Section \ref{sec:bumps}.

However, we emphasise that the \textsc{MOSFiT} model is used here primarily to reproduce the overall behaviour of the light curve, rather than to derive precise physical parameters. The model is based on hydrodynamical simulations and semi-analytic prescriptions of the tidal disruption process \citep{Mockler2019}, and therefore depends on simplifying assumptions about uncertain processes such as circularisation, energy dissipation, and reprocessing. In addition, the mechanisms responsible for the optical/UV emission in TDEs are still not fully understood, which further limits the physical interpretation of the fitted parameters.


A corner plot of the inferred parameter posteriors can be seen in Figure \ref{fig:mosfit_corner}. We find a black hole mass of $log(M_{\rm BH}/M_{\odot}) = 7.18^{+0.20}_{-0.22}$, incorporating the model systematics from \citet{Mockler2019} into the uncertainty. This value is consistent with the black hole mass found from the light curve timescales in Section \ref{sec:lc}, as well as the mass found from velocity dispersion of the host lines in Section \ref{sec:host}. 
The model also finds a best fit to the disruption of a star with mass $M_{\star} = 0.41^{+0.19}_{-0.11}$\,M$_{\odot}$, with an additional systematic uncertainty of $\pm0.66$ dex associated with the \textsc{MOSFiT} model \citep{Mockler2019}.. The low preferred stellar mass may also be related to the relatively low inferred impact parameter and the degeneracy between stellar mass and impact parameter within the \textsc{MOSFiT} TDE model \citep{Guillochon2013, Mockler2019, Angus2026}.

The peak of the bolometric light curve can also be modelled to constain the black hole and stellar masses using \textsc{TDEmass} \citep{Ryu2020}. \textsc{TDEmass} uses the peak bolometric luminosity and temperature to infer the properties of the disrupted star and black hole under a shock-powered model for the optical/UV emission. The high peak temperature of TDE\,2025aarm favours the disruption of a star with mass $M_{*}=0.72^{+0.15}_{-0.08}\,\rm{M}_{\odot}$ by a black hole with mass $log(M_{\rm BH}/M_{\odot}) = 5.96^{+0.56}_{-0.48}$. This black hole mass is substantially lower than those inferred from the host-galaxy velocity dispersion and \textsc{MOSFiT} modelling. This discrepancy likely reflects the different physical assumptions underlying the methods. \textsc{TDEmass} assumes that the observed optical/UV emission is powered by shocks associated with debris circularisation, such that the unusually high temperature of TDE\,2025aarm strongly influences the inferred masses.

\begin{figure*}
    \centering
    \includegraphics[width=1.9\columnwidth]{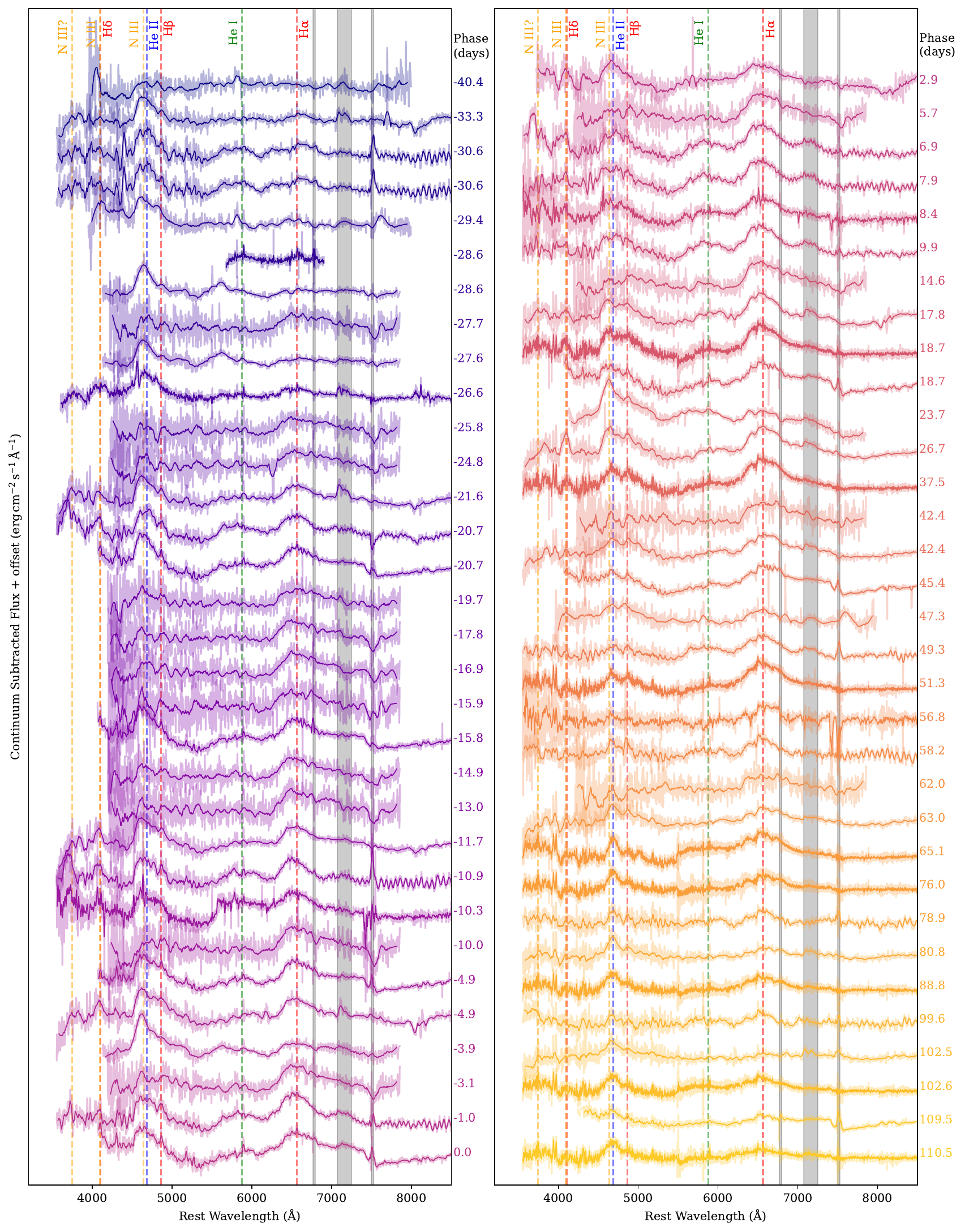}
    \caption{The optical spectral evolution of TDE\,2025aarm with the continuum subtracted using a cubic function. For visualisation the spectra have been smoothed using a Savitzky-Golay filter shown as a darker line. Includes data from SPRAT, ALFOSC, FLOYDS, B\&C, Mookodi, BlueChannel, LRS2, Binospec, SpUpNIC, WiFeS, and X-Shooter. Possible line identifications are denoted with vertical dashed lines at their rest wavelengths.}
    \label{fig:spectra}
\end{figure*}

\begin{figure}
    \centering
    \includegraphics[width=1\columnwidth]{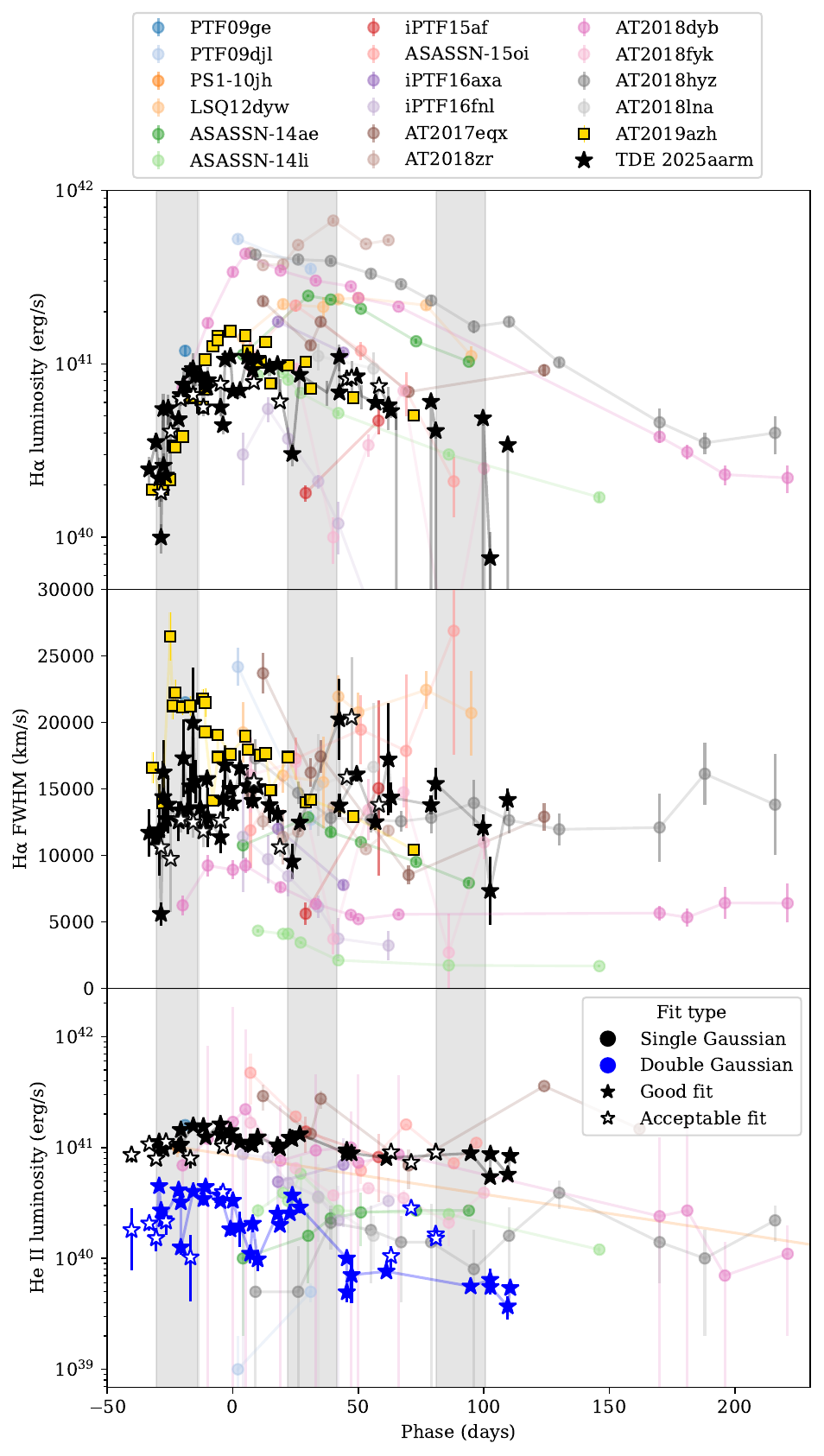}
    \caption{Evolution of line luminosities and widths. The quality of the fits are assessed by comparing the root-mean-square scatter of the residuals with the noise measured in continuum regions on either side of the profiles. Comparison objects are from \citet{Charalampopoulos2022} and \citet{Faris2024}. Grey shaded regions indicate the locations of Bump 1, 2, and 3 respectively. \textit{Top:} The H$\alpha$ line luminosity. \textit{Middle:} The FWHM of the Gaussian fit to the H$\alpha$ profile. \textit{Bottom:} The He II line luminosity if fit with a single Gaussian, or double Gaussian.}
    \label{fig:line_lums}
\end{figure}

\section{Spectral Evolution}\label{sec:spec_ev}

The spectral evolution of TDE\,2025aarm is shown in Figure \ref{fig:spectra}, with spectra spanning $-40$ to $+110$ days from maximum light. The most prominent feature is a broad emission complex centred around $\sim4600$\,\AA. Similar to \cite{Simongini2026} we attribute this feature to a blend of He II $\lambda$4686, H$\beta$ $\lambda$4861, and N III $\lambda$4640. We also detect the emergence and strengthening of H$\alpha$ at 6563\,\AA\ over time. The feature $\sim$4100\,\AA\ could be attributed to either H$\delta$ $\lambda$4102 or N III $\lambda\lambda$4097, 4104. However, as identified by \citet{Baldini2026}, the lack of H$\gamma$ emission makes the N III identification more likely. In addition, we identify an emission feature around $\sim3800$\,\AA. This could arise from Bowen fluorescence with N III $\lambda$3746, N III $\lambda$3754 and O III $\lambda$3760 close by. Broad H, He, and Bowen fluorescence features such as these are commonly observed in optical TDE spectra \citep{Blagorodnova2019, Leloudas2019, vanVelzen2021}. We find no strong evidence for He I emission throughout the spectroscopic sequence, although a weak blueshifted excess may be present near the expected wavelength of He I $\lambda5876$.

Line luminosities were calculated by first fitting a pseudo-continuum on either side of the line. Isolated lines such as H$\alpha$ and N III $\lambda$4100 were then fit with a single Gaussian. While a single Gaussian provides a good description of the H$\alpha$ profile at early times, from $+20$,d onwards excess emission emerges on both the blue and red sides of the line. These regions were masked when fitting the central H$\alpha$ component, with the masked regions shown in Figure \ref{fig:Halpha_cont}. We discuss the origin and evolution of this excess emission further in Section \ref{sec:disk}. We also tested a Lorentzian profile for the central H$\alpha$ component, but found that it did not provide an improved fit over the Gaussian profile. A Lorentzian profile was also fit but did not produce a better fit to the H$\alpha$ profile. The blended line region around He II was fit with a double Gaussian fit to account for the blending of H$\beta$ to the right of the feature. Although a triple-Gaussian model was fitted to separate the He II and N III features, the fit instead converged on two overlapping components with the same central wavelengths, indicating that the two features could not be reliably decomposed. The quality of the fits are assessed by comparing the root-mean-square (RMS) scatter of the residuals, with the noise measured in continuum regions on either side of the profiles. Fits with residual scatter less than 1.5 times the continuum noise were classified as good, while those with ratios between 1.5 and 3 were classified as acceptable.

The evolution of luminosity of the H$\alpha$ line is shown in the top panel of Figure \ref{fig:line_lums}. The line peaks at approximately $L_{\mathrm{H}\alpha} \sim 10^{41}$\,erg\,s$^{-1}$ around 10 days post-peak. Comparing the evolution to the sample of well-observed TDEs from \citet{Charalampopoulos2022}, the delayed H$\alpha$ peak relative to the optical light curve is consistent with other events that have sufficiently early spectroscopic coverage. At peak, the luminosity is comparable to events such as ASASSN-14li, although ASASSN-14li declines more rapidly \citep{Miller2015}. Several other TDEs, including AT\,2018zr, AT\,2018hyz, and AT\,2018fyk, \citep{Holoien2019b, Short2020, vanVelzen2021, Wevers2019} also exhibit non-monotonic line evolution, though their sparser spectroscopic sampling makes such variability more difficult to confirm. At later phases ($\gtrsim50$ days), the decline rate of the H$\alpha$ luminosity is broadly consistent with that seen in events such as ASASSN-14ae \citep{Holoien2014} and AT\,2018hyz.

The second panel of Figure \ref{fig:line_lums} shows the evolution of the H$\alpha$ full width at half maximum (FWHM). The line exhibits significant variability, initially broadening to $\sim15,000$\,km\,s$^{-1}$ by -15 days before narrowing to $\sim12,000$\,km\,s$^{-1}$ by -10 days. Following this, the FWHM increases up to a maximum of $16200 \pm 1100$\,km\,s$^{-1}$ around peak before decreasing again to $\sim10,000$\,km\,s$^{-1}$. A third broadening episode follows coinciding with the start of Bump 2. The line reaches a maximum width of $15,400\pm300$\,km\,s$^{-1}$ at $\sim50$ days, before gradually settling back to $\sim12,000$\,km\,s$^{-1}$ at later epochs. While other TDEs in the \citet{Charalampopoulos2022} sample show similarly broad lines, few display such pronounced short-timescale variability. The changes in line width also appear to coincide with bumps in the optical light curve, suggesting a possible connection between the continuum luminosity and the kinematics of the emitting material.

The He II feature is significantly blended with N III and H$\beta$, making a clean decomposition challenging. To estimate the He II luminosity, we fit the profile using both a single-Gaussian model and a two-component model that includes an additional Gaussian centred on H$\beta$. The resulting luminosity evolution is shown in Figure \ref{fig:line_lums}. We show that the He II luminosity peaks approximately 5 days before optical maximum, reaching a luminosity of $\sim1.5 \times 10^{41}$\,erg\,s$^{-1}$. Unlike H$\alpha$, there is no clear evidence for a systematic lag relative to the optical light curve. The He II luminosity declines steadily over time, remaining broadly consistent with values observed in other optically selected TDEs.


\begin{figure}
    \centering
    \includegraphics[width=1\columnwidth]{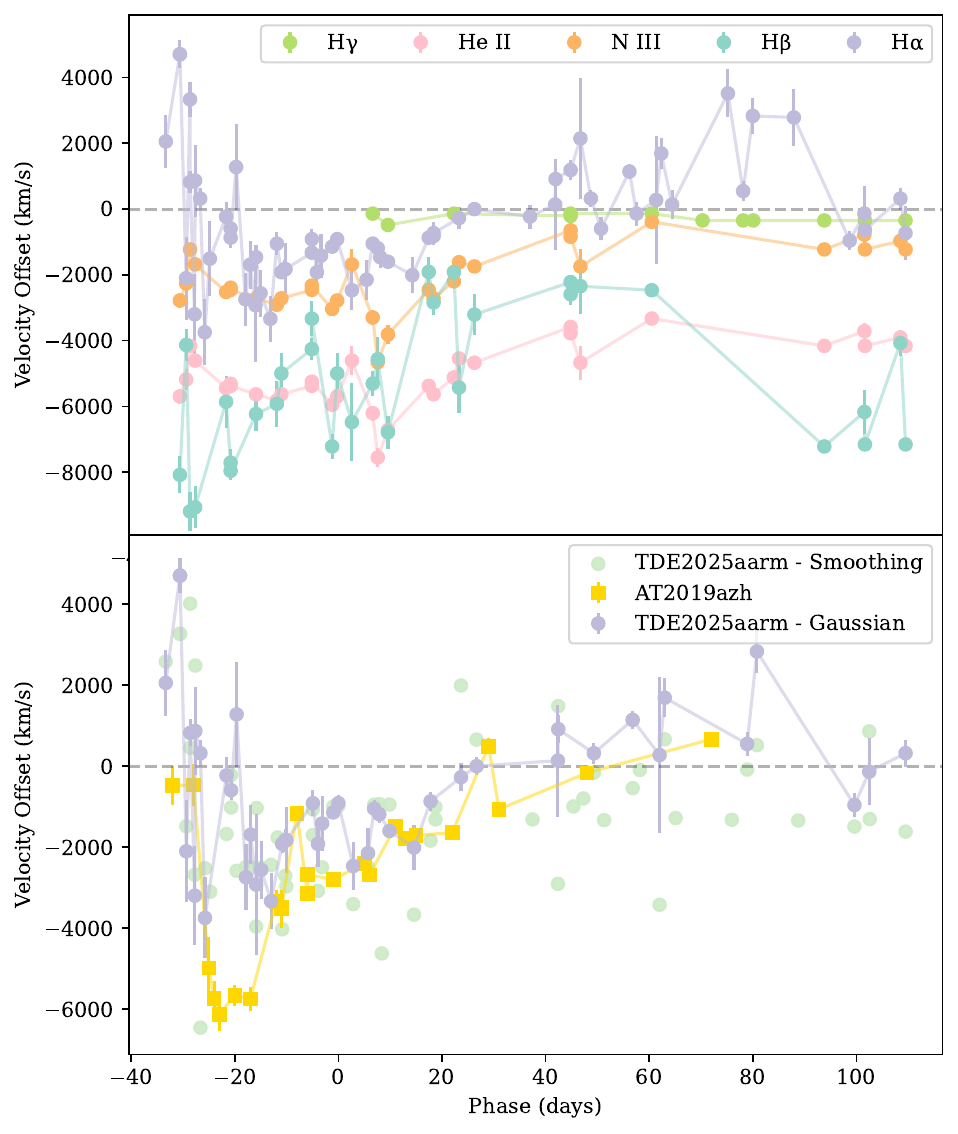}
    \caption{Evolution of the measured velocity offsets. \textit{Top:} The velocity offsets for the emission lines TDE\,2025aarm. The He II and N III velocities are derived from the same Gaussian fit, corresponding to the two possible identifications of the emission feature. All lines exhibit a gradual decrease in blueshift until approximately 70 days after peak, after which they show an increase in blueshift. \textit{Bottom:} A comparison of the H$\alpha$ blueshift velocity between TDE\,2025aarm and AT\,2019azh \citep{Faris2024}. For TDE\,2025aarm, the blueshift has been caluclated both by fitting a Gaussian for the profile, and by heavily smoothing the line profile and finding the maximum. Both objects show a very similar velocity evolution.}
    \label{fig:blueshifts}
\end{figure}


\begin{figure}
    \centering
    \includegraphics[width=1\columnwidth]{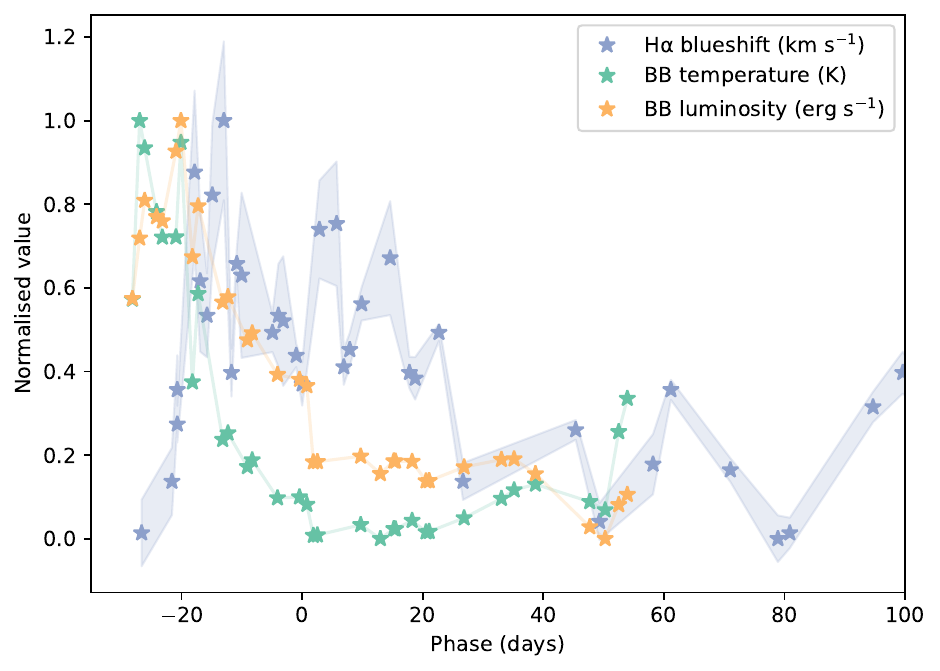}
    \caption{Evolution of H$\alpha$ blueshift, blackbody temperature, and blackbody luminosity. All values have been normalised between zero and one. All three parameters show a double peak, as well as a late time rise around 50 days.}
    \label{fig:blueshift_offsets}
\end{figure}

\subsection{Line Blueshifts and Continuum Properties} \label{sec:blueshifts_continuum}

From the Gaussian fitting described in Section \ref{sec:spec_ev}, we measured the blueshift of the line centres. The blueshifts of H$\alpha$, H$\beta$, H$\gamma$, and He II $\lambda$4686 are shown in Figure \ref{fig:blueshifts}.

The H$\alpha$ profile initially appears slightly redshifted at the earliest epochs before evolving toward progressively larger blueshifts. We identify three distinct increases in the H$\alpha$ blueshift: the first beginning around $-20$ days, the second around $+10$ days, and the last around 80 days relative to optical peak. 
The other Balmer lines broadly follow the same behaviour. The H$\beta$ measurements exhibit substantially larger scatter, making short-timescale variations more difficult to identify. Nevertheless, the overall trend in H$\beta$ is consistent with the variability seen in the H$\alpha$ velocity. There were fewer epochs where H$\gamma$ could be reliably fit however we see the same trend of the smallest blueshift velocity occurring around 60 days with an increase on either side.

We plot the blueshift of the line around 4800\,\AA\ under both assumptions that the line could originate from He II or N III in Figure \ref{fig:blueshifts}, although it is likely the feature is dominated by He II. In either interpretation, the line exhibits variability that broadly tracks the structure of the optical light curve, with the increase in blueshift up to +10 days appearing more pronounced than in H$\alpha$ and H$\gamma$. At later phases, however, the blueshift evolution is less extreme than that observed in the Balmer lines. 

More generally, the overall decrease in line blueshift over the first $\sim60$ days is broadly consistent with trends observed in other TDEs \citep{Charalampopoulos2022}. This behaviour can be explained by an outflowing line-emitting region that becomes increasingly optically thin with time, allowing a greater contribution from material on the receding side of the outflow and shifting the line centroid towards its rest wavelength. However, TDE\,2025aarm differs from this general trend through both its short-timescale blueshift variability and the increase in blueshift at late times.

To test whether the presence of a disk-like component affected the velocity offsets measured for TDE\,2025aarm, we used two independent methods. Our primary measurements use the centre of the fitted Gaussian to determine the velocity offset. As a sanity check, we also heavily smoothed each line profile using a Savitzky-Golay filter and measured the wavelength of the resulting maximum, avoiding any assumptions about the underlying line profile. Although this method provides a cruder estimate and can be affected by host-galaxy absorption features, both methods recover the same overall velocity evolution as shown in the bottom panel of Figure \ref{fig:blueshifts}.

The bottom panel of Figure \ref{fig:blueshifts} also highlights the velocity offset of the H$\alpha$ line in AT\,2019azh. The evolution is remarkably similar to that of TDE\,2025aarm, both in the magnitude and temporal evolution of the velocity offset. The main difference occurs around the maximum blueshift. While AT\,2019azh continues to a maximum blueshift of $\sim-6000$,km,s$^{-1}$, TDE\,2025aarm instead shows a temporary reversal towards rest velocity. Following this, both events tend towards zero velocity by 40 days. Although we see TDE\,2025aarm becomes increasingly blueshifted again at $\gtrsim80$ days, the spectroscopic coverage of AT\,2019azh does not extend to these phases.

Figure \ref{fig:blueshift_offsets} compares the evolution of the H$\alpha$ blueshift with the blackbody luminosity and temperature derived in Section \ref{sec:bb_params}. For visual clarity, all quantities have been normalised between 0 and 1, and the velocity offset has been inverted such that increasing values correspond to larger blueshifts. The first two velocity measurements, which show large inferred redshifts, have also been excluded from the figure.

Several phases of enhanced blueshift appear to coincide with structure in both the temperature and bolometric luminosity evolution, most notably the double peaked profile around maximum, and the late time rise after 50 days. To investigate this further, we performed a cross-correlation analysis between the H$\alpha$ blueshift evolution and the bolometric luminosity. Both parameters were first interpolated onto a uniformly sampled phase grid using one-dimensional linear interpolation, after which the \texttt{scipy} correlate function was used to compute the cross-correlation function. From this analysis, we find a preferred lag of $\sim -9.7$ days, implying that increases in luminosity precede increases in the H$\alpha$ blueshift by $\sim$-10 days. This may indicate that changes in the continuum emission drive subsequent changes in the velocity structure of the line-forming region. A similar analysis for the temperature evolution gives a preferred lag of -1.5 days, suggesting that variations in the temperature may also precede the observed increases in blueshift. 

In addition to the prominent features, all three parameters exhibit lower-amplitude variations between approximately $-20$ and $+40$ days that appear qualitatively similar. However, the significance of these fluctuations is difficult to assess, and they may be influenced by the intrinsic scatter and measurement uncertainties in the data.

Nevertheless, a qualitative correlation between the three quantities is apparent. In particular, the enhanced blueshift observed around $\sim-10$ days before optical peak and the late-time increase seen after $\sim50$ days coincide with periods of elevated luminosity and temperature (Bump 1 and Bump 3). Interestingly, AT\,2018fyk exhibits similar behaviour, with correlated evolution between the H$\alpha$ luminosity, velocity offsets, and broad structure in the continuum light curve \citep{Wevers2019,Charalampopoulos2022}. In contrast, while TDE\,2025aarm shows a comparable relationship between the continuum evolution and the H$\alpha$ velocity offsets, we find no corresponding variability in the H$\alpha$ luminosity. 

One possible interpretation is that these phases trace the presence of evolving outflows. Outflows have now been inferred in a growing number of TDEs through optical, UV, and radio observations. The optical peak of AT\,2019qiz has been interpreted as being powered by an outflow \citep{Nicholl2020}. Similarly, UV spectroscopy of AT\,2018zr revealed transient high-velocity absorption features resembling those seen in broad absorption line quasars, which \citet{Hung2019} interpreted as evidence for fast outflowing material generated during the disruption event. Additional support for outflow activity may come from radio observations of TDE\,2025aarm (Franz et al. in prep).

\begin{figure}
    \centering
    \includegraphics[width=1\columnwidth]{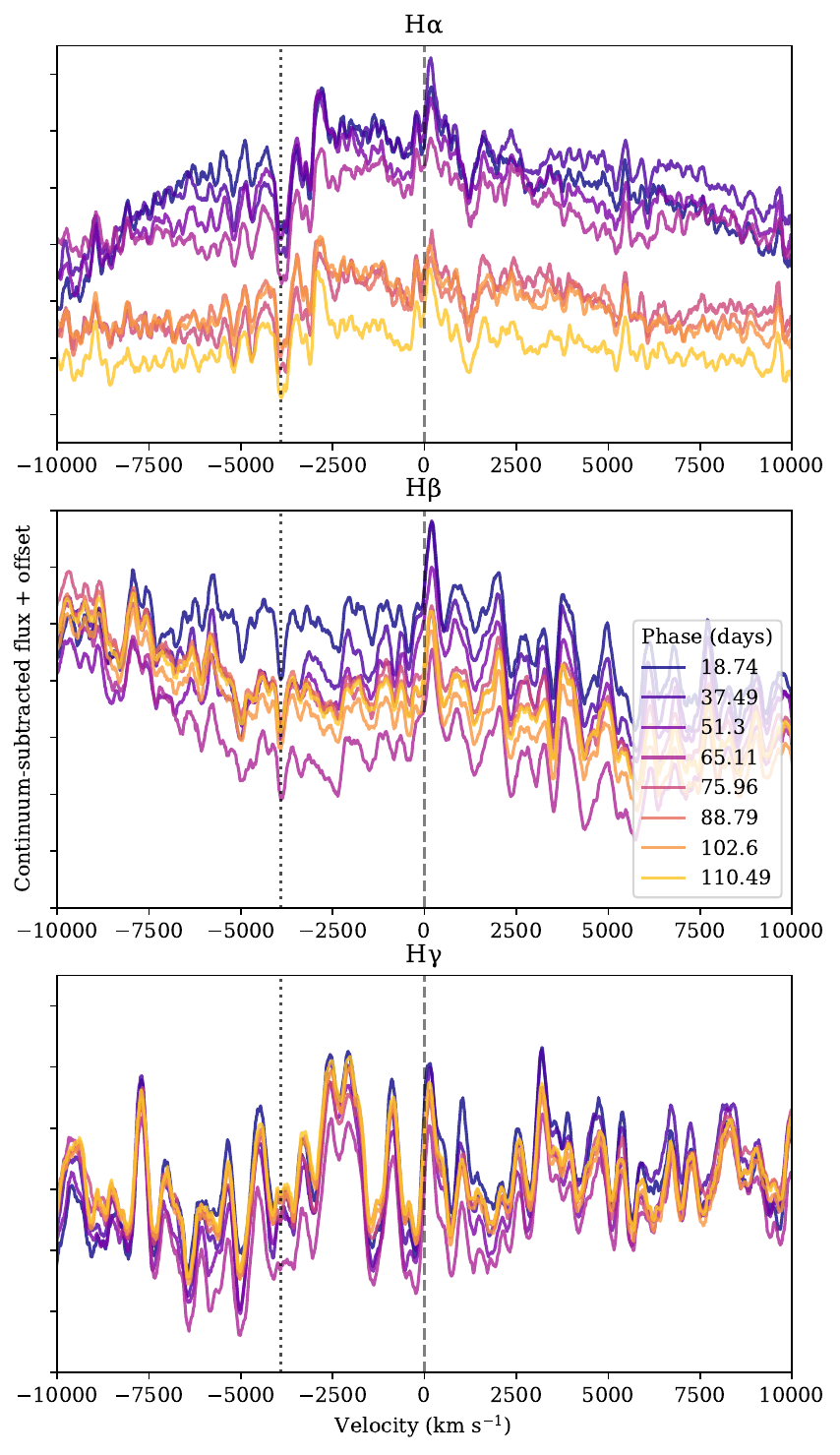}
    \caption{X-Shooter spectra of TDE\,2025aarm showing the velocity evolution of H$\alpha$, H$\beta$, and H$\gamma$. The dashed vertical line marks the rest-frame velocity, while the dotted line marks $-3900$\,km\,s$^{-1}$, highlighting the similar blueshifted feature present across the Balmer lines.}
    \label{fig:balmer_abs}
\end{figure}

Further evidence for outflowing material may be present in the Balmer line profiles themselves. An unidentified narrow absorption feature is visible within H$\alpha$ at a velocity of approximately $-3900$\,km\,s$^{-1}$, with no obvious corresponding host galaxy absorption line. To test whether this feature is associated with TDE\,2025aarm, we compare the Balmer lines in velocity space in Figure \ref{fig:balmer_abs}. A coincident dip is present at approximately the same velocity in H$\beta$ across all epochs in the X-Shooter spectra, with a possible counterpart in H$\gamma$, although the spectra are considerably noisier at these wavelengths. The presence of absorption at a common blueshift across multiple Balmer transitions suggests that the feature is intrinsic to the TDE and may trace material moving towards the observer, providing additional evidence for an outflow.

\begin{figure}
    \centering
    \includegraphics[width=1\columnwidth]{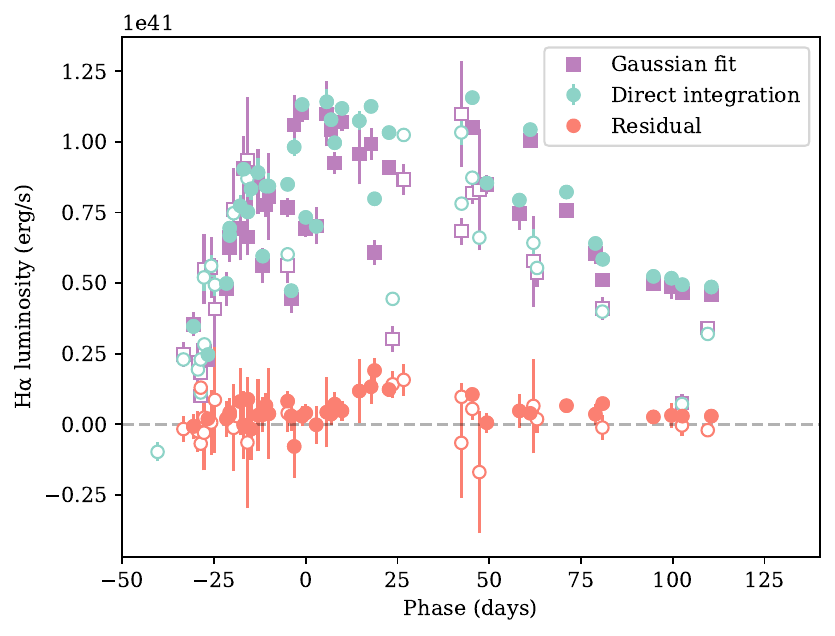}
    \caption{H$\alpha$ luminosity when directly integrating under the profile, and when fitting a Gaussian and integrating under that profile. In orange are the residuals when subtracting the Gaussian luminosity from the direct luminosity. Filled markers indicate epochs where the Gaussian fit was good, and open markers where the fit was acceptable. We can see that the Gaussian fits the profile well up to optical peak and then deviates significantly where the disk-like emission is most prominent.}
    \label{fig:direct_gaussian}
\end{figure}

\begin{figure}
    \centering
    \includegraphics[width=1\columnwidth]{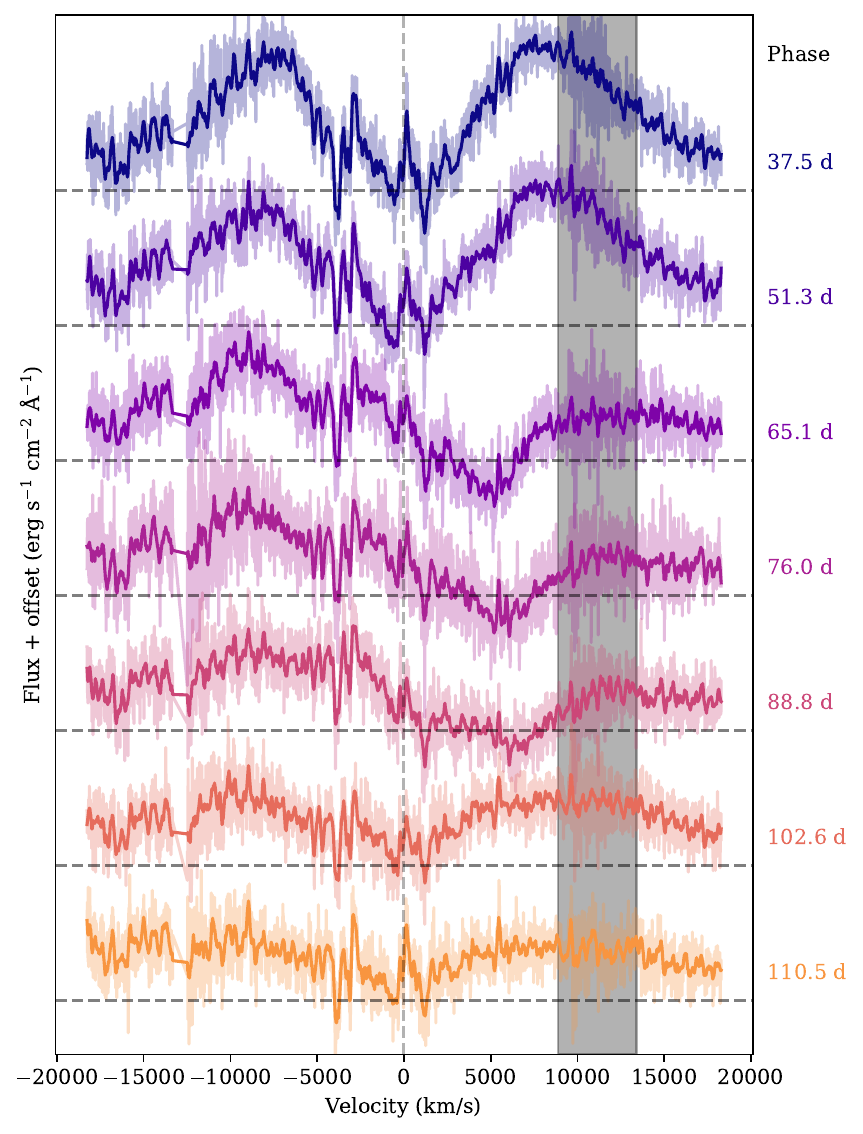}
    \caption{H$\alpha$ feature in the X-Shooter spectra with the best fit Gaussian subtracted, plotted in velocity space. The dashed grey horizontal lines under each feature indicate where the best fit Gaussian is. There is a clear blueshifted excess at all epochs, and a redshfited excess in the earlier epochs. The grey shaded region indicates the location of a telluric feature. A dashed vertical line is plotted at rest velocity.}
    \label{fig:Halpha_excess}
\end{figure}

\begin{figure}
    \centering
    \includegraphics[width=1\columnwidth]{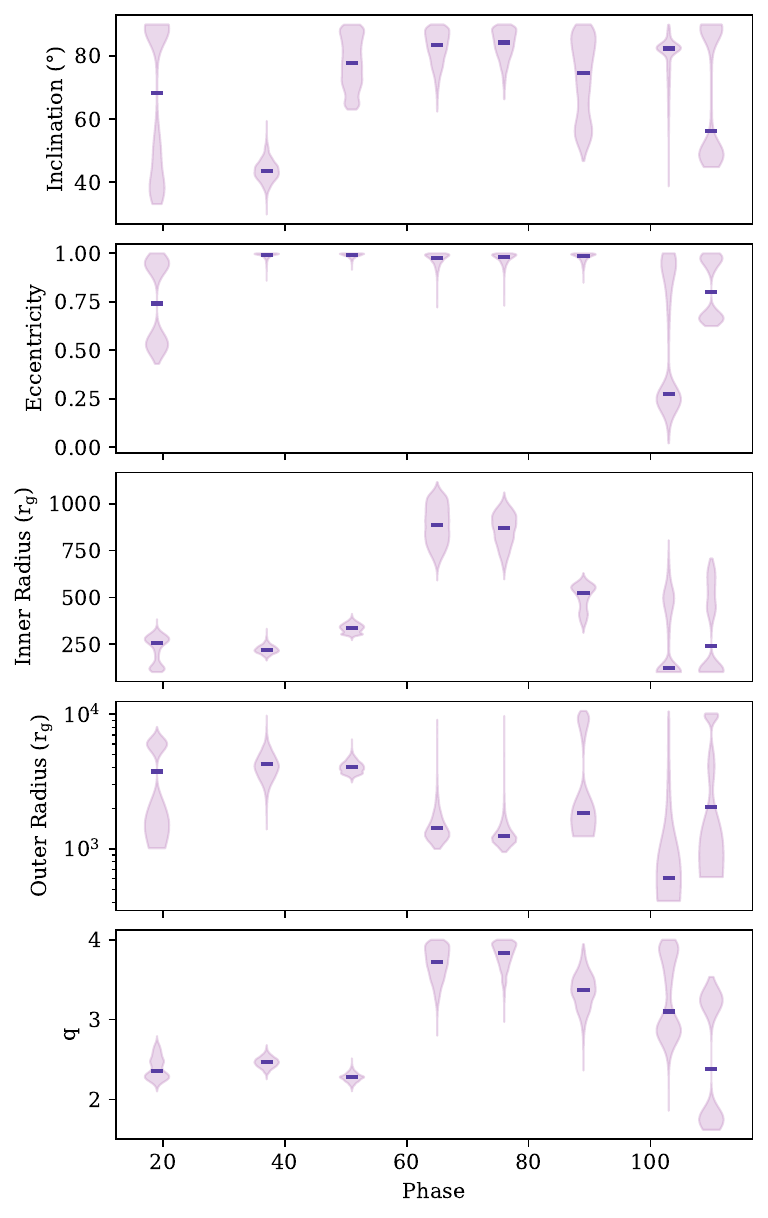}
    \caption{Violin plots showing the posterior distributions of the H$\alpha$ disk model parameters from \texttt{feadme} fits. The horizontal bar represents the median parameter value at each epoch.}
    \label{fig:disk_params}
\end{figure}

\subsection{Disk Formation and Evolution}\label{sec:disk}

\begin{deluxetable*}{ccccccc}
    \tablecaption{Best-fitting parameters from the accretion disk fits using \texttt{feadme}.\label{tab:feadme_params}}
    \tablehead{Phase & Inclination ($^{\circ}$) & Eccentricity & Inner Radius ($r_{\rm g}$) & Outer Radius ($r_{\rm g}$) & Apocenter ($^{\circ}$) & q }
    \startdata
        \hline
        19.0 & $68.3^{+21.1}_{-30.8}$ & $0.74^{+0.22}_{-0.23}$ & $255^{+37}_{-133}$ & $3747^{+2518}_{-2434}$ & $231.7^{+5.1}_{-5.5}$ & $2.36^{+0.20}_{-0.10}$ \\
        37.6 & $43.5^{+3.6}_{-3.2}$ & $0.99^{+0.01}_{-0.02}$ & $220^{+23}_{-19}$ & $4260^{+1081}_{-841}$ & $213.3^{+1.7}_{-1.7}$ & $2.46^{+0.07}_{-0.07}$ \\
        51.4 & $77.8^{+9.3}_{-9.2}$ & $0.99^{+0.01}_{-0.01}$ & $336^{+25}_{-31}$ & $4066^{+496}_{-425}$ & $216.2^{+1.6}_{-1.3}$ & $2.28^{+0.05}_{-0.06}$ \\
        65.1 & $83.5^{+4.6}_{-6.9}$ & $0.98^{+0.02}_{-0.04}$ & $889^{+117}_{-117}$ & $1428^{+587}_{-224}$ & $193.8^{+1.0}_{-0.8}$ & $3.73^{+0.19}_{-0.25}$ \\
        76.0 & $84.3^{+4.1}_{-5.7}$ & $0.98^{+0.01}_{-0.03}$ & $873^{+80}_{-105}$ & $1247^{+342}_{-130}$ & $196.3^{+1.1}_{-1.0}$ & $3.83^{+0.12}_{-0.21}$ \\
        88.8 & $74.7^{+11.5}_{-17.9}$ & $0.99^{+0.01}_{-0.03}$ & $524^{+42}_{-113}$ & $1833^{+7615}_{-244}$ & $198.9^{+2.0}_{-1.6}$ & $3.37^{+0.21}_{-0.24}$ \\
        102.6 & $82.4^{+0.6}_{-9.7}$ & $0.27^{+0.69}_{-0.03}$ & $123^{+396}_{-22}$ & $607^{+3602}_{-105}$ & $192.4^{+1.2}_{-160.6}$ & $3.10^{+0.78}_{-0.32}$ \\
        110.5 & $56.3^{+33.0}_{-8.3}$ & $0.80^{+0.18}_{-0.14}$ & $239^{+350}_{-132}$ & $2054^{+7698}_{-1234}$ & $213.7^{+35.0}_{-4.8}$ & $2.38^{+0.89}_{-0.65}$ \\
        \hline \hline
    \enddata
\end{deluxetable*}

As discussed in Section \ref{sec:spec_ev}, the H$\alpha$ profile shows excess emission on both the blue and red sides of the line at several epochs compared to a Gaussian profile at intermediate epochs. Figure \ref{fig:direct_gaussian} shows the luminosity when directly integrating under the line profile, and when integrating under the fitted Gaussian profile. The residuals are plotted when subtracting the Gaussian luminosity from the direct luminosity which shows that the Gaussian fits well up until optical peak, before it no longer captures the entire luminosity. After this point, all spectra show excess emission compared to the Gaussian. Figure \ref{fig:Halpha_excess} shows the residuals obtained after subtracting the best-fitting Gaussian profile from the observed line for our series of X-shooter spectra. \citet{Simongini2026} and \citet{Baldini2026} also identify this feature on the red side in their spectra and \citet{Simongini2026} attribute it to redshifted He I $\lambda$6678 emission. However, the X-Shooter spectra reveal a clear evolving double-sided excess, potentially consistent with emission from a disk-like structure. Furthermore, there is little to no evidence for emission from He I $\lambda$5876, making an identification of the red excess with He I $\lambda$6678 less compelling and suggesting instead that the feature is associated with H$\alpha$. Similar residual structure may also be present in spectra obtained at comparable phases, although the lower signal-to-noise ratio of those spectra makes the feature more difficult to confirm.

Broad and asymmetric Balmer line profiles have previously been predicted in models where line formation occurs in an optically thick outflow or reprocessing layer surrounding the accretion flow \citep{Roth2018}. These models can naturally produce extended red wings through electron scattering and radiative transfer effects, yielding profiles that deviate significantly from a simple Gaussian shape. Although \citet{Baldini2026} attribute the H$\alpha$ feature to this effect, the morphology observed in TDE\,2025aarm differs from these predictions. While the spectra do exhibit excess emission on the red side of the line, this excess does not appear as a smooth, continuously declining red wing. Instead, the X-Shooter spectra reveal a shoulder shape on the red and blue tails. Similar, double peak-like structures have been observed in a handful of TDEs where they have been interpreted as evidence for emission originating in an accretion disk (e.g. AT\,2018hyz \citep{Short2020, Hung2020} and AT\,2020zso \citep{Wevers2022}). The visibility of such double-peaked structures may also depend on viewing geometry, with the inclination of the disk affecting whether the double-peaked profile can be observed \citep{Hung2020,Charalampopoulos2022}.

Residual H$\alpha$ profiles for a larger subset of the spectroscopic sample are shown in Figure \ref{fig:Halpha_excess_all}. While these spectra have lower signal-to-noise ratios, we tentatively identify similar excess emission at earlier phases, particularly on the blue side of the line from approximately -10 days onwards. Motivated by this behaviour, we model the entire H$\alpha$ profile of the XShooter spectra using the Python package \texttt{feadme} \citep{Earl2025}. The model includes a broad Gaussian component, an eccentric accretion-disk component, and narrow absorption components associated with host-galaxy absorption features. The disk component is described by its inclination, eccentricity, inner radius, outer radius, apocentre orientation angle, and emissivity index $q$. The radii are given in units of gravitational radius, $r_{\rm g}=GM_{\rm BH}/c^2$, the inclination and apocentre angles are given in radians, and $\sigma$ is given in km\,s$^{-1}$. The median values of the fitted parameters are listed in Table \ref{tab:feadme_params}, with uncertainties corresponding to the 16th and 84th percentiles of the posterior distributions. The priors used for the fits are given in Appendix \ref{tab:feadme_priors}.

Continuum subtraction was performed both globally and locally. Global subtraction was performed by fitting a cubic function across the defined continuum regions across the spectrum as defined in Section \ref{sec:spec_ev}. Local subtraction was performed by defining a continuum region on either side of the line, randomly sampling 1000 points within that region, and using the median to calculate the line of best fit to subtract off. This process is visualised in Figure \ref{fig:Halpha_cont}. Figure \ref{fig:cont_sub} presents the H$\alpha$ line profile using both of these continuum subtraction methods from which we see very little difference between both.

Figure \ref{fig:disk_params} shows the evolution of the posterior distributions for each disk parameter as a function of phase. The violin plots are constructed from the full posterior samples produced by \texttt{feadme}, with the width of each violin indicating the relative probability density of a given parameter value. The top panel shows the inclination of the disk which remains relatively high throughout the epochs modelled. Most epochs are consistent with inclinations of $\sim70^{\circ}-90^{\circ}$, suggesting that the disk-like component is viewed close to edge-on. The only exception is the disk fit at 37.6 days has a best fit inclination of $43.5^{+3.6}_{-3.2}\,^{\circ}$. The predominantly high inferred inclination is also consistent with viewing-angle interpretations of the spectroscopic diversity of TDEs. In the unified model of \citet{Dai2018}, high-inclination sightlines intersect a larger column of optically thick material in the disk wind, resulting in more efficient reprocessing of high-energy radiation into optical/UV emission. \citet{Leloudas2019} suggested that such a geometry may also favour the production of strong Bowen fluorescence features, as ionising photons undergo multiple scatterings before escaping. The presence of Bowen features in TDE\,2025aarm is therefore broadly consistent with the high inclination inferred from our H$\alpha$ disk-profile modelling.

The second panel of Figure \ref{fig:disk_params} shows the evolution of the disk eccentricity. The fits consistently favour high eccentricities, with the posterior distributions extending towards $e\sim1$ at most epochs. Such highly eccentric geometries are qualitatively consistent with expectations for stellar debris following a tidal disruption, which initially occupies highly elongated orbits and must dissipate orbital energy before forming a circular accretion disk \citep{Piran2015, Hayasaki2016}. The persistently high eccentricities may therefore suggest that the line-emitting material has not yet fully circularised.

In comparison, the radial structure of the H$\alpha$ disk changes with time. The inner radius is initially small, starting at $r_{\rm in}=255^{+37}_{-133}\,r_{\rm g}$ ($\sim6\times10^{12}$\,cm) at 19.0 days, and remaining roughly constant up to 51.4 days. It then increases sharply up to $\sim900\,r_{\rm g}$ ($\sim2\times10^{13}$\,cm) at phases 65.1 and 76.0 days, before decreases gradually back to $239^{+350}_{-132}\,r_{\rm g}$ by 110.5 days. This evolution suggests that the inner boundary of the H$\alpha$-emitting region shifts outward between approximately $+60$ and $+80$\,days. The evolution of the outer radius is less clear, with the median values decreasing over this period but broad posterior distributions spanning $\sim10^{3}$--$10^{4}\,r_{\rm g}$ at most epochs. Despite the apparent change in the location of the H$\alpha$-emitting region between $+60$ and $+80$\,days, the blackbody radius remains approximately constant (Figure \ref{fig:BB_fits}), suggesting that the continuum-emitting region does not undergo a corresponding change in scale.

Finally, the bottom panel of Figure \ref{fig:disk_params} shows the emissivity index $q$. It remains near $q\sim2.2$--2.5 between 19.0 and 51.4 days, before rising sharply to $q\sim4$ between 65 and 76 days and subsequently decreasing to $q\sim2$ at later phases. Since the emissivity scales approximately as $\epsilon(r)\propto r^{-q}$, larger values of $q$ imply that a greater fraction of the H$\alpha$ emission originates from the inner regions of the emitting structure. The temporary increase in $q$ therefore indicates a more centrally concentrated H$\alpha$ emissivity profile between approximately 65 and 76 days, before returning to a distribution similar to that observed at earlier phases.

The increase in $q$ occurs simultaneously with changes in the inferred disk radii. Rather than tracing a physical outward movement of the inner disk edge, the temporary increases in $r_{\rm in}$ and $q$ may indicate that the H$\alpha$ emission becomes concentrated over a narrower range of radii. As the model only includes a single disk, the evolution of these parameters may reflect how the model reproduces this period of increased emission from a ring of material. This epoch also coincides with the increase in blueshift of all emission lines shown in Figure \ref{fig:blueshifts}. Interestingly, this epoch is also contemporaneous with the sharp rise in soft X-rays reported by \citet{Baldini2026}, although the H$\alpha$ and X-ray emission probe different temperature and likely different radii of the system. These changes raise the question of what could produce a temporary increase in the H$\alpha$ emission over a restricted range of radii.

\begin{deluxetable*}{ccccccc}
    \tablecaption{Eddington ratios based on black hole masses. The ratios are calculated for each bump in the light curve, as well as the optical peak. \label{tab:BH_masses}}
    \tablehead{Method & $\log_{10}(M_{\rm BH}/M_{\odot})$ & $L_{\rm Edd}$ (10$^{45}$ erg s$^{-1}$) & $\frac{L_{\rm Bump1}}{L_{\rm Edd}}$ & $\frac{L_{\rm{peak}}}{L_{\rm{Edd}}}$  & $\frac{L_{\rm{Bump2}}}{L_{\rm{Edd}}}$ & $\frac{L_{\rm{Bump3}}}{L_{\rm{Edd}}}$ }
    \startdata
        \hline
        $M_{\rm BH}$-$\sigma_{\ast}$ & 7.45 $\pm$ 0.29 & 3.55$^{+3.37}_{-1.73}$ & 0.020$^{+0.021}_{-0.011}$ & 0.010$^{+0.010}_{-0.005}$ & 0.007$^{+0.008}_{-0.004}$ & 0.001$^{+0.002}_{-0.001}$ \\
        t$_{1/2}$ & $7.44 \pm 0.53$ & 4.79$^{+12.5}_{-3.44}$ & 0.014$^{+0.003}_{-0.002}$ & 0.007$^{+0.002}_{-0.001}$ & 0.005$^{+0.001}_{-0.001}$ & 0.001$^{+0.001}_{-0.001}$ \\
        \textsc{MOSFiT} & 7.18$^{+0.20}_{-0.22}$ & 1.91$^{+1.12}_{-0.76}$ & 0.036$^{+0.021}_{-0.014}$ & 0.017$^{+0.009}_{-0.007}$ & 0.013$^{+0.008}_{-0.005}$ & 0.003$^{+0.004}_{-0.001}$\\  
        \hline \hline
    \enddata
\end{deluxetable*}

\section{Discussion} \label{sec:discussion}

TDE\,2025aarm exhibits several contemporaneously evolving features, including multiple bumps in the optical/UV light curve, changes in the emission-line blueshifts, and short lags between the continuum and spectroscopic evolution. Modelling of the H$\alpha$ profile additionally reveals changes in the inferred disk parameters, including the characteristic emitting radii and emissivity index, suggesting that the region dominating the line emission evolves throughout the event.

\subsection{Eddington Induced Outflows}

One possible explanation for the observed bumps in the light curve and changes in the H$\alpha$-emitting region is the presence of radiation-driven outflows, which can be launched when the accretion rate approaches or exceeds the Eddington limit \citep{Strubbe2009,Metzger2016,Dai2018,Roth2018}. Changes of such an outflow could alter both the observed continuum emission and the distribution of material contributing to the emission lines. We therefore investigate whether the evolution of TDE\,2025aarm occurs close to the Eddington limit. We derived the black hole mass of TDE\,2025aarm using several independent methods, summarised in Table \ref{tab:BH_masses}. Using these mass estimates, we calculated the corresponding Eddington luminosities in order to estimate the Eddington ratios during the main optical peak and the rebrightening episodes discussed in Section \ref{sec:bumps}. The bolometric luminosity at each epoch was estimated by selecting the closest blackbody luminosity measurement from the SED fits described in Section \ref{sec:bb_params}. These luminosities are also listed in Table \ref{tab:BH_masses}.

Using the larger black hole masses inferred from the $M$-$\sigma$ relation and the $t_{1/2}$ method, we find that the inferred Eddington ratios are relatively small, typically $\lesssim 0.01$. Adopting instead the lower black hole mass inferred from the \textsc{MOSFiT} modelling yields Eddington ratios of $\sim0.03$--$0.4$ during both the primary peak and the rebrightening episodes. These values are much more consistent with Eddington ratios measured from the optical emission for the population. However, these values remain formally below the Eddington limit.

The sub-Eddington ratios inferred from the optical/UV blackbody luminosities suggest that the observed thermal emission alone is unlikely to require strongly super-Eddington accretion. Although these luminosities do not account for emission outside the near-IR/optical/UV SED, the magnitude of the missing contribution is unlikely to be sufficient to raise the inferred luminosity to the Eddington limit. Mid-infrared dust echoes have been detected in a growing number of TDEs \citep[e.g.][]{vanVelzen2016,Jiang2016IR,Mattila2018,Kool2020,Jiang2021,Reynolds2026}. However, optically selected TDEs typically have inferred dust covering factors of only $\sim1\%$ \citep[e.g.]{Jiang2021}, suggesting that the associated reprocessed IR emission represents only a small fraction of the total radiated energy. Furthermore, \citet{Baldini2026} constrain the X-ray luminosity near optical maximum to less than $\sim0.02\%$ of the optical/UV luminosity. These additional observed components are therefore insufficient to account for the factor required for TDE\,2025aarm to reach the Eddington luminosity.



\subsection{Accretion State Transition}

Another origin of the contemporaneous changes observed in the properties of TDE\,2025aarm is a change in the structure of the accretion flow. Black hole X-ray binaries (XRBs) undergo transitions between distinct accretion states as their accretion rates evolve. These are accompanied by substantial changes in the relative contributions of the accretion disk and corona. Similar behaviour has also been proposed in TDEs, suggesting that accretion flows around black holes across a wide range of masses may undergo transitions at characteristic Eddington ratios \citep[e.g.][]{Wevers2021, Alexander2026, Goodwin2026}.

This interpretation is particularly relevant for TDE\,2025aarm. \citet{Baldini2026} find that the initially faint, hard X-ray emission evolves into a brighter, softer, disk-dominated state before hardening again. They interpret this evolution as a transition analogous to the low-hard to high-soft state transition observed in XRBs. The evolution of the H$\alpha$-emitting region may also be connected to this change in the inner accretion flow. Our disk modelling in Section \ref{sec:disk} indicates that the H$\alpha$ emitting region originates at larger radii after $\sim+60$ days. Rather than requiring the material to physically move outwards, the emergence of a brighter soft X-ray/UV emitting disk could alter the ionisation structure of the material present. Gas at a smaller radius may become sufficiently ionised that is no longer efficiently produces H$\alpha$. Meanwhile the increase in the ionising radiation could illuminate material at a larger radius, shifting the apparent H$\alpha$ emitting region.

Simultaneously, the increase in the emissivity index $q$ indicates that the H$\alpha$ emission becomes more strongly concentrated towards the inner edge of this disk. This is more difficult to explain in this picture as it is not a obvious consequence of the increase in ionising radiation.

\subsection{Stream–Stream Collisions}

An alternative interpretation for the observed light-curve structure is emission powered by shocks during circularisation. In this picture, collisions between the stellar debris streams produce shocks that dissipate orbital kinetic energy and power optical/UV emission \citep[e.g.][]{Piran2015,Dai2015,Lu2020,Bonnerot2021a}. Radiation pressure generated by these shocks can additionally accelerate material into an outflow \citep{Jiang2016,Bonnerot2021a}. Radiation-hydrodynamic simulations by \citet{Bonnerot2021a} show that, despite the generation of this outflow, the returning debris is rapidly circularised through a series of secondary shocks close to the black hole, leading to the prompt formation of an accretion disk. 

\begin{figure*}
    \centering
    \includegraphics[trim={0 4cm 0 2cm},clip,width=2\columnwidth]{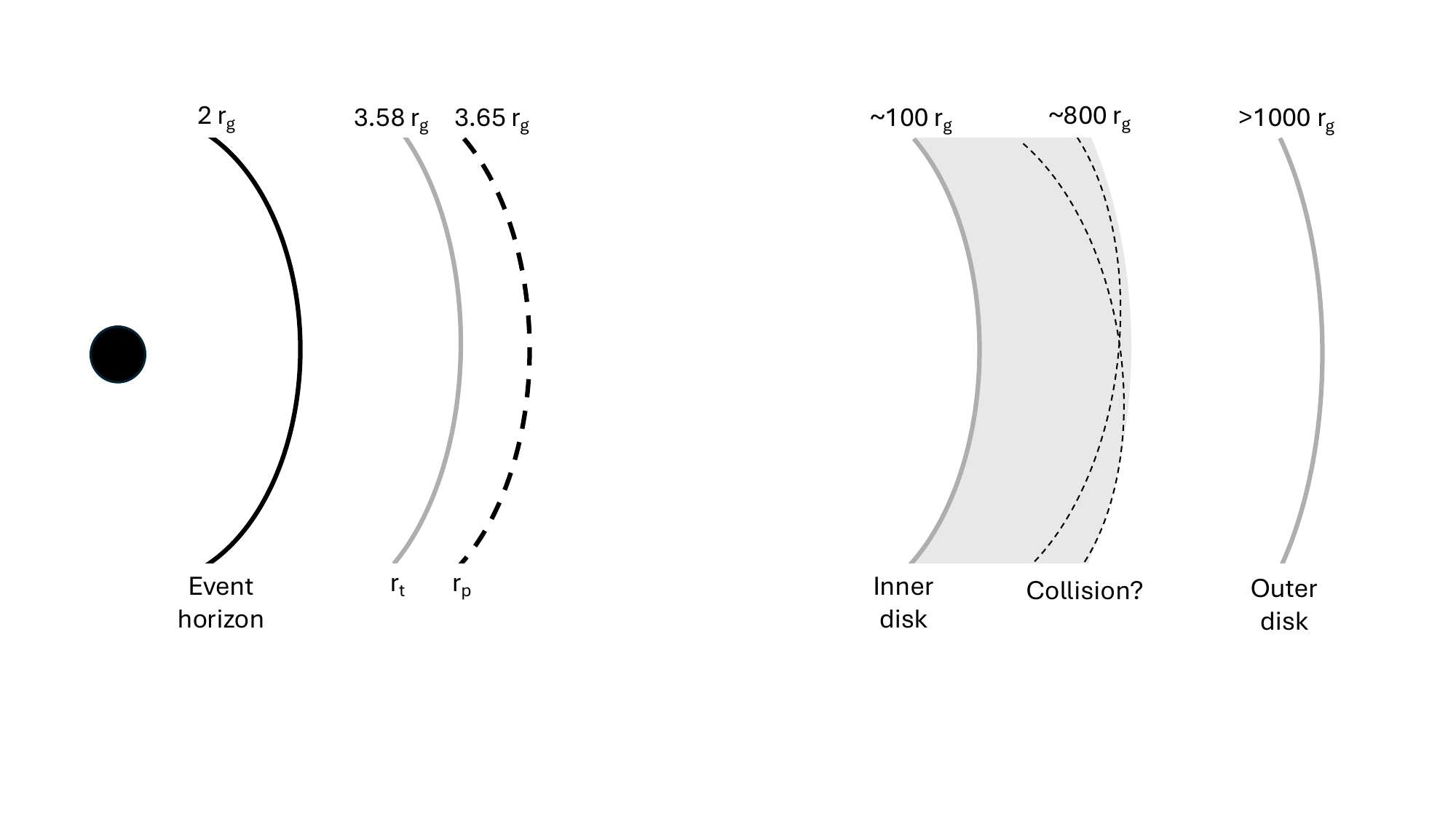}
    \caption{Schematic illustrating the characteristic radii for TDE\,2025aarm, including the event horizon, $r_{t}$, $r_{p}$, and the inner and outer radii of the H$\alpha$ emitting disk inferred from the \texttt{feadme} fits.}
    \label{fig:radii}
\end{figure*}

This picture may be broadly consistent with the H$\alpha$ line-profile modelling presented in Section \ref{sec:disk}, which indicates the presence of a disk-like emitting region by $\sim20$ days post peak. Figure \ref{fig:radii} illustrates the characteristic radii relevant to TDE\,2025aarm. The \textsc{MOSFiT} parameters give a black hole with mass $\log_{10}(M_{\rm BH}/M_{\odot}) = 7.18^{+0.20}_{-0.22}$ disrupting a star of mass $\log_{10}(M_{*}/M_{\odot}) = 0.41^{+0.19}_{-0.11}$ with an impact parameter of $\beta=0.98^{+0.25}_{-0.15}$. Using the empirical mass-radius relation from \citet{Eker2018}, we find the star has a radius of $R_{*}=0.35\pm0.22\,R_{\odot}$ corresponding to a tidal radius $r_{t}=3.58\pm2.02\,r_{g}$ using the unscaled $\beta$ from \textsc{MOSFiT}, and a pericenter distance of $r_{p}=3.65\pm2.19\,r_{g}$. The disk inferred from the H$\alpha$ profile extends from $\sim100\,r_{g}$ to $\gtrsim1000\,r_{g}$ for the first $\sim70$ days. After this the sharp change in properties pushes the apparent inner disk to $\sim600\,r_{g}$ and the emissivity profile becoming substantially more centrally concentrated. Simulations predict that stream-stream collisions in deeply penetrating encounters can occur over a broad range of radii ($\sim10$-$1000\,r_{g}$; \citealt{Lu2020}), overlapping the inferred H$\alpha$ disk region. It is therefore plausible that shocks associated with continued stream collisions contribute to the changing disk-like emission observed. 

It is also interesting to compare the characteristic timescales of stream-stream simulations with those observed in TDE\,2025aarm. For this event, the inferred fallback time is relatively short, $t_{\rm min}\sim8$ days, estimated using the standard fallback timescale \citep[e.g.][]{Rees1988}. We use the inferred black hole and stellar masses, with the stellar radius determined from the empirical main-sequence mass-radius relation of \citet{Eker2018}. The simulations of \citet{Bonnerot2021a} therefore predict the strongest shock-powered luminosity variations and peak mass outflow rates to occur within only a few days of disruption. This is significantly earlier than the prominent UV/optical bump observed near peak brightness, suggesting that the earliest stream-stream collision emission from the model is unlikely to directly explain that feature in the light curve. However, the simulations also exhibit non-monotonic luminosity evolution and continued interaction between the returning debris, outflow, and circularising material. Consequently, later time fluctuations in the bolometric luminosity could still contribute to the bumps seen in the light curve.

\section{Conclusions} \label{sec:conclusions}

This papers present optical photometry and spectra for TDE\,2025aarm, one of the closest TDEs discovered to date. The comprehensive dataset provides coverage across 13 bands and spectral coverage running from -40.4 days pre-peak to +110.5 post-peak. The principal observational results of TDE\,2025aarm can be summarised as follows:

\begin{itemize}

\item The light curve exhibits multiple deviations from smooth evolution, including a prominent UV bump near peak ($\sim-20$ days), a secondary blue rebrightening at $\sim+20$ days, and a later rise at $\sim+80$ days.

\item These photometric bumps are present both in individual filters and in the bolometric light curve, demonstrating that they correspond to genuine increases in luminosity rather than changes in spectra alone.

\item Blackbody modelling shows that the rebrightening episodes are associated with temperature increases of approximately $5000-10000$\,K. In contrast, the inferred photospheric radius follows its overall evolution and shows no significant deviations associated with the rebrightening episodes.

\item The spectra show relatively little overall evolution throughout the observing campaign and are dominated by broad H$\alpha$, He II $\lambda4686$, H$\beta$, and potentially N III $\lambda4640$ emission.

\item Increases in the blueshift and FWHM of the H$\alpha$ line coincide with the photometric rebrightening episodes.

\item Cross-correlation analysis indicates a preferred lag of approximately 9.7 days between the bolometric luminosity evolution and the H$\alpha$ blueshift, with luminosity variations preceding changes in the line forming region.

\item A persistent narrow absorption feature is visible at $\sim-3900$\,km\,s$^{-1}$ in the broad Balmer lines indicative of outflowing material.

\item H$\alpha$ emission consistent with a disk-like component is tentatively visible from around optical peak onwards in several lower signal-to-noise spectra, with the clearest detection occurring in the higher-quality X-Shooter observations from $\sim+20$ to 110 days.

\item Modelling of the disk-like component shows significant changes in the inferred disk parameters occurring between $60 - 80$ days post peak.

\end{itemize}

Overall, the observations suggest that the photometric bumps are associated with changes in the physical conditions of the emitting and reprocessing material. The rebrightening episodes are accompanied by increases in the inferred blackbody temperature, but without a corresponding increase in the photospheric radius, indicating that the luminosity enhancements are not primarily driven by an expansion of the emitting surface. At the same time, the H$\alpha$ profile becomes increasingly blueshifted, implying changes in the geometry, kinematics, or optical depth of the line-forming region.

A similar coincidence is seen in the X-ray evolution of TDE\,2025aarm, where \citet{Baldini2026} report a transition from a hard to a soft spectral state at approximately the same epoch that we observe significant changes in the inferred H$\alpha$ disk parameters, an increase in temperatures, and an increase in the blueshift of the emission lines. However, the X-ray and optical emission probe different temperature, and therefore regions of material surrounding the system. The correlation between these observables heavily hints at a broader change in the structure of the accretion flow or the surrounding environment.

One possible interpretation is that episodes of enhanced stream-stream collisions drive the observed rebrightening events. Increased shock heating during these episodes could account for the higher continuum temperatures, while simultaneously altering the structure of the circularising debris and surrounding reprocessing layers. Such changes may influence both the properties of the forming accretion flow and our view of it, potentially explaining the contemporaneous evolution of the H$\alpha$ profile, the inferred disk parameters (including the characteristic radii and emissivity profile), and the varying prominence of the disk-like H$\alpha$ emission observed in the X-Shooter spectra. In this scenario, the observed line evolution may reflect changes in our view of a disk that is progressively revealed or obscured as the reprocessing layer evolves.

While the precise mechanism responsible for the rebrightenings remains uncertain, the observations of TDE\,2025aarm demonstrate that these features are accompanied by measurable changes in both the continuum and line-forming regions. The dense multi-wavelength coverage obtained for this event was crucial for identifying these short-timescale variations and their spectroscopic counterparts, which would likely have been missed in more sparsely sampled datasets. Although these observations provide much stronger constraints on the timescales over which these changes occur, their physical origin remains difficult to determine given the current uncertainties in models for TDE emission. TDE\,2025aarm therefore highlights the importance of high-cadence photometric and spectroscopic monitoring, alongside further theoretical modelling, for understanding the evolving relationship between accretion, outflows, and line formation in TDEs.

\begin{acknowledgments}

Based on observations made with the Nordic Optical Telescope, owned in collaboration by the University of Turku and Aarhus University, and operated jointly by Aarhus University, the University of Turku and the University of Oslo, representing Denmark, Finland, and Norway, the University of Iceland and Stockholm University at the Observatorio del Roque de los Muchachos, La Palma, Spain, of the Instituto de Astrofisica de Canarias. The data presented here were obtained with ALFOSC, which is provided by the Instituto de Astrofisica de Andalucia (IAA) under a joint agreement with the University of Copenhagen and NOT. Observations from the Nordic Optical Telescope were obtained through the NUTS2 collaboration which is supported in part by the Instrument Centre for Danish Astrophysics (IDA), and the Finnish Centre for Astronomy with ESO (FINCA) via Academy of Finland grant nr 306531. The Liverpool Telescope is operated on the island of La Palma by Liverpool John Moores University in the Spanish Observatorio del Roque de los Muchachos of the Instituto de Astrofisica de Canarias with financial support from the UK Science and Technology Facilities Council.

Based in part on data acquired at the ANU 2.3-metre telescope under program 2520021 and 2520184. The automation of the telescope was made possible through an initial grant provided by the Centre of Gravitational Astrophysics and the Research School of Astronomy and Astrophysics at the Australian National University and through a grant provided by the Australian Research Council through LE230100063. The Lens proposal system is maintained by the AAO Research Data \& Software team as part of the Data Central Science Platform. We acknowledge the traditional custodians of the land on which the telescope stands, the Gamilaraay people, and pay our respects to elders past and present.

Observations reported here were obtained at the MMT Observatory, a joint facility of the University of Arizona and the Smithsonian Institution.

AA and KDA acknowledge support provided by the NSF through award AST-2307668. KDA gratefully acknowledges support from the Alfred P. Sloan Foundation.

CRA and MN were supported by the European Research Council (ERC) under the European Union's Horizon 2020 research and innovation programme (grant agreement no. 948381).

IA acknowledges support from the European Research Council (ERC) under the European Union’s Horizon 2020 research and innovation program (grant agreement number 852097), from the Israel Science Foundation (grant number 2752/19), from the United States - Israel Binational Science Foundation (BSF; grant number 2024812), and from the Pazy foundation (grant number 216312).

GL and SDW were supported by a research grant (VIL60862) from VILLUM FONDEN.

YZC is supported by the National Natural Science Foundation of China (No. 12303054), the Yunnan Fundamental Research Projects (Grant Nos. 202401AU070063, 202501AS070078), the National Key Research and Development Program of China (Grant No. 2024YFA1611603), and the International Centre of Supernovae (ICESUN), Yunnan Key Laboratory of Supernova Research (No. 202505AV340004). Y.-Z. Cai acknowledges financial support from the SOXS project (PI S. Campana).

MGB acknowledges financial support from the Spanish Ministerio de Ciencia e Innovación (MCIN) and the Agencia Estatal de Investigación (AEI) 10.13039/501100011033 under the PID2023-151307NB-I00 SNNEXT project, from Centro Superior de Investigaciones Científicas (CSIC) under projects PIE 20215AT016, ILINK23001, COOPB2304, and the program Unidad de Excelencia María de Maeztu CEX2020-001058-M, and from the Departament de Recerca i Universitats de la Generalitat de Catalunya through the 2021-SGR-01270 grant.
MGB’s work has been carried out within the framework of the doctoral program in Physics of the Universitat Autònoma de Barcelona.

NF acknowledges support from the National Science Foundation Graduate Research Fellowship Program under Grant No. DGE-2137419.

PJG is partly supported by NRF SARChI grant 111692.
The Lesedi-Mookodi observations were obtained under the SAAO "Transients
and Variables" program, PI Groot.

CPG acknowledges financial support from grant RYC2024-050959-I, funded by MICIU/AEI/10.13039/501100011033, the FSE+ and FEDER, UE, as well as from projects PID2025-172154NA-I00, PID2023-151307NB-I00, PIE 20215AT016, and CEX2020-001058-M, and the MaX-CSIC Excellence Award MaX4-SOMMA-ICE.

TLK acknowledges support via a Warwick Astrophysics prize post-doctoral fellowship made possible thanks to a generous philanthropic donation.

SM acknowledges financial support from the Research Council of Finland project 350458.

FO acknowledges support from the INAF-Large Grant 2024:"Envisioning Tomorrow: prospects and challenges for multimessenger astronomy in the era of Rubin and Einstein Telescope"; the INAF-GO Large Grant: "Exploitation of optical and near-infrared followup data of Gamma-ray Bursts" and the INAF-MINIGRANT (2023): "SeaTiDE - Searching for Tidal Disruption Events with ZTF: the Tidal Disruption Event population in the era of wide field surveys".

MP acknowledges support from a UK Research and Innovation Fellowship (UKRI1062).

Parts of this research were supported by the Australian Research Council Discovery Early Career Researcher Award (DECRA) through project number DE230101069.

\end{acknowledgments}

\begin{contribution}



\end{contribution}

%
\facilities{Swift(XRT and UVOT), MMTO(Binospec, MMIRS and BlueChannel), Bok(B\&C), SALT, LCOGT, FTN, FTS, NOT, VLT:Kueyen, PS1, ATLAS, PO:1.2m, SAAO:1m(Mookodi), Liverpool:2m, ATT}



\appendix

\renewcommand{\thefigure}{\thesection.\arabic{figure}}
\counterwithin{figure}{section}

\renewcommand{\thetable}{\thesection.\arabic{table}}
\counterwithin{table}{section}

\section{Spectral Observations} \label{sec:spec_obs}

\startlongtable
\begin{deluxetable*}{ccccccc}
    \tablecaption{Spectroscopic observations of TDE\,2025aarm. Phase is given in rest frame days with respect to the time of maximum light in the $g$-band.\label{tab:spectra}}

        \tablehead{Date & Time & MJD & Phase & Disperser/Grism & Telescope & Instrument \\
                  &  &  & (days) &  &  & }
        \startdata
            \hline
            19/10/2025 & 01:55:12 & 60967.08 & --40.37 & -- &  LT & SPRAT \\
            26/10/2025 & 05:16:48 & 60974.22 & --33.32 & \#4 & NOT & ALFOSC \\
            29/10/2025 & 10:26:06 & 60977.44 & --30.15 & -- & OGG 2m & FLOYDS \\
            30/10/2025 & 05:31:12 & 60978.23 & --29.37 & -- & LT & SPRAT \\
            31/10/2025 & 08:03:55 & 60979.34 & --28.28 & 300\,g/mm & Bok & B\&C \\
            31/10/2025 & 11:08:49 & 60979.46 & --28.15 & Clear/LP-530 & MMT & BlueChannel \\
            31/10/2025 & 21:31:54 & 60979.90 & --27.72 &  -- & Lesedi & Mookodi \\
            01/11/2025 & 07:48:16 & 60980.33 & --27.30 & 300\,g/mm & Bok & B\&C \\
            02/11/2025 & -- & -- & -- & -- & Hobby-Eberly & LRS2 \\
            02/11/2025 & 21:25:15 & 60981.89 & --25.76 & -- & Lesedi & Mookodi \\
            03/11/2025 & 21:47:02 & 60982.91 & --24.75 & -- & Lesedi & Mookodi \\
            07/11/2025 & 02:24:00 & 60986.10 & --21.60 & \#4 & NOT & ALFOSC \\
            08/11/2025 & 02:29:27 & 60987.10 & --20.61 & 270\,g/mm & MMT & Binospec \\
            08/11/2025 & 13:46:53 & 60987.57 & --20.15 & -- & OGG 2m & FLOYDS \\
            09/11/2025 & 01:55:13 & 60988.08 & --19.65 & -- & Lesedi & Mookodi \\
            10/11/2025 & 21:46:17 & 60989.91 & --17.85 & -- & Lesedi & Mookodi \\
            11/11/2025 & 20:22:11 & 60990.85 & --16.92 & -- & Lesedi & Mookodi \\
            12/11/2025 & 21:36:59 & 60991.90 & --15.88 & -- & Lesedi & Mookodi \\
            13/11/2025 & 00:26:14 & 60992.02 & --15.77 & 270\,g/mm & MMT & Binospec \\
            13/11/2025 & 21:25:25 & 60992.89 & --14.91 & -- & Lesedi & Mookodi \\
            15/11/2025 & 19:58:07 & 60994.83 & --12.99 & -- & Lesedi & Mookodi \\
            17/11/2025 & 02:38:24 & 60996.11 & --11.73 & \#4 & NOT & ALFOSC \\
            18/11/2025 & 14:05:01 & 60997.59 & --10.27 & -- & COJ 2m & FLOYDS \\
            18/11/2025 & 14:33:57 & 60997.61 & --10.25 & B3000/R3000 & WiFeS & ANU 2.3m \\
            18/11/2025 & 19:30:26 & 60997.81 & --10.05 & -- & Lesedi & Mookodi \\
            24/11/2025 & 00:14:24 & 61003.01 & --4.92 & \#4 & NOT & ALFOSC \\
            24/11/2025 & 07:41:28 & 61003.32 & --4.62 & 300\,g/mm & Bok & B\&C \\
            24/11/2025 & 02:56:46 & 61003.12 & --4.81 & 270\,g/mm & MMT & Binospec \\
            25/11/2025 & 19:28:59 & 61004.81 & --3.15 & -- & Lesedi & Mookodi \\
            28/11/2025 & 10:47:44 & 61007.45 & --0.54 & -- & OGG 2m & FLOYDS \\
            29/11/2025 & 23:41:30 & 61008.99 & 0.97 & 270\,g/mm & MMT & Binospec \\
            01/12/2025 & 21:29:49 & 61010.90 & 2.86 & -- & 1.9 SAAO & SpUpNIC \\
            04/12/2025 & 19:00:47 & 61013.79 & 5.71 & -- & Lesedi & Mookodi \\
            06/12/2025 & 11:13:12 & 61015.47 & 7.37 & -- & OGG 2m & FLOYDS \\
            07/12/2025 & 11:42:38 & 61016.49 & 8.37 & B3000/R3000 & WiFeS & ANU 2.3m \\
            07/12/2025 & 11:53:14 & 61016.50 & 8.38 & -- & OGG 2m & FLOYDS \\
            09/12/2025 & 11:13:54 & 61018.47 & 10.33 & -- & OGG 2m & FLOYDS \\
            13/12/2025 & 19:08:18 & 61022.80 & 14.60 & -- & Lesedi & Mookodi \\
            17/12/2025 & 00:57:36 & 61026.04 & 17.80 & \#4 & NOT & ALFOSC \\
            18/12/2025 & 05:25:07 & 61031.23 & 18.97 & -- & VLT & X-Shooter \\
            18/12/2025 & 22:43:07 & 61027.95 & 19.68 & 270\,g/mm & MMT & Binospec \\
            23/12/2025 & 03:51:35 & 61032.16 & 23.83 & 300\,g/mm &  Bok & B\&C \\
            26/12/2025 & 01:12:00 & 61035.05 & 26.68 & \#4 & NOT & ALFOSC \\
            06/01/2026 & 03:44:56 & 61046.16 & 37.64 & -- & VLT & X-Shooter \\
            10/01/2026 & 23:00:33 & 61050.96 & 42.38 & -- & Lesedi & Mookodi \\
            11/01/2026 & 00:43:12 & 61051.03 & 42.45 & \#4 & NOT & ALFOSC \\
            14/01/2026 & 21:39:13 & 61054.90 & 46.27 & 270\,g/mm & MMT & Binospec \\
            15/01/2026 & 22:33:36 & 61055.94 & 47.29 & -- & LT & SPRAT \\
            18/01/2026 & 06:59:35 & 61058.29 & 49.61 & -- & OGG 2m & FLOYDS \\
            20/01/2026 & 01:49:25 & 61060.08 & 51.37 & -- & VLT & X-Shooter \\
            25/01/2026 & 14:14:04 & 61065.59 & 56.82 & B3000/R3000 & WiFeS & ANU 2.3m \\
            27/01/2026 & 12:26:57 & 61067.52 & 58.72 & -- & COJ 2m & FLOYDS \\
            30/01/2026 & 20:40:14 & 61070.86 & 62.01 & -- & Lesedi & Mookodi \\
            31/01/2026 & 20:52:48 & 61071.87 & 63.01 & \#4 & NOT & ALFOSC \\
            03/02/2026 & 00:29:52 & 61074.02 & 65.13 & -- & VLT & X-Shooter \\
            14/02/2026 & 01:03:16 & 61085.04 & 76.00 & -- & VLT & X-Shooter \\
            17/02/2026 & 11:46:19 & 61088.49 & 79.40 & -- & COJ 2m & FLOYDS \\
            18/02/2026 & 21:50:24 & 61089.91 & 80.80 & \#4 & NOT & ALFOSC \\
            27/02/2026 & 00:39:19 & 61098.03 & 88.81 & -- & VLT & X-Shooter \\
            10/03/2026 & 06:11:24 & 61109.26 & 99.89 & -- & COJ 2m & FLOYDS \\
            12/03/2026 & 21:21:36 & 61111.89 & 102.49 & \#4 & NOT & ALFOSC \\
            13/03/2026 & 00:19:40 & 61112.01 & 102.61 & -- & VLT & X-Shooter \\
            20/03/2026 & 19:58:52 & 61119.83 & 110.32 & 270 & MMT & Binospec \\
            21/03/2026 & 01:06:24 & 61120.05 & 110.53 & - & VLT & X-Shooter \\
		\hline \hline
	\enddata
\end{deluxetable*}
\twocolumngrid

\section{Additional Figures}

\begin{figure}
    \centering
    \includegraphics[trim={5em 0em 6em 3em},clip,width=1\columnwidth]{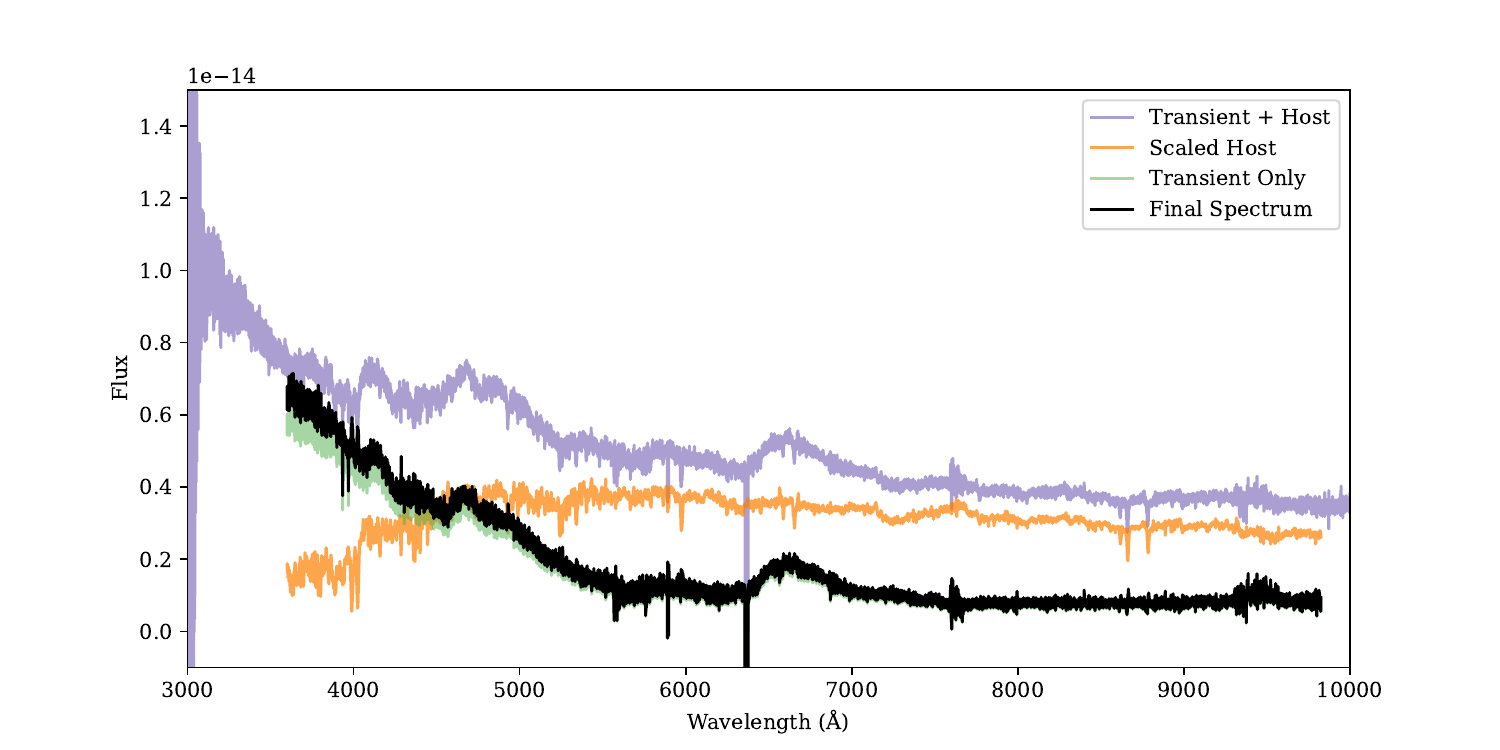}
    \caption{An example of the host galaxy subtraction of the spectra using the DESI archival spectrum. The transient spectrum has already been processed and corrected for MW extinction. The host spectrum has been scaled to archival Pan-STARRS photometry. The final spectrum has then been corrected for host galaxy extinction.}
    \label{fig:host_sub}
\end{figure}

\begin{figure*}
    \centering
    \includegraphics[trim={0.2cm 0 0.2cm 0.2cm},clip,width=2\columnwidth]{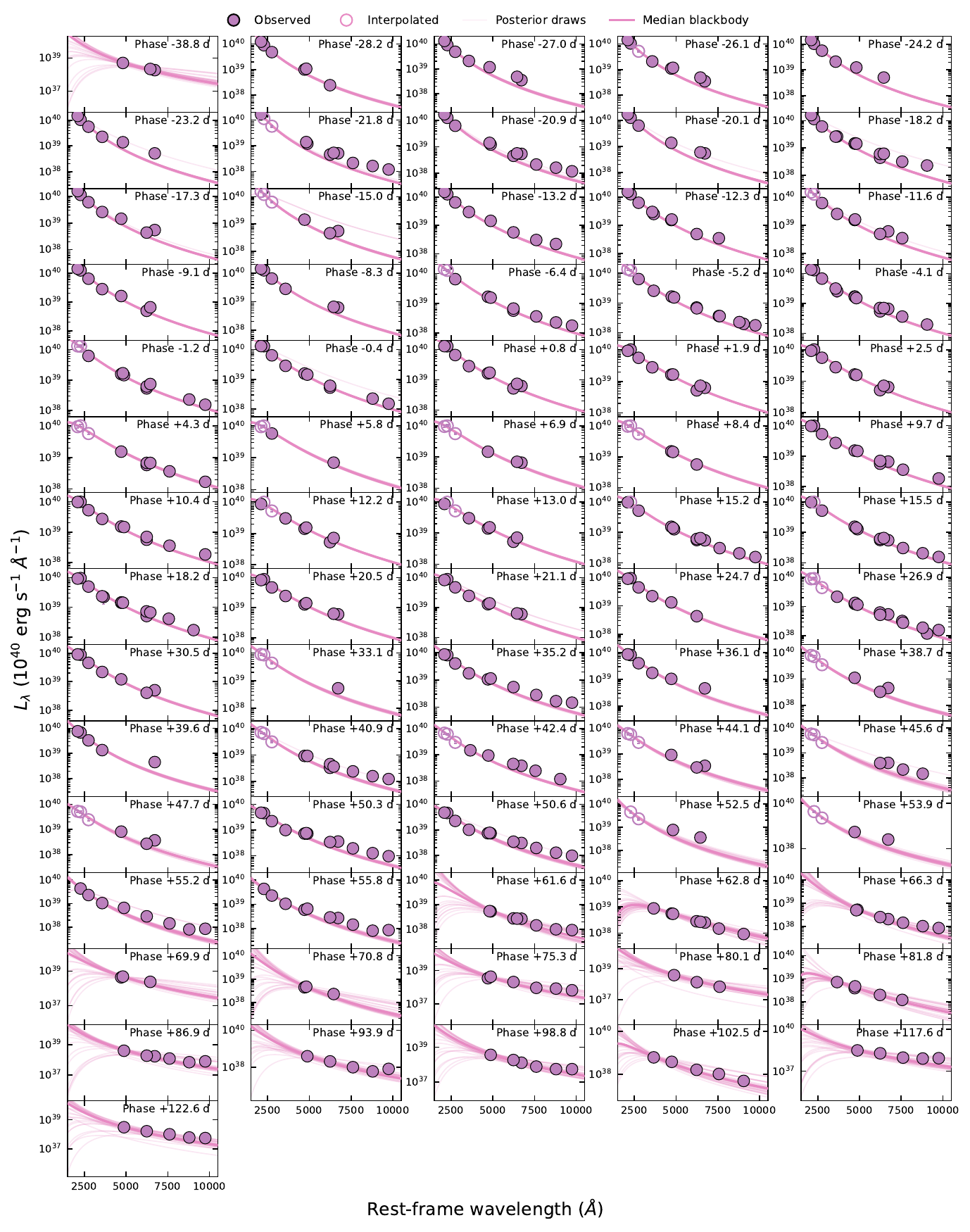}
    \caption{Full grid of blackbody fits to the SEDs. Filled markers are data points, and empty markers are interpolated points. A sample of 100 posterior draws are shown for each epoch, as well as the median blackbody fit.}
    \label{fig:BB_fits_grid}
\end{figure*}


\begin{figure}
    \centering
    \includegraphics[width=1\columnwidth]{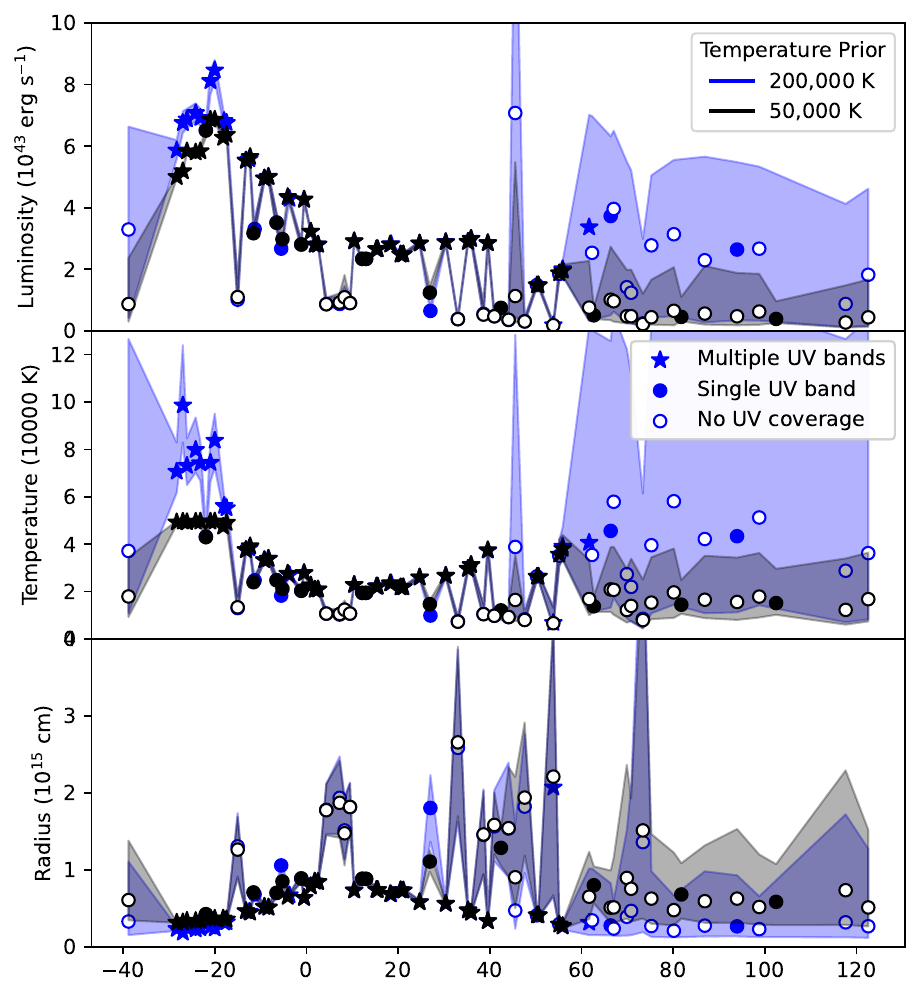}
    \caption{Blackbody parameters using different upper bounds for the temperature prior and no interpolation of filters used. Fitting only optical part of the SED where UV coverage is missing results in significantly lower temperatures and luminosities.}
    \label{fig:BB_fits_high}
\end{figure}



\begin{figure*}
    \centering

    \begin{subfigure}{2\columnwidth}
        \centering
        \includegraphics[width=1\columnwidth]{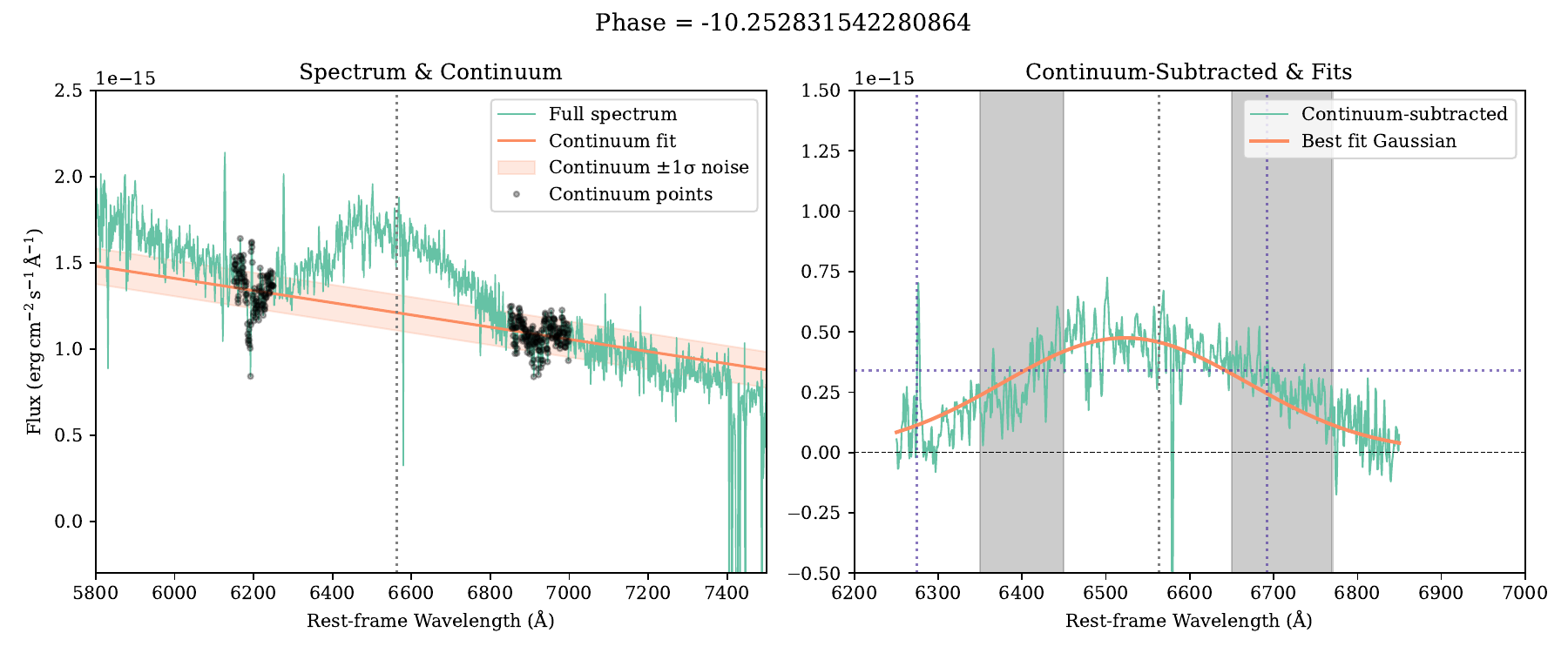}
    \end{subfigure}

    \smallskip

    \begin{subfigure}{2\columnwidth}
    \centering
        \includegraphics[width=1\columnwidth]{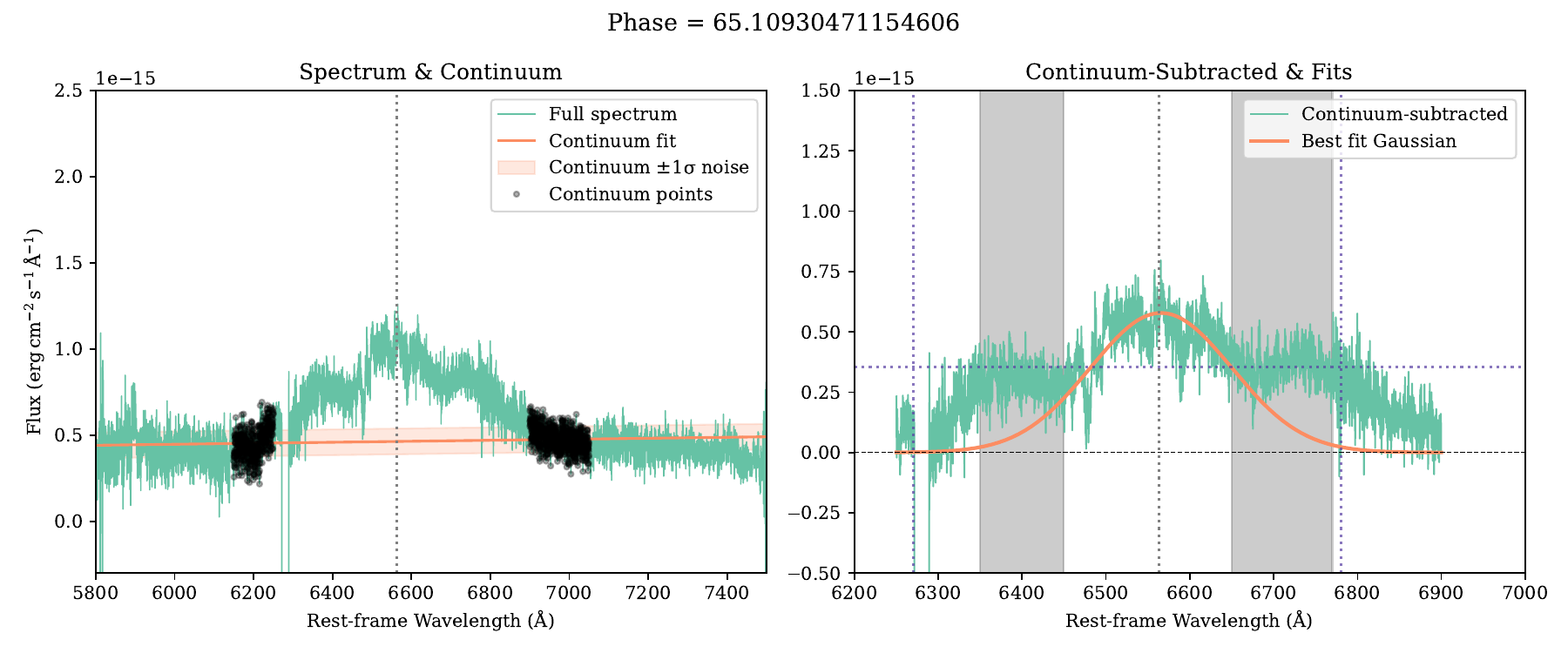}
    \end{subfigure}

    \caption{Examples of the local continuum subtracting procedure and Gaussian fitting around the H$\alpha$ profile. \textit{Left:} Local continuum subtraction around the H$\alpha$ line. The black points are randomly sampled from a defined continuum region and used to create a median line of best fit for the underlying continuum. Regions were selected to avoid interference from telluric features. \textit{Right:} Best fitting Gaussian profiles after continuum subtraction. At early times, a single Gaussian provides a good description of the H$\alpha$ profile. At intermediate epochs, clear excess emission develops on both sides of the line. These regions, shown by the grey shaded areas, were therefore masked when fitting the central Gaussian component.}
    \label{fig:Halpha_cont}

\end{figure*}


\begin{figure}
    \centering
    \includegraphics[width=1\columnwidth]{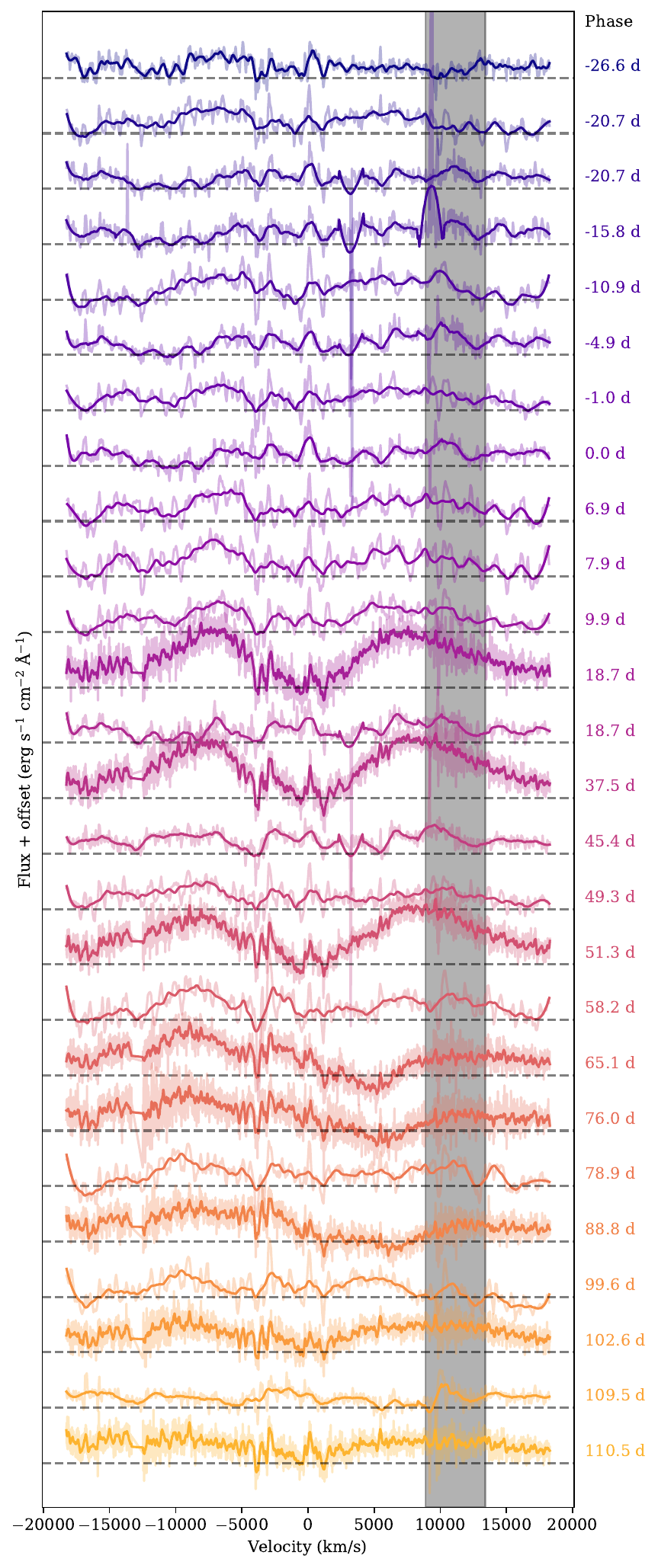}
    \caption{H$\alpha$ excess for a subset of spectra similar to Figure \ref{fig:Halpha_excess}. There may be evidence for an accretion disk as early as $\sim-10$ days with an excess potentially visible on either side of the rest wavelength. The grey shaded region indicates the location of a telluric feature.}
    \label{fig:Halpha_excess_all}
\end{figure}

\begin{figure*}
    \centering
    \includegraphics[width=2\columnwidth]{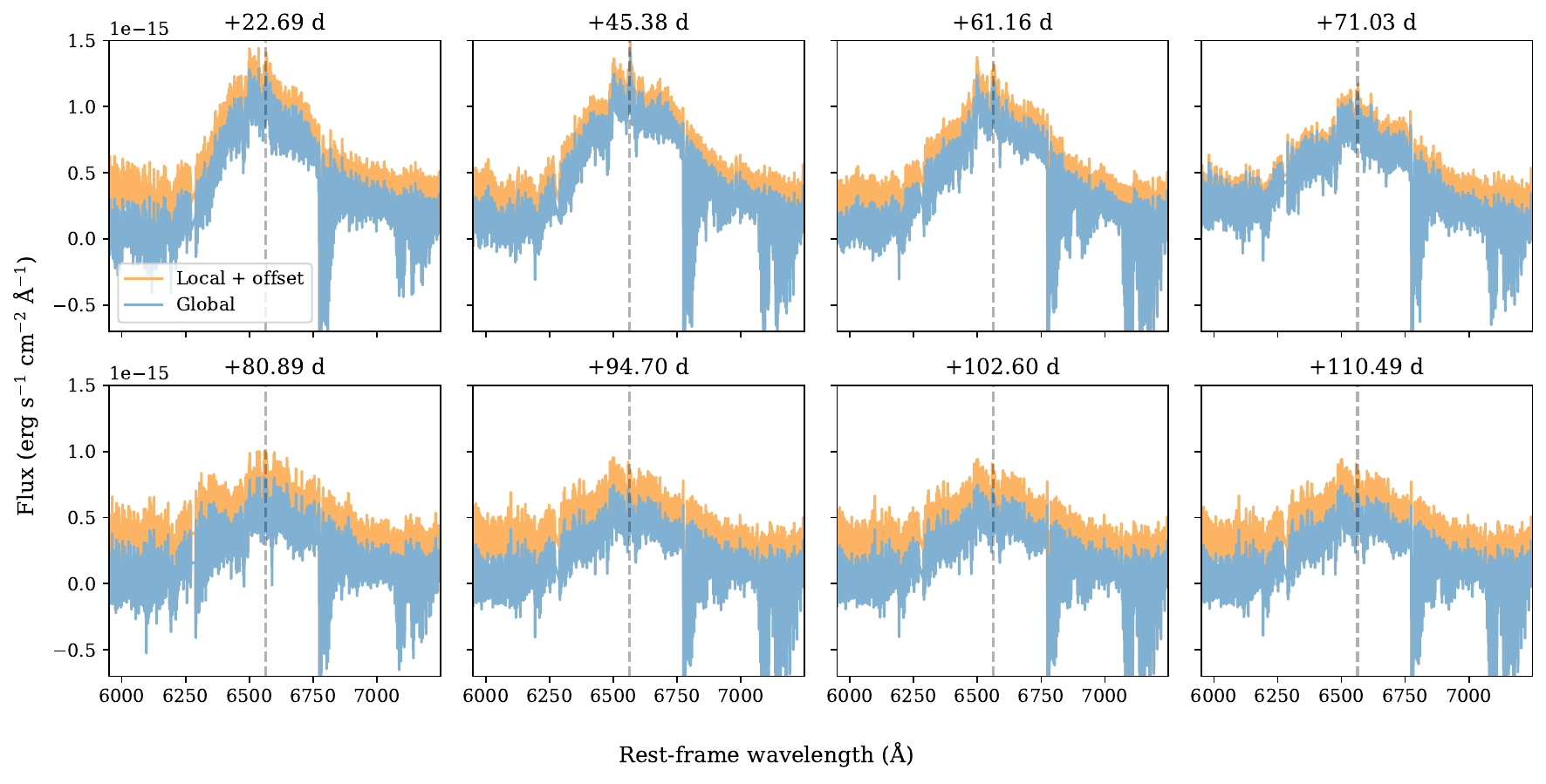}
    \caption{Global and local continuum subtraction around the H$\alpha$ profile for the X-Shooter spectra as described in Section \ref{sec:disk}. We can see both methods produce very similar line profiles.}
    \label{fig:cont_sub}
\end{figure*}

\begin{figure*}
    \centering

    \begin{subfigure}{1\columnwidth}
        \centering
        \includegraphics[width=1\linewidth]{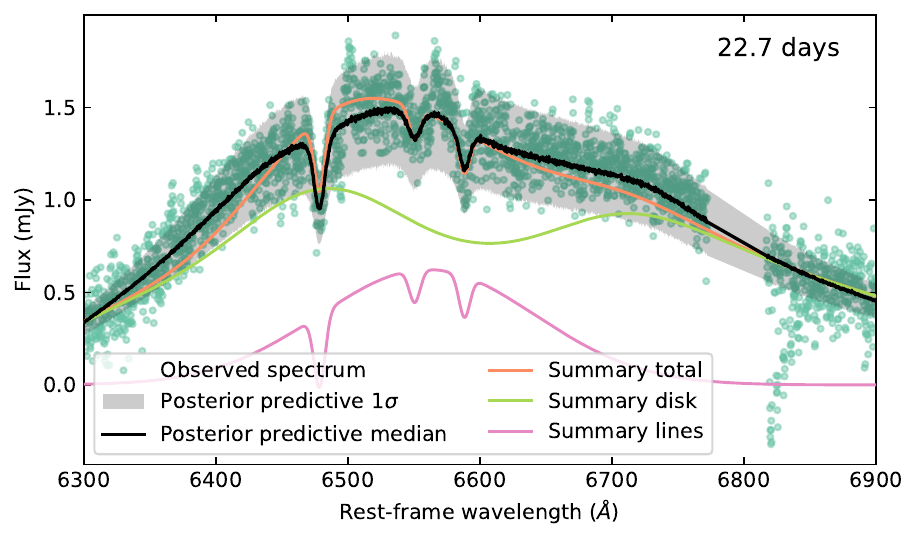}
    \end{subfigure}
    \hfill
    \begin{subfigure}{1\columnwidth}
    \centering
        \includegraphics[width=1\linewidth]{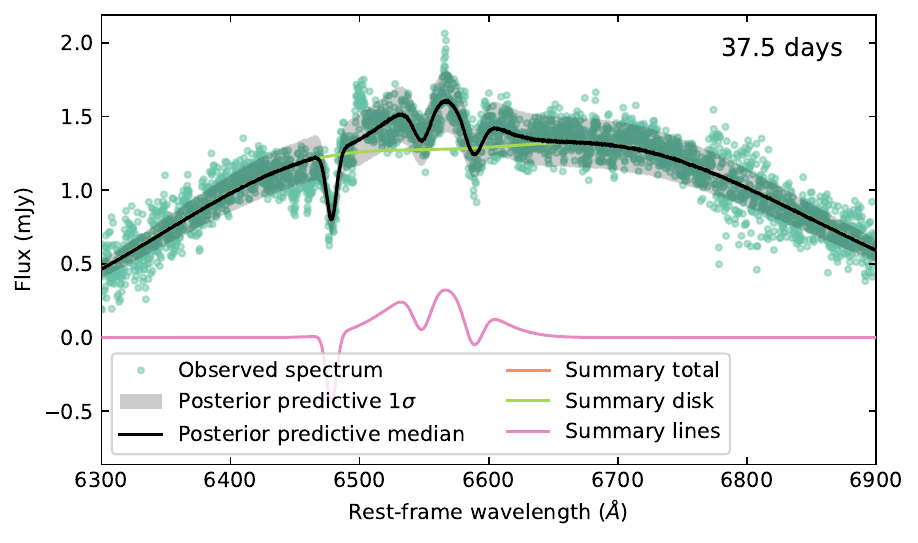}
    \end{subfigure}

    \smallskip

    \begin{subfigure}{1\columnwidth}
    \centering
        \includegraphics[width=1\linewidth]{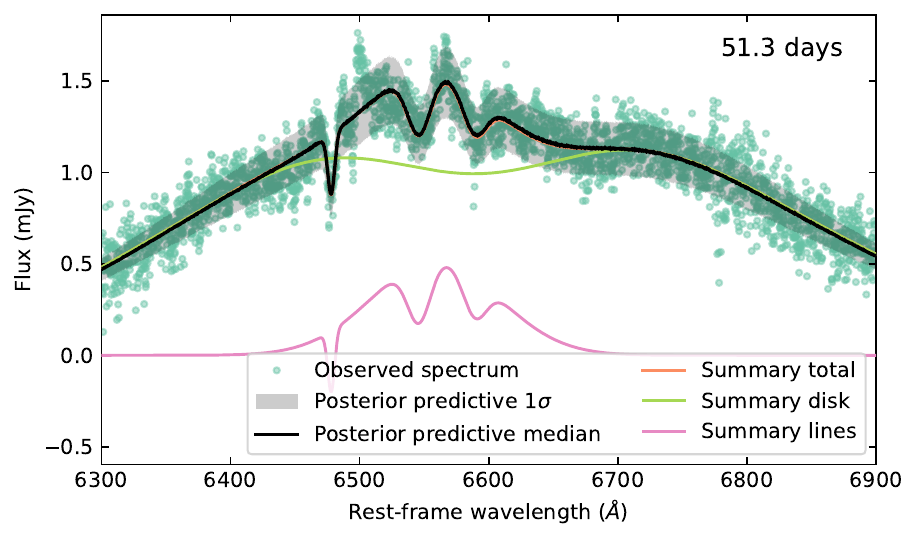}
    \end{subfigure}
    \hfill
    \begin{subfigure}{1\columnwidth}
    \centering
        \includegraphics[width=1\linewidth]{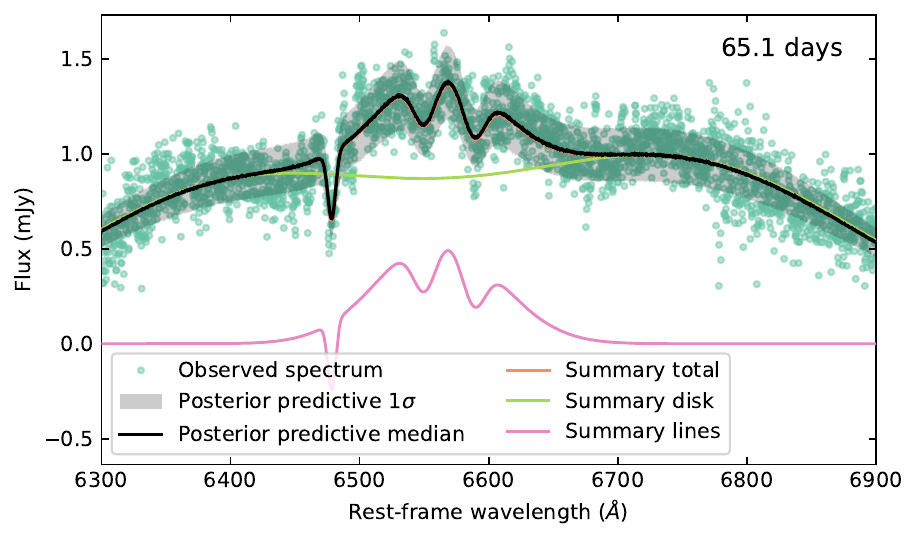}
    \end{subfigure}

    \smallskip

    \begin{subfigure}{1\columnwidth}
    \centering
        \includegraphics[width=1\linewidth]{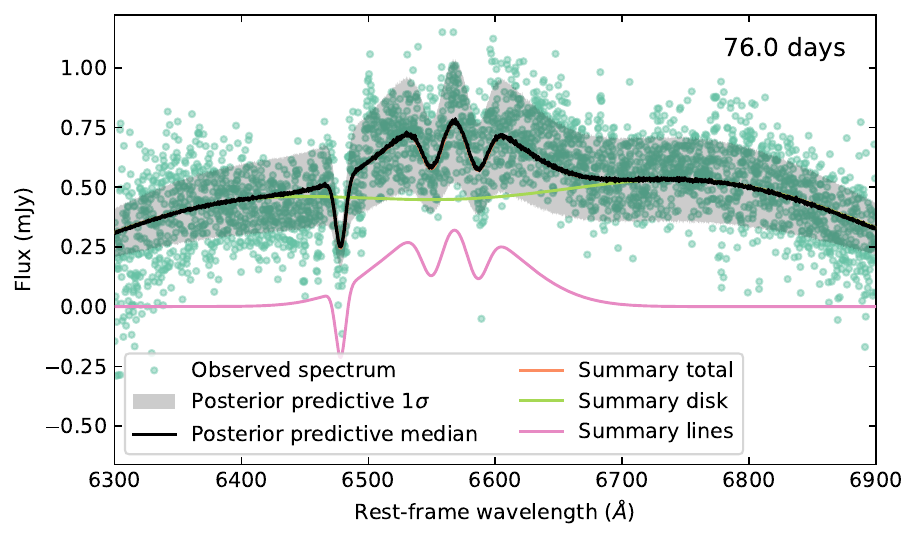}
    \end{subfigure}
    \hfill
    \begin{subfigure}{1\columnwidth}
    \centering
        \includegraphics[width=1\linewidth]{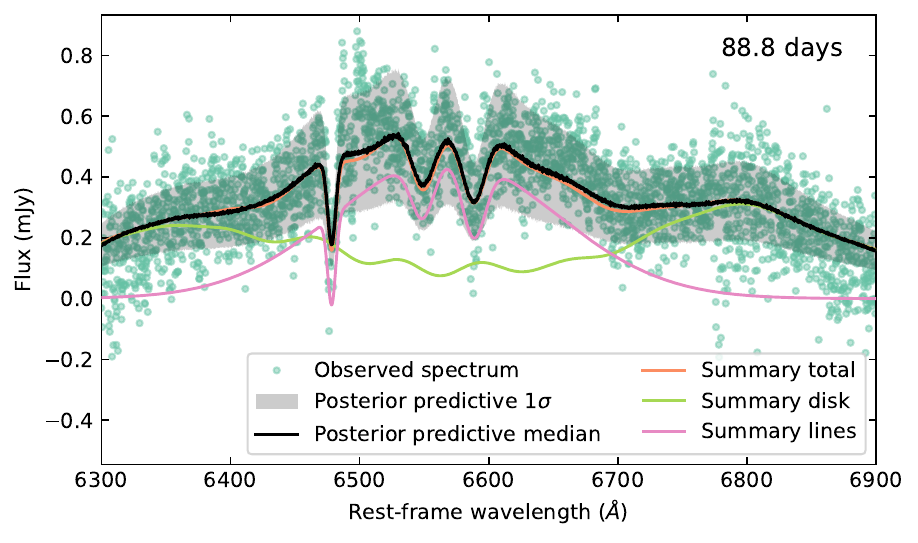}
    \end{subfigure}

    \smallskip

    \begin{subfigure}{1\columnwidth}
    \centering
        \includegraphics[width=1\linewidth]{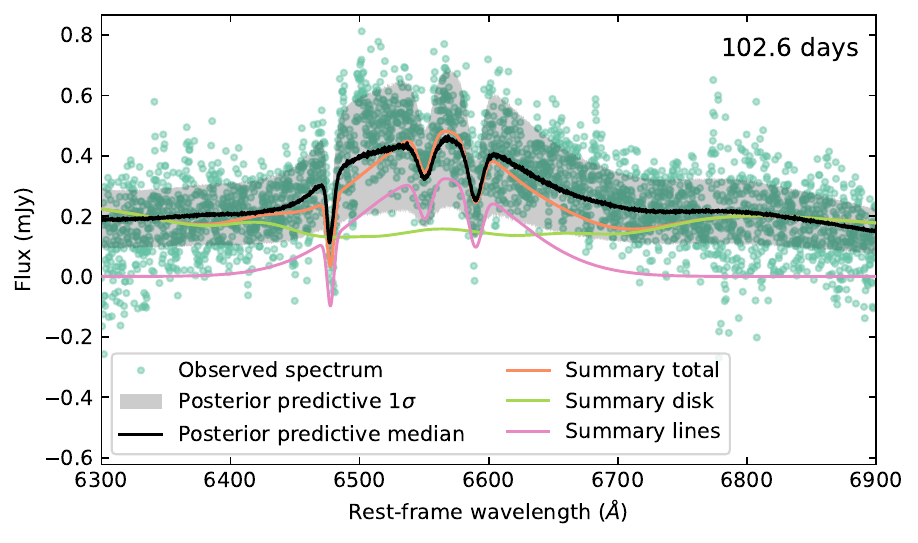}
    \end{subfigure}
    \hfill
    \begin{subfigure}{1\columnwidth}
    \centering
        \includegraphics[width=1\linewidth]{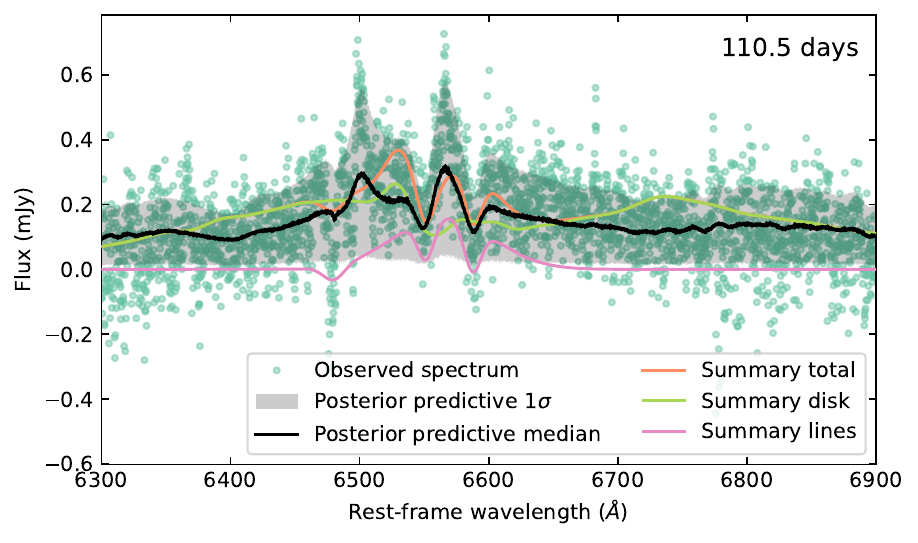}
    \end{subfigure}

    \caption{Spectral disk models for the H$\alpha$ profile in the X-Shooter spectra. The fits include a broad H$\alpha$ component, a disk component, and narrow absorption components from N II $\lambda\lambda$6548, 6583, and Fe $\lambda$6481. The teal points show the observed spectrum used in the fit, while the solid black curve and grey shaded region represent the median posterior predictive model and the corresponding 16th–84th percentile credible interval, respectively. The coloured curves show the individual model components evaluated using the posterior median parameter values.}
    \label{fig:disk_fits}
\end{figure*}

\begin{table*}
    \centering
    \caption{Priors adopted for the \texttt{feadme} modelling of the 
    H$\alpha$ region. Uniform priors are denoted by $\mathcal{U}(a,b)$, log-uniform priors by $\log\mathcal{U}(a,b)$, and Gaussian priors by $\mathcal{N}(\mu,\sigma)$. Shared priors are indicated.}
    \label{tab:feadme_priors}
    \begin{tabular}{llll}
        \hline
        Component & Parameter & Prior type & Prior \\
        \hline

        H$\alpha$ disk
        & Apocenter angle, $\phi$ & Uniform & $\mathcal{U}(0,2\pi)$ rad \\
        & Line centre, $\lambda_0$ & Uniform & $\mathcal{U}(6557.819,6567.819)$ \AA \\
        & Radial width, $\Delta r$ & Log-uniform & $\log\mathcal{U}(100,10000)\,r_{\rm g}$ \\
        & Eccentricity, $e$ & Uniform & $\mathcal{U}(0,1)$ \\
        & Inclination, $i$ & Uniform & $\mathcal{U}(0,\pi/2)$ rad \\
        & Inner radius, $r_{\rm in}$ & Log-uniform & $\log\mathcal{U}(100,5000)\,r_{\rm g}$ \\
        & Emissivity index, $q$ & Gaussian & $\mathcal{N}(2,1)$, $0.5<q<4$ \\
        & Scale & Uniform & $\mathcal{U}(0,1.601)$ \\
        & Velocity dispersion, $\sigma$ & Log-uniform & $\log\mathcal{U}(200,10000)$ km\,s$^{-1}$ \\

        \hline

        Broad H$\alpha$
        & Amplitude & Uniform & $\mathcal{U}(0,1.601)$ \\
        & Line centre, $\lambda_0$ & Shared & H$\alpha$ disk \\
        & Velocity width & Log-uniform & $\log\mathcal{U}(1000,15000)$ km\,s$^{-1}$ \\

        \hline

        [N\,II] $\lambda6583$ absorption
        & Amplitude & Uniform & $\mathcal{U}(-0.5,0.05)$ \\
        & Line centre, $\lambda_0$ & Uniform & $\mathcal{U}(6573,6593)$ \AA \\
        & Velocity width & Log-uniform & $\log\mathcal{U}(30,1000)$ km\,s$^{-1}$ \\

        \hline

        [N\,II] $\lambda6548$ absorption
        & Amplitude & Uniform & $\mathcal{U}(-0.5,0.05)$ \\
        & Line centre, $\lambda_0$ & Uniform & $\mathcal{U}(6538,6558)$ \AA \\
        & Velocity width & Shared & [N\,II] $\lambda6583$ \\

        \hline

        6480\,\AA\ absorption
        & Amplitude & Uniform & $\mathcal{U}(-0.5,0.05)$ \\
        & Line centre, $\lambda_0$ & Uniform & $\mathcal{U}(6475,6485)$ \AA \\
        & Velocity width & Log-uniform & $\log\mathcal{U}(30,1000)$ km\,s$^{-1}$ \\

        \hline
    \end{tabular}
\end{table*}



\bibliography{lib}{}
\bibliographystyle{aasjournalv7}



\end{document}